\pdfoutput=1

\documentclass[12pt,a4paper]{article}
\usepackage{ifthen} 
\newboolean{pdflatex}
\setboolean{pdflatex}{true} 

\newboolean{articletitles}
\setboolean{articletitles}{true} 

\newboolean{uprightparticles}
\setboolean{uprightparticles}{false} 

\newboolean{inbibliography}
\setboolean{inbibliography}{false} 

\usepackage{afterpage}

\usepackage{microtype}
\usepackage{lineno}  
\usepackage{xspace} 
\usepackage{caption} 

\usepackage{graphicx}  
\usepackage{color}
\usepackage{colortbl}
\graphicspath{{./figs/}} 

\usepackage{amsmath} 
\usepackage{amssymb}
\usepackage{amsfonts}
\usepackage{upgreek} 

\newcommand*\patchAmsMathEnvironmentForLineno[1]{%
\expandafter\let\csname old#1\expandafter\endcsname\csname #1\endcsname
\expandafter\let\csname oldend#1\expandafter\endcsname\csname
end#1\endcsname
 \renewenvironment{#1}%
   {\linenomath\csname old#1\endcsname}%
   {\csname oldend#1\endcsname\endlinenomath}%
}
\newcommand*\patchBothAmsMathEnvironmentsForLineno[1]{%
  \patchAmsMathEnvironmentForLineno{#1}%
  \patchAmsMathEnvironmentForLineno{#1*}%
}
\AtBeginDocument{%
\patchBothAmsMathEnvironmentsForLineno{equation}%
\patchBothAmsMathEnvironmentsForLineno{align}%
\patchBothAmsMathEnvironmentsForLineno{flalign}%
\patchBothAmsMathEnvironmentsForLineno{alignat}%
\patchBothAmsMathEnvironmentsForLineno{gather}%
\patchBothAmsMathEnvironmentsForLineno{multline}%
\patchBothAmsMathEnvironmentsForLineno{eqnarray}%
}

\usepackage{hyperref}    
\usepackage[all]{hypcap} 

\def\lhcb {\mbox{LHCb}\xspace}

\def\MagUp {\mbox{\em Mag\kern -0.05em Up}\xspace}

\ifthenelse{\boolean{uprightparticles}}%
{

 \def\Pmu         {\ensuremath{\upmu}\xspace}

 \def\Ppsi        {\ensuremath{\uppsi}\xspace}

 \def\PDelta      {\ensuremath{\Delta}\xspace}                 
 \def\PXi      {\ensuremath{\Xi}\xspace}                 
 \def\PLambda      {\ensuremath{\Lambda}\xspace}                 
 \def\PSigma      {\ensuremath{\Sigma}\xspace}                 
 \def\POmega      {\ensuremath{\Omega}\xspace}                 
 \def\PUpsilon      {\ensuremath{\Upsilon}\xspace}                 
 
 \def\PB      {\ensuremath{\mathrm{B}}\xspace}                 
                  
 \def\PD      {\ensuremath{\mathrm{D}}\xspace}

 \def\PJ      {\ensuremath{\mathrm{J}}\xspace}                 
 \def\PK      {\ensuremath{\mathrm{K}}\xspace}

 \def\Pb      {\ensuremath{\mathrm{b}}\xspace}

 \def\Pi      {\ensuremath{\mathrm{i}}\xspace}

 \def\Ps      {\ensuremath{\mathrm{s}}\xspace}

}
{

 \def\Pmu         {\ensuremath{\mu}\xspace}

 \def\Ppsi        {\ensuremath{\psi}\xspace}                 
                  
 \mathchardef\PDelta="7101
 \mathchardef\PXi="7104
 \mathchardef\PLambda="7103
 \mathchardef\PSigma="7106
 \mathchardef\POmega="710A
 \mathchardef\PUpsilon="7107
                  
 \def\PB      {\ensuremath{B}\xspace}                 
                  
 \def\PD      {\ensuremath{D}\xspace}

 \def\PJ      {\ensuremath{J}\xspace}                 
 \def\PK      {\ensuremath{K}\xspace}

 \def\Pb      {\ensuremath{b}\xspace}

 \def\Pi      {\ensuremath{i}\xspace}

 \def\Ps      {\ensuremath{s}\xspace}

}

\makeatletter
\ifcase \@ptsize \relax
  \newcommand{\miniscule}{\@setfontsize\miniscule{4}{5}}
\or
  \newcommand{\miniscule}{\@setfontsize\miniscule{5}{6}}
\or
  \newcommand{\miniscule}{\@setfontsize\miniscule{5}{6}}
\fi
\makeatother

\DeclareRobustCommand{\optbar}[1]{\shortstack{{\miniscule (\rule[.5ex]{1.25em}{.18mm})}
  \\ [-.7ex] $#1$}}

\def\mup        {{\ensuremath{\Pmu^+}}\xspace}
\def\mumu       {{\ensuremath{\Pmu^+\Pmu^-}}\xspace}

\def\H      {{\ensuremath{\PH^0}}\xspace}

\def\squark    {{\ensuremath{\Ps}}\xspace}

\def\bquark    {{\ensuremath{\Pb}}\xspace}

  \def\Kbar    {{\kern 0.2em\overline{\kern -0.2em \PK}{}}\xspace}

\def\KorKbar    {\kern 0.18em\optbar{\kern -0.18em K}{}\xspace}

  \def\Dbar    {{\kern 0.2em\overline{\kern -0.2em \PD}{}}\xspace}
\def\D       {{\ensuremath{\PD}}\xspace}

\def\DorDbar    {\kern 0.18em\optbar{\kern -0.18em D}{}\xspace}
\def\Dz      {{\ensuremath{\D^0}}\xspace}
\def\Dzb     {{\ensuremath{\Dbar{}^0}}\xspace}

\def\B       {{\ensuremath{\PB}}\xspace}
\def\Bbar    {{\ensuremath{\kern 0.18em\overline{\kern -0.18em \PB}{}}}\xspace}

\def\BorBbar    {\kern 0.18em\optbar{\kern -0.18em B}{}\xspace}

\def\Bzb     {{\ensuremath{\Bbar{}^0}}\xspace}

\def\Bs      {{\ensuremath{\B^0_\squark}}\xspace}
\def\Bsb     {{\ensuremath{\Bbar{}^0_\squark}}\xspace}
\def\Bdb     {{\ensuremath{\Bbar{}^0}}\xspace}

\def\jpsi     {{\ensuremath{{\PJ\mskip -3mu/\mskip -2mu\Ppsi\mskip 2mu}}}\xspace}

  \def\Y#1S{\ensuremath{\PUpsilon{(#1S)}}\xspace}

\def\Xires       {{\ensuremath{\PXi}}\xspace}

\def\Lz          {{\ensuremath{\PLambda}}\xspace}
\def\Lbar        {{\ensuremath{\kern 0.1em\overline{\kern -0.1em\PLambda}}}\xspace}
\def\LorLbar    {\kern 0.18em\optbar{\kern -0.18em \PLambda}{}\xspace}

\def\Lb      {{\ensuremath{\Lz^0_\bquark}}\xspace}
\def\Lbbar   {{\ensuremath{\Lbar{}^0_\bquark}}\xspace}

\def\Xib     {{\ensuremath{\Xires_\bquark}}\xspace}

\newcommand{\decay}[2]{\ensuremath{#1\!\to #2}\xspace}         

\def\to                 {\ensuremath{\rightarrow}\xspace}

\def\CP                {{\ensuremath{C\!P}}\xspace}

\def\AT#1     {\ensuremath{A_{\mathrm{T}}^{#1}}\xspace}           

\def\C#1      {\ensuremath{\mathcal{C}_{#1}}\xspace}                       
\def\Cp#1     {\ensuremath{\mathcal{C}_{#1}^{'}}\xspace}                    
\def\Ceff#1   {\ensuremath{\mathcal{C}_{#1}^{\mathrm{(eff)}}}\xspace}        
\def\Cpeff#1  {\ensuremath{\mathcal{C}_{#1}^{'\mathrm{(eff)}}}\xspace}       
\def\Ope#1    {\ensuremath{\mathcal{O}_{#1}}\xspace}                       
\def\Opep#1   {\ensuremath{\mathcal{O}_{#1}^{'}}\xspace}                    

\newcommand{\tev}{\ifthenelse{\boolean{inbibliography}}{\ensuremath{~T\kern -0.05em eV}\xspace}{\ensuremath{\mathrm{\,Te\kern -0.1em V}}}\xspace}
\newcommand{\gev}{\ensuremath{\mathrm{\,Ge\kern -0.1em V}}\xspace}
\newcommand{\mev}{\ensuremath{\mathrm{\,Me\kern -0.1em V}}\xspace}
\newcommand{\kev}{\ensuremath{\mathrm{\,ke\kern -0.1em V}}\xspace}
\newcommand{\ev}{\ensuremath{\mathrm{\,e\kern -0.1em V}}\xspace}
\newcommand{\gevc}{\ensuremath{{\mathrm{\,Ge\kern -0.1em V\!/}c}}\xspace}
\newcommand{\mevc}{\ensuremath{{\mathrm{\,Me\kern -0.1em V\!/}c}}\xspace}
\newcommand{\gevcc}{\ensuremath{{\mathrm{\,Ge\kern -0.1em V\!/}c^2}}\xspace}
\newcommand{\gevgevcccc}{\ensuremath{{\mathrm{\,Ge\kern -0.1em V^2\!/}c^4}}\xspace}
\newcommand{\mevcc}{\ensuremath{{\mathrm{\,Me\kern -0.1em V\!/}c^2}}\xspace}

\def\invfb   {\ensuremath{\mbox{\,fb}^{-1}}\xspace}

\def\gsim{{~\raise.15em\hbox{$>$}\kern-.85em
          \lower.35em\hbox{$\sim$}~}\xspace}
\def\lsim{{~\raise.15em\hbox{$<$}\kern-.85em
          \lower.35em\hbox{$\sim$}~}\xspace}

\def\PDF {PDF\xspace}

\def\pt         {\mbox{$p_{\rm T}$}\xspace}

\def\geant      {\mbox{\textsc{Geant4}}\xspace}

\def\pythia     {\mbox{\textsc{Pythia}}\xspace}

\def\tell1  {TELL1\xspace}
\def\ukl1   {UKL1\xspace}

\newcommand{\eg}{\mbox{\itshape e.g.}\xspace}
\newcommand{\ie}{\mbox{\itshape i.e.}\xspace}

\usepackage{cite} 
\usepackage{mciteplus}

\usepackage{longtable} 

\def\ZP{P_c}
\def\LambdaStar{{\Lz^*}}
\def\LambdaStarn{{\Lz^*_{\!n}}}
\def\H{{\cal H}}
\def\F#1{\{#1\}}
\def\BA#1#2#3{{#1}_{{#2}}^{\,\,\F{\!#3\!}}}
\def\Pars{\overrightarrow{\omega}}

\def\PDF{\mathcal{P}}
\def\Mat{\mathcal{M}}
\def\Like{\mathcal{L}}
\def\Dll{\Delta(\!-2\ln\Like)}

\begin{document}
\pagestyle{plain} 
\setcounter{page}{1}
\pagenumbering{arabic}


\begin{titlepage}
\pagenumbering{roman}

\vspace*{-1.5cm}
\centerline{\large EUROPEAN ORGANIZATION FOR NUCLEAR RESEARCH (CERN)}
\vspace*{1.5cm}
\hspace*{-0.5cm}
\begin{tabular*}{\linewidth}{lc@{\extracolsep{\fill}}r}
\vspace*{-4.3cm}\mbox{\!\!\!\includegraphics[width=.14\textwidth]{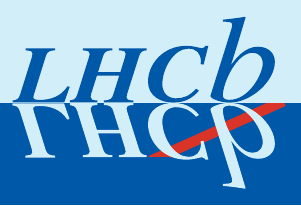}} & &\\
& &\\
& &\\
& &\\
& &\\
& &\\
 & & CERN-PH-EP-2015-153 \\  
 & & LHCb-PAPER-2015-029 \\  
 & & July 13, 2015 \\ 
 & & \\
\end{tabular*}

\vspace*{2.8cm}

{\bf\huge
\begin{center}
\boldmath Observation of $\jpsi p$ resonances consistent with pentaquark states  in  ${\Lb\to J/\psi K^-p}$ decays
\end{center}
}
\vspace*{1.6cm}

\begin{center}
The LHCb collaboration\footnote{Authors are listed at the end of this Letter.}
\end{center}

\vspace{\fill}
\begin{abstract}
  \noindent
Observations of exotic structures in the $\jpsi p$ channel, which we refer to as charmonium-pentaquark states, in $\Lb\to\jpsi K^- p$ decays are presented. The data sample corresponds to an integrated luminosity of 3~\invfb acquired with the LHCb detector from 7 and 8~TeV $pp$ collisions. An amplitude analysis of the three-body final-state reproduces the two-body mass and angular distributions. To obtain a satisfactory fit of the  structures seen in the $\jpsi p$ mass spectrum, it is necessary to include two Breit-Wigner amplitudes that each describe a resonant state. 
The significance of each of these resonances is more than 9 standard deviations. One has a mass of $4380\pm 8\pm 29$~MeV and a width of $205\pm 18\pm 86$ MeV, while the second is narrower, with a mass of $4449.8\pm 1.7\pm 2.5$~MeV and a width of $39\pm 5\pm 19$ MeV.  The preferred $J^P$ assignments
are of opposite parity,  with one state having spin 3/2 and the other 5/2.
\end{abstract}

\vspace*{1.0cm}
\begin{center}
  Submitted to Phys.~Rev.~Lett. 
\end{center}

\vspace{\fill}

{\footnotesize 
\centerline{\copyright~CERN on behalf of the \lhcb collaboration, license \href{http://creativecommons.org/licenses/by/4.0/}{CC-BY-4.0}.}}
\vspace*{2mm}

\end{titlepage}

\pagestyle{empty}  


\newpage
\setcounter{page}{2}
\mbox{~}

\cleardoublepage



\pagestyle{plain} 
\setcounter{page}{1}
\pagenumbering{arabic}


%
\section*{Introduction and summary}
The prospect of hadrons with more than the minimal quark content ($q\overline{q}$ or $qqq$) was proposed by Gell-Mann in 1964 \cite{GellMann:1964nj} and Zweig \cite{Zweig:1964}, followed
by a quantitative model for two quarks plus two antiquarks developed by Jaffe in 1976 \cite{Jaffe:1976ig}. The idea was expanded upon  \cite{Strottman:1979qu,*Hogaasen:1978jw} to include  baryons composed of four quarks plus one antiquark; the name pentaquark was coined by Lipkin \cite{Lipkin:1987sk}.
Past claimed observations of pentaquark states have been shown to be spurious \cite{Hicks:2012zz}, although there is at least one viable tetraquark candidate, the $Z(4430)^+$  observed in $\Bzb\to\psi'K^-\pi^+$ decays \cite{Choi:2007wga,Chilikin:2013tch,Aaij:2014jqa},
implying that the existence of pentaquark baryon states would not be surprising. States that decay into charmonium may have particularly distinctive signatures \cite{Li:2014gra}.

\begin{figure}[h]
\begin{center}
\includegraphics[width=0.99\textwidth]{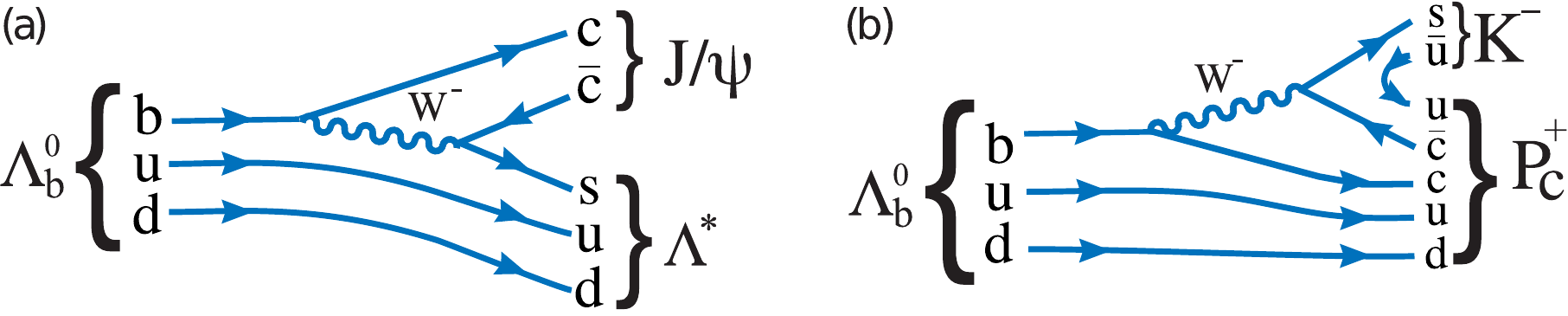}
\end{center}
\vskip -0.3cm
\caption{Feynman diagrams for (a) $\Lb\to \jpsi \Lz^*$ and (b) $\Lb\to P_c^+ K^-$ decay.}
\label{Feynman-Pc}
\end{figure}

\begin{figure}[b]
\vskip -0.7cm
\begin{center}
\includegraphics[width=0.42\textwidth]{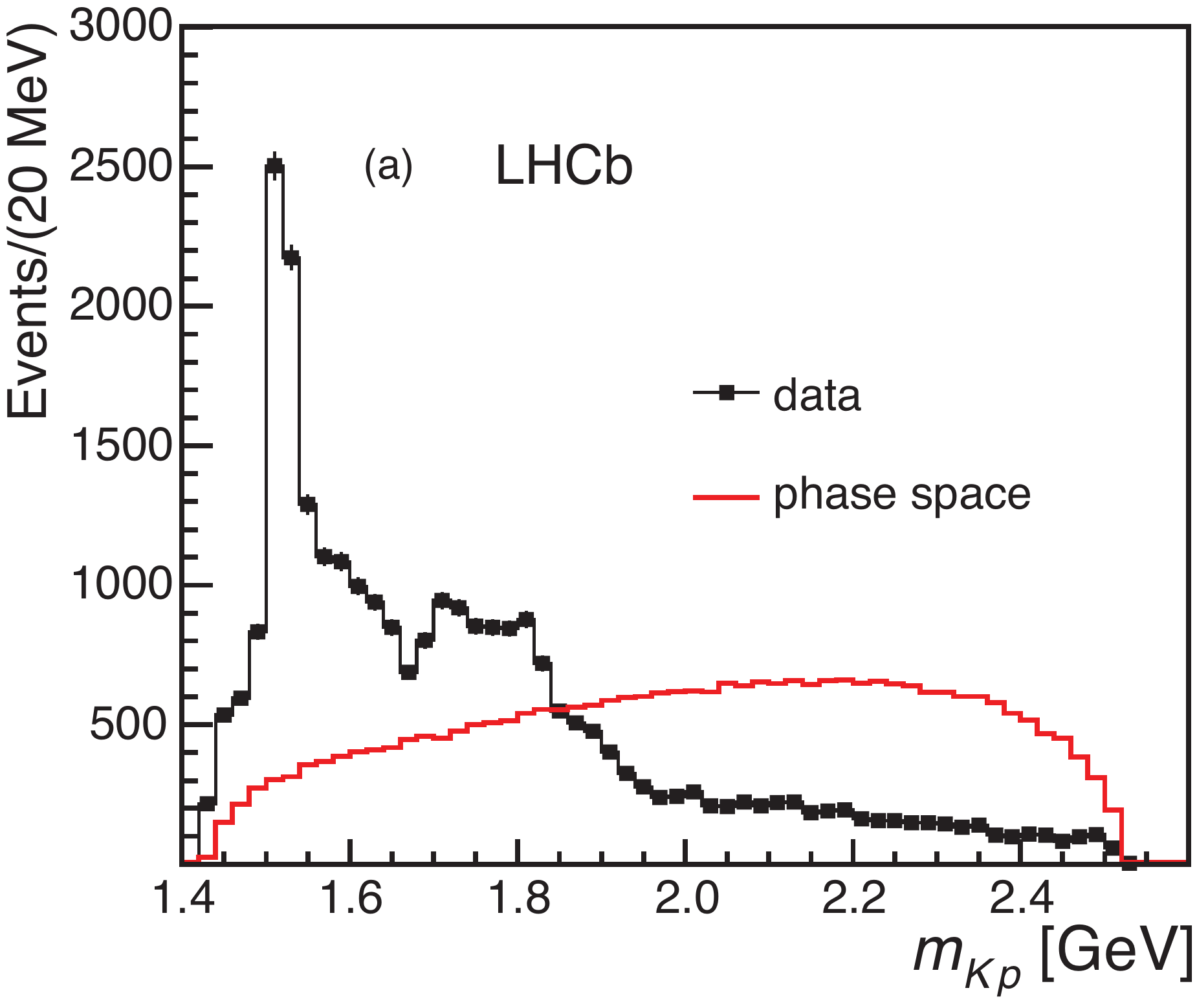}\hspace*{0.99cm}\includegraphics[width=0.42\textwidth]{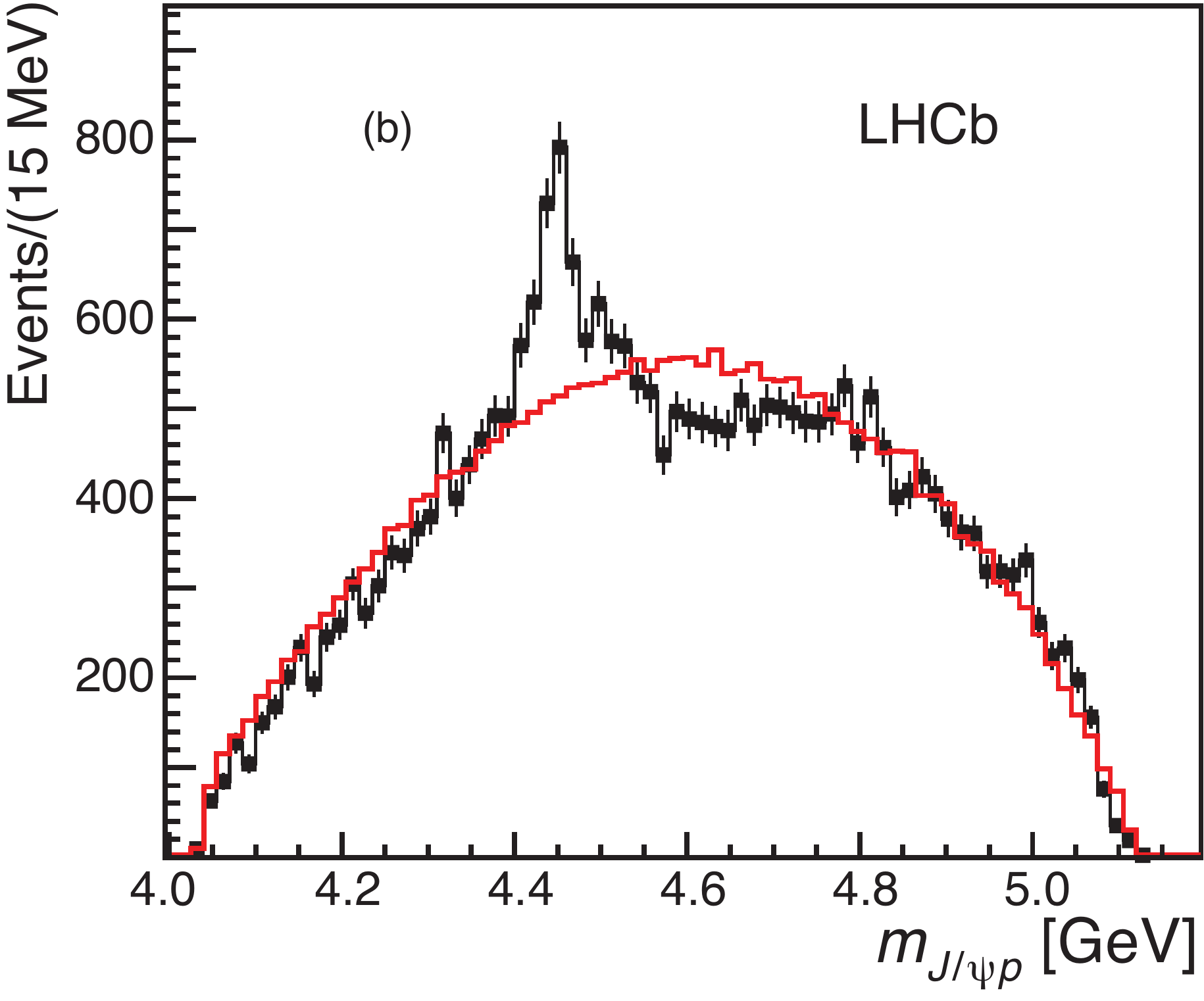}
\end{center}
\vskip -0.2cm
\caption{ Invariant mass of (a) $K^-p$  and (b) $\jpsi p$ combinations from $\Lb\to\jpsi K^-p$ decays. The solid (red) curve is the expectation from phase space. The background has been subtracted.}
\label{mpk-mjpsi}
\end{figure}

Large yields of 
 $\Lb\to\jpsi K^- p$ decays are available at LHCb and have been used for the precise measurement of the $\Lb$ lifetime \cite{Aaij:2014zyy,*Aaij:2013oha}. (In this Letter mention of a particular mode implies use of its charge conjugate as well.) This decay can proceed by the diagram shown in Fig.~\ref{Feynman-Pc}(a), and is expected to be dominated by $\Lz^*\to K^-p$ resonances, as are evident in our data shown in Fig.~\ref{mpk-mjpsi}(a). It could also have exotic contributions, as indicated by the diagram in Fig.~\ref{Feynman-Pc}(b), that could result in resonant structures in the $\jpsi p$ mass spectrum shown in Fig.~\ref{mpk-mjpsi}(b).
\clearpage

In practice resonances decaying strongly into $\jpsi p$ must have a minimal quark content of $c\overline{c}uud$, and thus are charmonium-pentaquarks;
we label such states $P_c^+$, irrespective of the internal binding mechanism.
In order to ascertain if the structures seen in Fig.~\ref{mpk-mjpsi}(b)  are resonant in nature and not due to reflections generated by the $\Lz^*$ states, it is necessary to perform a full amplitude analysis, allowing for interference effects between both decay sequences.

\begin{figure}[b]
\begin{center}
\includegraphics[width=0.49\textwidth]{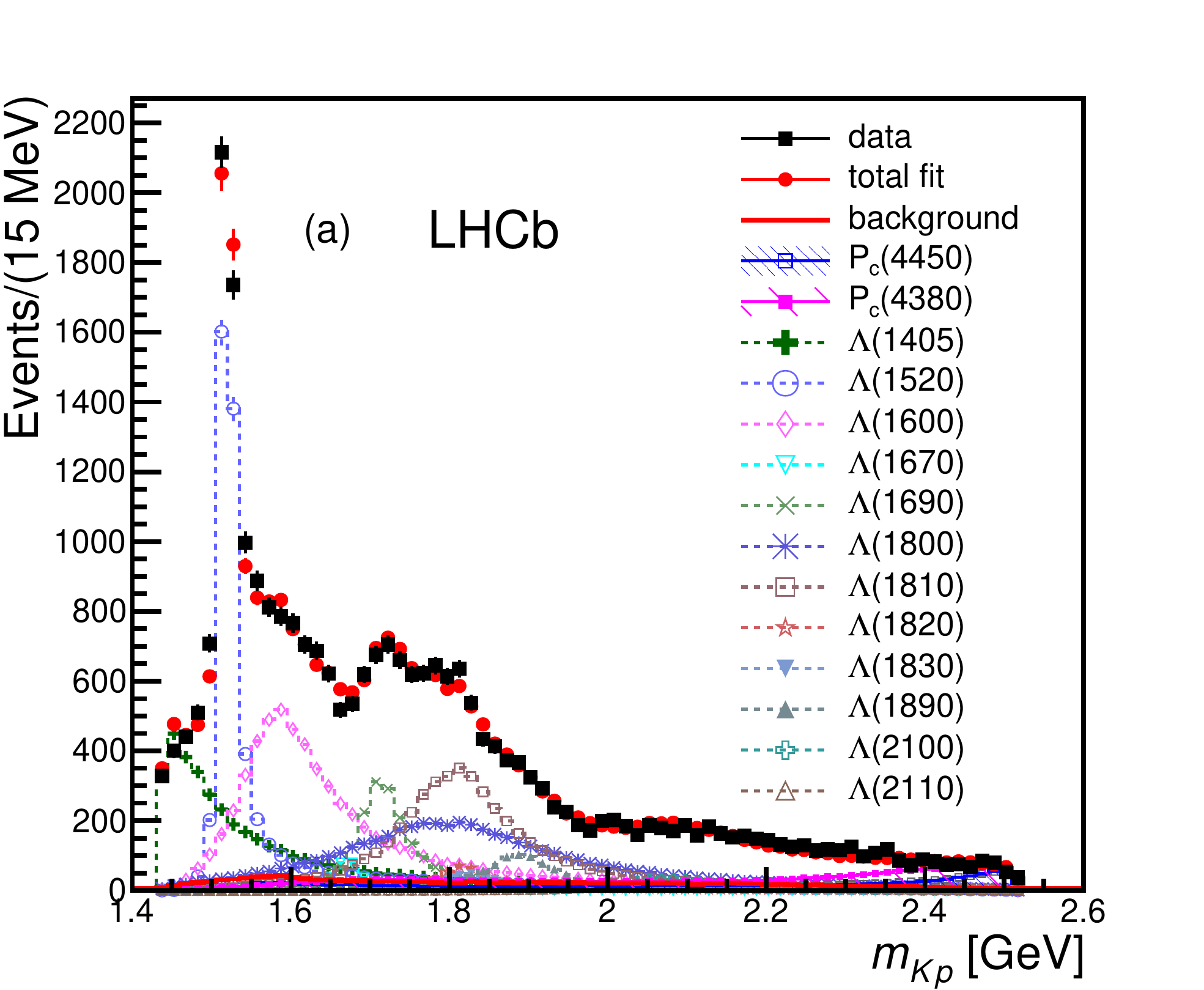}\includegraphics[width=0.49\textwidth]{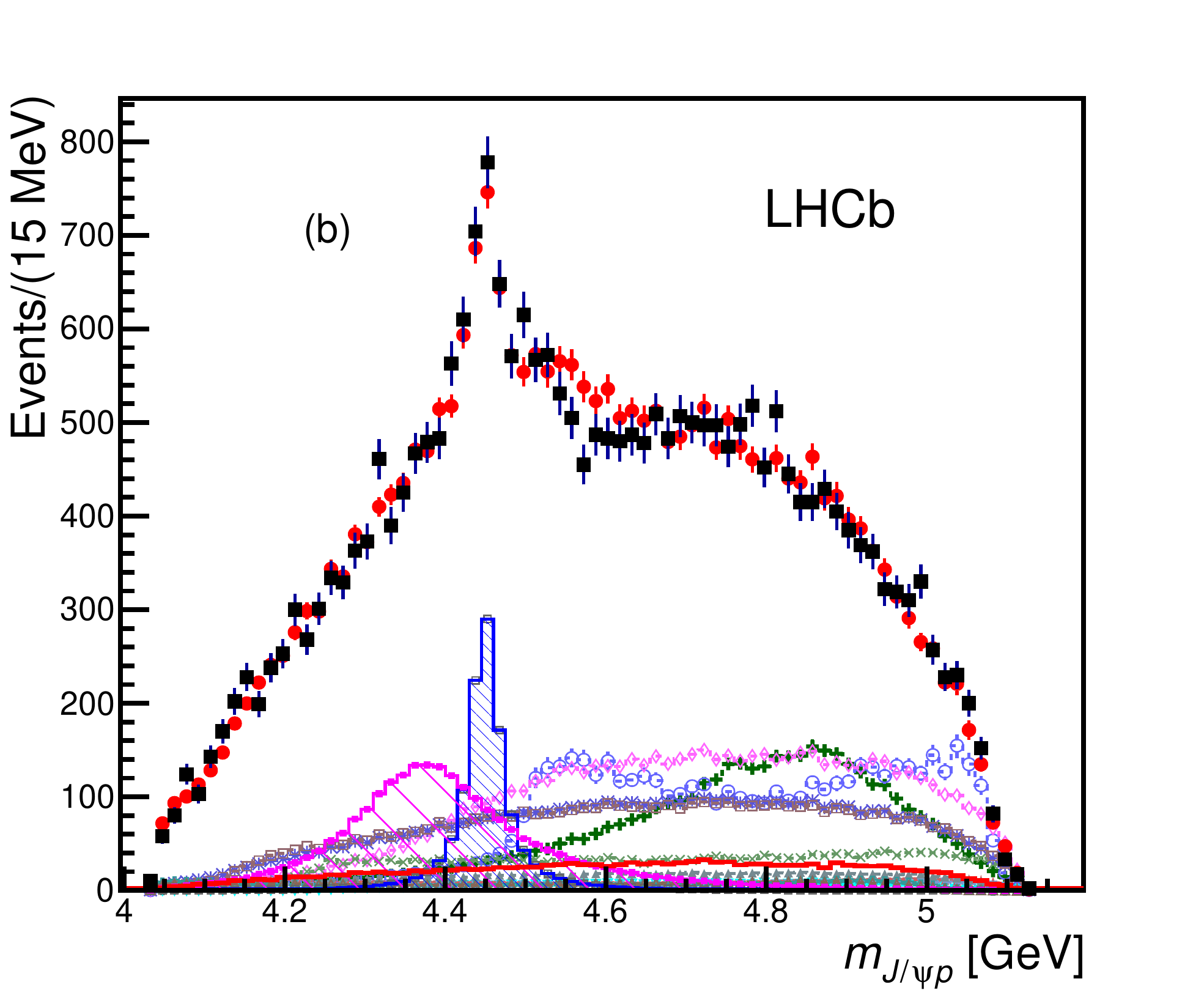}
\end{center}
\vskip -0.5cm
\caption{Fit projections for (a) $m_{Kp}$  and (b) $m_{\jpsi p}$ for the reduced $\Lz^*$ model with two $P_c^+$ states (see Table~\ref{tab:Lstar}). The data are shown as solid (black) squares, while the solid (red) points show the results of the fit.  The solid (red) histogram shows the background distribution. The (blue) open squares with the shaded histogram represent the $P_c(4450)^+$ state, and the shaded histogram topped with (purple) filled squares represents the $P_c(4380)^+$ state. Each $\Lz^*$ component is also shown. The error bars on the points showing the fit results are due to simulation statistics.}
\label{Pc2d}
\end{figure}

The  fit uses five decay angles and the $K^-p$ invariant mass $m_{Kp}$ as independent variables.
First we tried to fit the data with an amplitude model that contains  14 $\Lz^*$ states listed by the Particle Data Group \cite{PDG}.  As this did not give a satisfactory description of the data, we added one $P_c^+$ state, and when that was not sufficient
we included a second state. The two  $P_c^+$ states are found to have masses of  $4380\pm 8\pm 29$~MeV and $4449.8\pm 1.7\pm 2.5$~MeV, with corresponding widths of  $205\pm 18\pm 86$ MeV and $39\pm 5\pm19$ MeV. (Natural units are used throughout this Letter. Whenever two uncertainties are quoted the first is statistical and the second systematic.)  
The fractions of the total sample due to the lower mass and higher mass states are ($8.4\pm0.7\pm4.2$)\% and ($4.1\pm0.5\pm 1.1)$\%, respectively. The best fit solution has spin-parity $J^P$ values of ($3/2^-$, $5/2^+$). Acceptable solutions are  also found for additional cases with opposite parity, either ($3/2^+$, $5/2^-$) or ($5/2^+$, $3/2^-$). 
The best  fit projections are shown in Fig.~\ref{Pc2d}. Both  $m_{Kp}$ and the peaking structure in $m_{\jpsi p}$ are reproduced by the fit. The significances of the lower mass and higher mass states are 9 and 12 standard deviations, respectively.

\section*{Analysis and results}

\label{sec:selection}
We use data corresponding to 1~fb$^{-1}$ of integrated luminosity acquired by the LHCb experiment in $pp$ collisions at 7~TeV center-of-mass energy, and  2~fb$^{-1}$ at 8~TeV.
The \lhcb detector~\cite{LHCb-det} is a single-arm forward
spectrometer covering the \mbox{pseudorapidity} range $2<\eta <5$. The detector includes a high-precision tracking system
consisting of a silicon-strip vertex detector surrounding the $pp$
interaction region~\cite{LHCb-DP-2014-001}, a large-area silicon-strip detector located
upstream of a dipole magnet with a bending power of about
$4{\rm\,Tm}$, and three stations of silicon-strip detectors and straw
drift tubes~\cite{LHCb-DP-2013-003} placed downstream of the magnet.
Different types of charged hadrons are distinguished using information
from two ring-imaging Cherenkov detectors~\cite{LHCb-DP-2012-003}.
Muons are identified by a
system composed of alternating layers of iron and multiwire
proportional chambers~\cite{LHCb-DP-2012-002}.

Events are triggered by a \decay{\jpsi}{\mumu} decay, requiring two identified muons with opposite charge, each with transverse momentum, $\pt$, greater than 500\mev. The dimuon system is required to  form a vertex with a fit $\chi^2 <16$, to be significantly displaced from the nearest $pp$ interaction vertex, and to have
an invariant mass within 120\mev of the \jpsi mass~\cite{PDG}. After applying these requirements, there is a large \jpsi signal over a small background \cite{Aaij:2011fx}. Only candidates with dimuon invariant mass between $-$48\mev and +43\mev relative to the observed $\jpsi$ mass peak are selected, the asymmetry accounting for final-state electromagnetic radiation. 
 
Analysis preselection requirements are imposed prior to using a gradient Boosted Decision Tree, BDTG~\cite{Breiman,*2007physics...3039H},  that separates the \Lb signal from backgrounds. Each track is required to be of good quality and multiple reconstructions of the same track are removed. Requirements on the individual particles include $\pt>550$~MeV for muons, and $\pt>250$~MeV for hadrons. Each hadron must have an impact parameter $\chi^2$ with respect to the primary $pp$ interaction vertex larger than 9, and must be positively identified in the particle identification system. The $K^-p$ system must form a vertex with $\chi^2<16$, as must the  two muons from the \jpsi decay. 
Requirements on the \Lb candidate include a vertex $\chi^2<50$  for 5 degrees of freedom,  and a flight distance of greater than 1.5~mm. The vector from the primary vertex to the \Lb vertex must align with the \Lb momentum so that the cosine of the angle between them is larger than $0.999$.  Candidate $\mu^+\mu^-$ combinations are  constrained to the \jpsi mass for subsequent use in event selection.

\begin{figure}[t]
\begin{center}
    \includegraphics[width=0.7\textwidth]{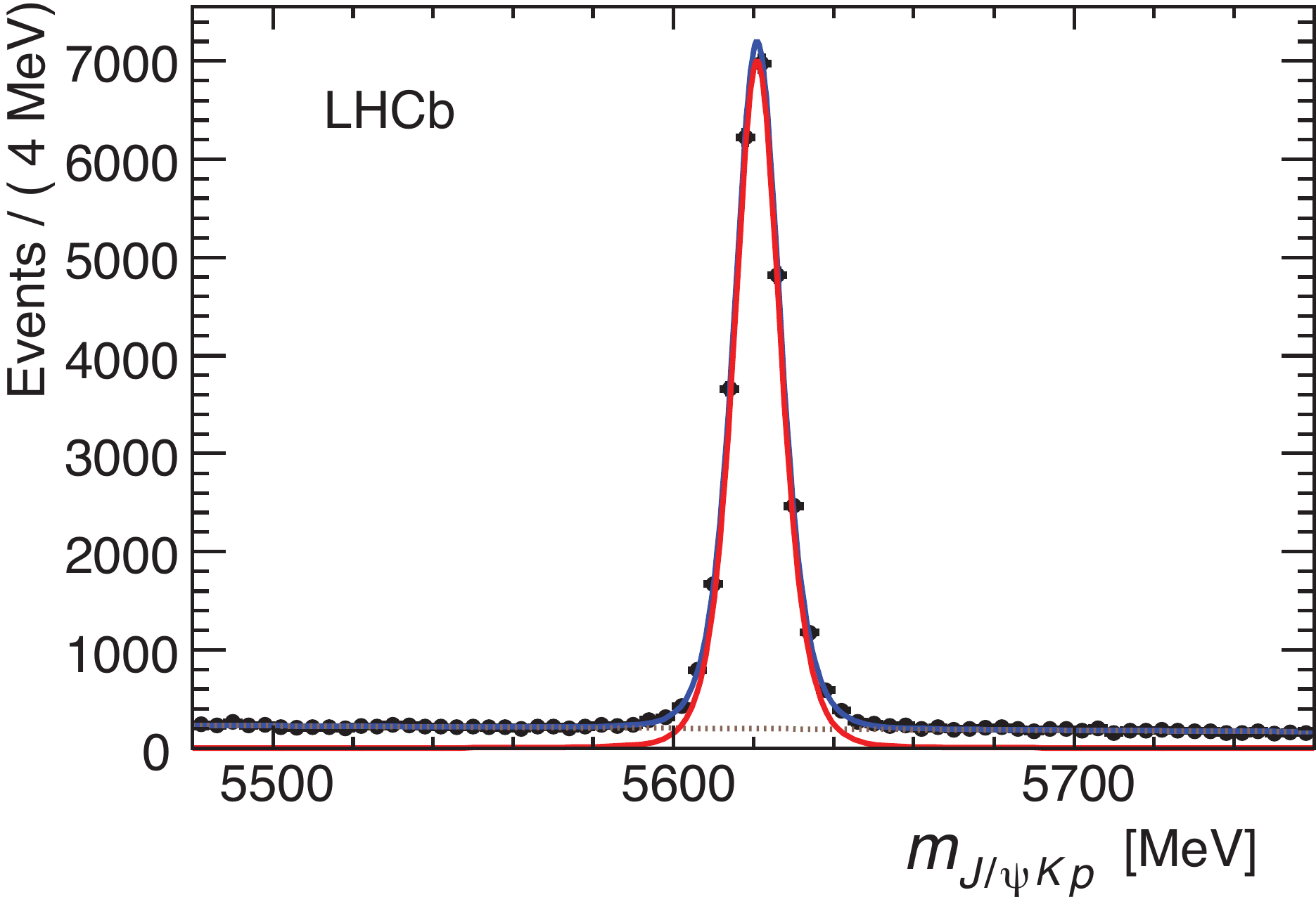} \end{center}
\vskip -0.5cm
\caption{Invariant mass spectrum of $\jpsi K^-p$ combinations, with the total fit, signal and background components shown as solid (blue), solid (red) and dashed lines, respectively.}
\label{fig:rawLb2JpsipK}
\end{figure}

The BDTG technique involves a  ``training" procedure using sideband data background and simulated signal samples. (The variables used are listed in the supplementary material.)  We use  $2\times10^6$ $\Lb\to \jpsi K^-p$ events with $\jpsi \to \mup\mu^{-}$  that are generated uniformly in phase space  in the LHCb acceptance, using \pythia~\cite{Sjostrand:2006za,*Sjostrand:2007gs} with a special \lhcb parameter tune~\cite{LHCb-PROC-2010-056}, and the \lhcb detector simulation based on \geant~\cite{Agostinelli:2002hh,*Allison:2006ve}, described in Ref.~\cite{LHCb-PROC-2011-006}. 
 The product of the reconstruction and trigger efficiencies within the LHCb geometric acceptance is about 10\%.  
In addition, specific backgrounds from  $\Bsb$ and $\Bzb$ decays are vetoed. This is accomplished by removing  combinations that when interpreted as $\jpsi K^+K^-$ fall within $\pm$30~MeV of the \Bsb mass or when interpreted as $\jpsi K^-\pi^+$ fall within $\pm$30~MeV of the \Bdb mass. This requirement effectively eliminates background from these sources  and causes only smooth changes in the detection efficiencies across the \Lb decay phase space.
Backgrounds from $\Xib$ decays cannot contribute significantly to our sample. We choose a relatively tight cut on the BDTG output variable that leaves  $26 \,007\pm$166 signal candidates containing 5.4\% background within $\pm 15$~MeV ($\pm2\,\sigma$) of the $\jpsi K^- p$ mass peak, 
as determined by the unbinned extended likelihood fit shown in Fig.~\ref{fig:rawLb2JpsipK}. 
The combinatorial background is modeled 
with an exponential function and the $\Lb$ signal shape is parameterized by a double-sided Hypatia function \cite{Santos:2013gra}, where the signal radiative
tail parameters are fixed to values obtained from simulation. For subsequent analysis we constrain the $\jpsi K^- p$ four-vectors to give the \Lb invariant mass and the \Lb momentum vector to be aligned with the measured direction from the primary  to the \Lb vertices \cite{Hulsbergen:2005pu}.

\begin{figure}[hbt]
\begin{center}
 \includegraphics[width=0.7\textwidth]{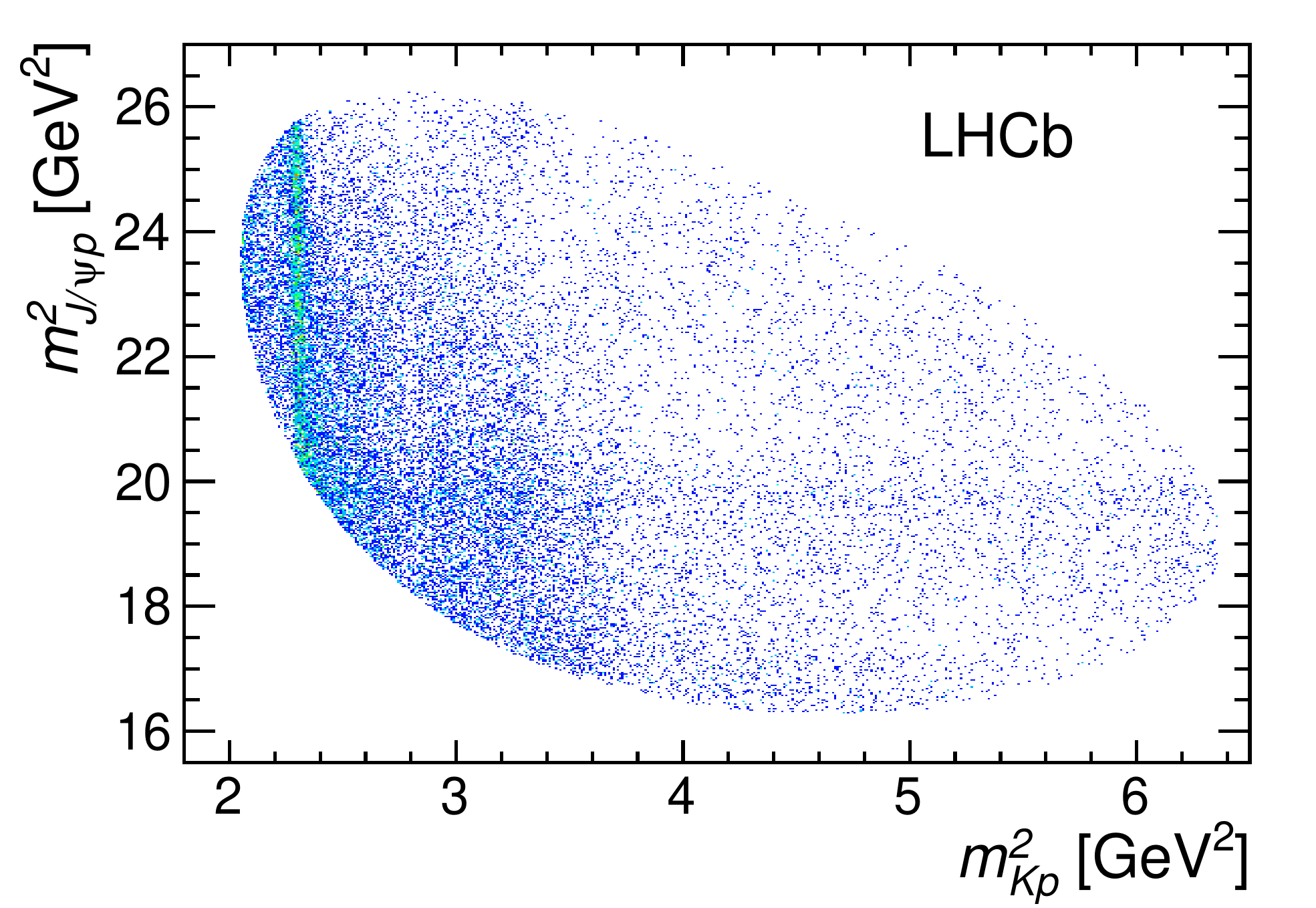}
\end{center}
\vskip -0.5cm
\caption{Invariant mass  squared of $K^-p$  versus $\jpsi p$ for candidates within $\pm15$~MeV of the \Lb mass.}
\label{dlz}
\end{figure}

In Fig.~\ref{dlz} we show the ``Dalitz" plot \cite{Dalitz:1953cp} using the $K^-p$  and $\jpsi p$ invariant masses-squared as independent variables.
A distinct vertical band is observed in the $K^-p$ invariant mass distribution near 2.3~GeV$^2$ corresponding to  the $\Lz(1520)$ resonance. There is also a distinct horizontal band near 19.5~GeV$^2$.
As we  see structures in both $K^-p$ and $\jpsi p$ mass distributions we perform a full amplitude analysis, using the available angular variables in addition to the mass distributions, in order to determine the resonances present.
No structure is seen in the $\jpsi K^-$ invariant mass.

We consider the two interfering processes shown in Fig.~\ref{Feynman-Pc}, which produce
two distinct decay sequences: $\Lb\to \jpsi\LambdaStar$, $\LambdaStar\to K^- p$ and
$\Lb\to \ZP^+ K^-$, $\ZP^+\to \jpsi p$, with $\jpsi\to\mu^+\mu^-$ in both cases.
We use the helicity formalism \cite{Chung:1971ri,*Richman:1984gh,*Jacob:1959at} in which each sequential decay $A\to B\,C$ contributes 
to the amplitude a term
\begin{equation}
\label{eq:hgeneric}
\H^{A\to B\,C}_{\lambda_{B},\,\lambda_{C}} \,\,
D^{\,\,J_A}_{\lambda_{A},\,\lambda_B-\lambda_C}(\phi_B,\theta_A,0)^*R_A(m_{BC})
= 
\H^{A\to B\,C}_{\lambda_{B},\,\lambda_{C}} \,\,
e^{i\,\lambda_A\,\phi_B}\,\,d^{\,\,J_A}_{\lambda_{A},\,\lambda_B-\lambda_C}(\theta_A)R_A(m_{BC}),
\end{equation}
where $\lambda$ is the quantum number related to 
the projection of the spin of the particle onto its momentum vector (helicity)
and $\H^{A\to B\,C}_{\lambda_{B},\,\lambda_{C}}$ are 
complex helicity-coupling amplitudes describing the decay dynamics. Here
$\theta_A$ and $\phi_B$ are the polar and azimuthal angles of $B$ in the rest frame of $A$ ($\theta_A$ is known as the ``helicity angle" of $A$).
The three arguments of Wigner's $D$-matrix are Euler angles describing the rotation of the initial coordinate system 
with the $z$-axis along the helicity axis of $A$ 
to the coordinate system with the $z$-axis along the helicity axis of $B$ \cite{PDG}. 
We choose  the convention in which the third Euler angle is zero. In Eq.~(\ref{eq:hgeneric}), $d^{J_A}_{\lambda_A,\lambda_B-\lambda_C}(\theta_A)$ is the Wigner small-$d$ matrix.
If $A$ has a non-negligible natural width, the invariant mass distribution of the $B$ and $C$ daughters
is described by the complex function $R_A(m_{BC})$ discussed below, otherwise $R_A(m_{BC})=1$.      

Using Clebsch-Gordan coefficients,
we express the helicity couplings in terms of $LS$ couplings ($B_{L,S}$), 
where $L$ is the orbital angular momentum
in the decay, and $S$ is the total spin of $A$ plus $B$:
\begin{equation}
\H_{\lambda_B,\lambda_C}^{A\to B\,C}=\sum_{L} \sum_{S} 
\sqrt{ \tfrac{2L+1}{2J_A+1} } B_{L,S} 
\left( 
\begin{array}{cc|c}
 J_{B} & J_{C} & S \\
 \lambda_{B} & -\lambda_{C} & \lambda_{B}-\lambda_{C} 
\end{array}
\right)
\times
\left( 
\begin{array}{cc|c}
 L  & S & J_A \\
 0 & \lambda_{B}-\lambda_{C} & \lambda_{B}-\lambda_{C}   
\end{array}
\right), 
\label{eq:LS}
\end{equation}
where the expressions in parentheses are the standard Wigner 3j-symbols.
For strong decays, possible $L$ values are constrained 
by the conservation of parity ($P$): $P_A=P_B\,P_C\,(-1)^L$.

Denoting $\jpsi$ as $\psi$,
the matrix element for the $\Lb\to\jpsi\LambdaStar$ decay sequence is 
\begin{eqnarray}
\Mat_{\lambda_{\Lb},\,\lambda_p,\,\Delta\lambda_\mu}^{\LambdaStar} \equiv 
\sum\limits_{n}
\sum\limits_{\lambda_{\LambdaStar}}
\sum\limits_{\lambda_{\psi}} 
\,\,
\H^{\Lb\to \LambdaStarn \psi}_{\lambda_{\LambdaStar},\,\lambda_{\psi}} 
D^{\,\,\frac{1}{2}}_{\lambda_{\Lb},\,\lambda_\LambdaStar-\lambda_\psi}(0,\theta_{\Lb},0)^*
& \notag\\
\H^{\LambdaStarn\to K p}_{\lambda_p,\,0} 
D^{\,\,J_{\LambdaStarn}}_{\lambda_{\LambdaStar},\,\lambda_p}(
\phi_{K},\theta_{\LambdaStar},0)^*
R_{\LambdaStarn}(m_{Kp})
&\!\!\!\!
D^{\,\,1}_{\lambda_{\psi},\,\Delta\lambda_\mu}(\phi_{\mu},\theta_{\psi},0)^*,
\label{eq:lbtolpsi}
\end{eqnarray}
where the $x$-axis, in the coordinates describing the $\Lb$ decay, is chosen 
to fix $\phi_{\LambdaStar}=0$.
The sum over $n$ is due to many different $\LambdaStarn$ resonances 
contributing to the amplitude. Since the $\jpsi$ decay is electromagnetic, the values of 
$\Delta\lambda_\mu \equiv \lambda_{\mu^+}-\lambda_{\mu^-}$ are restricted to $\pm1$.

There are 4 (6) independent complex
$\H^{\Lb\to \LambdaStarn \psi}_{\lambda_{\LambdaStar},\,\lambda_{\psi}}$ 
couplings to fit for each $\LambdaStarn$ resonance for $J_{\LambdaStarn}=\frac{1}{2}$ ($>\frac{1}{2}$). 
They can be reduced to only 1 (3) free $B_{L,S}$ couplings to fit if only the lowest
(the lowest two) values of $L$ are considered.  
The mass $m_{Kp}$, together with all decay angles entering Eq.~(\ref{eq:lbtolpsi}),
$\theta_{\Lb}$,  $\theta_{\LambdaStar}$, $\phi_{K}$, $\theta_{\psi}$ and $\phi_{\mu}$
(denoted collectively as $\POmega$),
constitute the six independent dimensions of the $\Lb\to\jpsi p K^-$ decay phase space. 

Similarly, the matrix element for the $\ZP^+$ decay chain is given by
\begin{eqnarray}
\Mat_{\lambda_{\Lb},\,\lambda_p^{\ZP},\,\Delta\lambda_\mu^{\ZP}}^{\ZP} \equiv 
\sum\limits_{j}
\sum\limits_{\lambda_{\ZP}}
\sum\limits_{\lambda_{\psi}^{\ZP}} 
\,\,
\H^{\Lb\to {\ZP}_j K}_{\lambda_{\ZP},\,0} 
D^{\,\,\frac{1}{2}}_{\lambda_{\Lb},\,\lambda_{\ZP}}(
\phi_{\ZP},\theta_{\Lb}^{\ZP},0)^*
& \notag\\
\H^{{\ZP}_j\to \psi p}_{\lambda_\psi^{\ZP},\lambda_p^{\ZP}} 
D^{\,\,J_{{\ZP}_j}}_{\lambda_{\ZP},\,\lambda_\psi^{\ZP}-\lambda_p^{\ZP}}(
\phi_{\psi},\theta_{\ZP},0)^*
R_{{\ZP}_j}(m_{\psi p}) \!
& \!
D^{\,\,1}_{\lambda_{\psi}^{\ZP},\,\Delta\lambda_\mu^{\ZP}}(
\phi_{\mu}^{\ZP},\theta_{\psi}^{\ZP},0)^*,
\label{eq:lbtopck}
\end{eqnarray}
where the angles and helicity states carry the superscript or subscript $\ZP$ to 
distinguish them from those defined for the $\LambdaStar$ decay chain. The sum over $j$ allows for the possibility of contributions from more than one $P_c^+$ resonance.
There are 2 (3) independent helicity couplings 
$\H^{{\ZP}_j\to \psi p}_{\lambda_\psi^{\ZP},\lambda_p^{\ZP}}$ 
for $J_{{\ZP}_j}=\frac{1}{2}$ ($>\frac{1}{2}$), 
and a ratio of the two $\H^{\Lb\to {\ZP}_j K}_{\lambda_{\ZP},\,0}$ couplings, 
to determine from the data.

The mass-dependent $R_{\LambdaStarn}(m_{Kp})$ and $R_{{\ZP}_j}(m_{\jpsi p})$ terms are given by  
\begin{equation}
R_X(m) =
B'_{L_{\Lb}^X}(p,p_0,d) \left(\frac{p}{M_{\Lb}}\right)^{L_{\Lb}^X}\, 
{\rm BW}(m | M_{0X}, \Gamma_{0X})\, B'_{L_{X}}(q,q_0,d) 
\left(\frac{q}{M_{0X}}\right)^{L_X}.
\label{eq:resshape}
\end{equation}
Here $p$ is the $X=\LambdaStar$ or $\ZP^+$ momentum in the $\Lb$ rest frame, and $q$ is the momentum of either decay product of $X$ in the $X$ rest frame.
The symbols $p_0$ and $q_0$ denote values of these quantities at the resonance peak 
($m=M_{0X}$). 
The orbital angular momentum between the decay products of $\Lb$ is denoted as $L_{\Lb}^{X}$.
Similarly, $L_{X}$ is the orbital angular momentum between the decay products of $X$.
The orbital angular momentum barrier factors, $p^L\,B'_{L}(p,p_0,d)$, involve the
Blatt-Weisskopf functions \cite{VonHippel:1972fg},
and account for the difficulty in creating larger orbital angular momentum $L$, 
which depends on the momentum of 
the decay products $p$ 
and on the size of the decaying particle, 
given by the $d$ constant.
We set $d=3.0~{\rm GeV}^{-1}$ $\sim 0.6$~fm.  
The relativistic Breit-Wigner amplitude is given by
\begin{equation}   
{\rm BW}(m | M_{0X}, \Gamma_{0X} ) = \frac{1}{{M_{0X}}^2-m^2 - i M_{0X} \Gamma(m)} \,,
\label{eq:breitwigner}
\end{equation}
where
\begin{equation}   
\Gamma(m)=\Gamma_{0X} \left(\frac{q}{q_0}\right)^{2\,L_{X}+1} \frac{M_{0X}}{m} B'_{L_{X}}(q,q_0,d)^2,
\label{eq:mwidth}
\end{equation}
is the mass dependent width of the resonance.
For the $\Lz(1405)$ resonance, which peaks below the $K^-p$ threshold, we use a two-component Flatt\'e-like
parameterization \cite{Flatte:1976xu} (see the supplementary material). 
The couplings for the allowed channels, $\PSigma\pi$ and $Kp$, are taken to be equal and to correspond to the nominal value of the width \cite{PDG}.
For all resonances we assume minimal values of $L_{\Lb}^X$ and of $L_X$ in $R_X(m)$.
For nonresonant (NR) terms we set ${\rm BW}(m)=1$ and $M_{0\,{\rm NR}}$ to the midrange mass.

Before the matrix elements for the two decay sequences can be added coherently,  
the proton and muon helicity states in the $\LambdaStar$ decay chain must be 
expressed in the basis of helicities in the $\ZP^+$  decay chain,
\begin{equation} 
\left| \Mat \right|^2 = 
\sum\limits_{\lambda_{\Lb}}
\sum\limits_{\lambda_{p}}
\sum\limits_{\Delta\lambda_{\mu}}
\left|
\Mat_{\lambda_{\Lb},\,\lambda_p,\,\Delta\lambda_\mu}^{\LambdaStar} 
+ 
e^{i\,{\Delta\lambda_\mu}\alpha_{\mu}}\,
\sum\limits_{\lambda_p^{\ZP}} 
d^{\,\,\frac{1}{2}}_{\lambda_p^{\ZP},\,\lambda_p}(\theta_p)\,
\Mat_{\lambda_{\Lb},\,\lambda_p^{\ZP},\,\Delta\lambda_\mu}^{\ZP} 
\right|^2,
\label{eq:matrixelement}
\end{equation}  
where $\theta_p$ is the polar angle in the $p$ rest frame between the boost directions from
the $\LambdaStar$ and $\ZP^+$ rest frames, 
and $\alpha_{\mu}$ is the azimuthal angle correcting for the difference between the muon helicity states 
in the two decay chains. 
Note that $m_{\psi p}$,
$\theta_{\Lb}^{\ZP}$, $\phi_{\ZP}$,
$\theta_{\ZP}$, $\phi_{\psi}$, 
$\theta_{\psi}^{\ZP}$, $\phi_{\mu}^{\ZP}$,
$\theta_p$ and $\alpha_\mu$ 
can all be derived from the values of $m_{Kp}$ and $\POmega$, and
thus do not constitute independent dimensions in the $\Lb$ decay phase space. 
(A detailed prescription for calculation of all the angles entering the matrix element
is given in the supplementary material.)

Strong interactions, which dominate $\Lb$ production at the LHC,
conserve parity and cannot produce longitudinal $\Lb$ polarization \cite{Soffer:1991am}. 
Therefore, $\lambda_{\Lb}=+1/2$ and $-1/2$ values are equally likely, which is
reflected in Eq.~(\ref{eq:matrixelement}). 
If we allow the $\Lb$ polarization to vary, the data are consistent with a  polarization of zero. 
Interferences between various $\LambdaStarn$ and $P^+_{cj}$ resonances vanish in the integrated rates unless the
resonances belong to the same decay chain and have the same quantum numbers.

The matrix element given by Eq.~(\ref{eq:matrixelement})
is a 6-dimensional function of $m_{Kp}$ and $\POmega$ and
depends on the fit parameters, $\Pars$, which represent 
independent helicity or $LS$ couplings, and masses and widths of resonances (or Flatt\'e parameters),
$\Mat=\Mat(m_{Kp},\POmega|\Pars)$.
After accounting for the selection efficiency to obtain the signal probability density function  (PDF) an unbinned maximum likelihood fit is used to determine the amplitudes.
Since the efficiency does not depend on $\Pars$, it  is needed
only in the normalization integral, which is carried out numerically by summing $\left|\Mat(m_{Kp},\POmega|\Pars)\right|^2$
over the simulated events generated uniformly in phase space and passed through the selection. (More details are given in the supplementary material.)

We use two fit algorithms, which were independently coded and which differ in the approach used for background subtraction.
In the first approach, which we refer to as cFit, the signal region is defined as $\pm2\,\sigma$ around the $\Lb$ mass peak.
The total PDF used in the fit to the candidates in the signal region, $\PDF(m_{Kp},\POmega|\Pars)$,
includes a background component with normalization fixed to be 5.4\%\ of the total.
The background PDF is found to factorize into five two-dimensional functions of $m_{Kp}$ and of each independent angle, which are
estimated using sidebands extending from $5.0\,\sigma$ to $13.5\,\sigma$ on both sides of the peak.     

In the complementary approach, called sFit, no explicit background parameterization is needed. The PDF consists of only the signal component, with the background subtracted using the sPlot technique \cite{Pivk:2004ty,*Xie:2009rka} applied
to the log-likelihood sum. All candidates shown in Fig.~\ref{fig:rawLb2JpsipK} are included in the sum with weights, $W_i$,
dependent on $m_{\jpsi Kp}$.  The weights are set according
to the signal and the background probabilities determined by the fits to the $m_{\jpsi p K}$ distributions,
similar to the fit displayed in Fig.~\ref{fig:rawLb2JpsipK}, but performed in 32 different bins of the two-dimensional plane of 
$\cos\theta_{\Lb}$ and $\cos\theta_{\jpsi}$ to account for correlations with the mass shapes of the signal and background components.
This quasi-log-likelihood sum is scaled by a constant factor, $s_W\equiv \sum_i W_i/\sum_i {W_i}^2$, 
to account for the effect of the background subtraction on the statistical uncertainty.
(More details on the cFit and sFit procedures are given in the supplementary material.)

In each approach, we minimize   
$-2\ln\Like(\Pars) =-2 s_W \sum_i W_i \ln \PDF(m_{{Kp}_ {i}},\POmega_i|\Pars)$,
 which gives the estimated values of the fit parameters,
${\Pars}_{\rm min}$, together with their covariance matrix ($W_i=1$ in cFit). 
The difference of $-2\ln\Like({\Pars}_{\rm min})$ between different
amplitude models, $\Dll$, allows  their discrimination.
For two models representing separate hypotheses, \eg\ when discriminating between
different $J^P$ values assigned to a $P_c^+$ state, 
the assumption of a $\chi^2$ distribution with one degree of freedom
for $\Dll$ under the disfavored $J^P$ hypothesis 
allows the calculation of a lower limit on the significance of its 
rejection, \ie\ the p-value \cite{james2006statistical}.
Therefore, it is convenient to express $\Dll$ values as
${n^2_\sigma}$, where $n_\sigma$ corresponds to the  number of  standard
deviations in the normal distribution with the same p-value. 
For nested hypotheses, \eg\ when discriminating between models without and with $\ZP^+$ states,
$n_\sigma$ overestimates the p-value by a modest amount. 
Simulations are used to obtain better estimates of the significance of the
$\ZP^+$ states.

Since the isospin of both the \Lb and the \jpsi particles are zero, we expect that the dominant contributions in the $K^-p$ system are $\Lz^*$ states, which would be produced via a $\Delta I=0$ process. It is also possible that $\PSigma^*$ resonances contribute, but these would have $\Delta I=1$. By analogy with kaon decays the $\Delta I=0$ process should be dominant \cite{Donoghue:1979mu}. The list of $\Lz^*$ states considered is shown in Table~\ref{tab:Lstar}. 

\begin{table}[htb]
\centering
\caption{The $\Lz^*$ resonances used in the different fits. Parameters are taken from the PDG \cite{PDG}. We take $5/2^-$ for the $J^P$ of the $\Lz(2585)$.
The number of $LS$ couplings is also listed for both the ``reduced'' and ``extended'' models. 
To fix overall phase and magnitude conventions, which otherwise are arbitrary, we set $B_{0,\frac{1}{2}}=(1,0)$ for $\Lz(1520)$.  
A zero entry means the state is excluded from the fit. }
\vspace{0.2cm}
\begin{tabular}{lccccc}
\hline\\[-2.5ex] 
State & $J^P$ & $M_0$ (MeV) & $\Gamma_0$ (MeV)& \# Reduced  & \# Extended \\
\hline \\[-2.5ex] 
$\Lz(1405)$ &1/2$^-$ & $1405.1^{+1.3}_{-1.0}$ & $50.5\pm 2.0$ & 3 & 4 \\
$\Lz(1520)$ &3/2$^-$ &$1519.5\pm 1.0$ & $15.6\pm 1.0$& 5 & 6 \\
$\Lz(1600)$ &1/2$^+$ &1600 & 150 &3 & 4 \\
$\Lz(1670)$ &1/2$^-$ & 1670 & 35 & 3 & 4\\
$\Lz(1690)$ &3/2$^-$ & 1690 & 60 & 5 & 6\\
$\Lz(1800)$ &1/2$^-$ & 1800 & 300 &4 &4 \\
$\Lz(1810)$ &1/2$^+$ &1810& 150&3&4\\
$\Lz(1820)$ &5/2$^+$ & 1820 & 80 &1&6\\
$\Lz(1830)$ &5/2$^-$ & 1830 & 95& 1&6 \\
$\Lz(1890)$ &3/2$^+$ & 1890 & 100 &3&6 \\
$\Lz(2100)$ &7/2$^-$ &2100 & 200&1 & 6\\
$\Lz(2110)$ &5/2$^+$ & 2110 & 200  &1 &6\\
$\Lz(2350)$ &9/2$^+$ & 2350 & 150 &0 &6\\
$\Lz(2585)$ &?& $\approx$2585 & 200 &0 & 6\\\hline
\end{tabular}
\label{tab:Lstar}
\end{table}

Our strategy is to first try to fit the data with a model that can describe the mass and angular distributions including only $\Lz^*$ resonances, allowing all possible known states and decay amplitudes. 
We call this the ``extended" model. It has 146 free parameters from the helicity couplings alone. The masses and widths of the $\Lz^*$ states are fixed to their PDG values, since allowing them to float prevents the fit from converging. Variations in these parameters are considered in the systematic uncertainties.

\begin{figure}[b]
\begin{center}
\vskip 8mm
\includegraphics[width=0.50\textwidth]{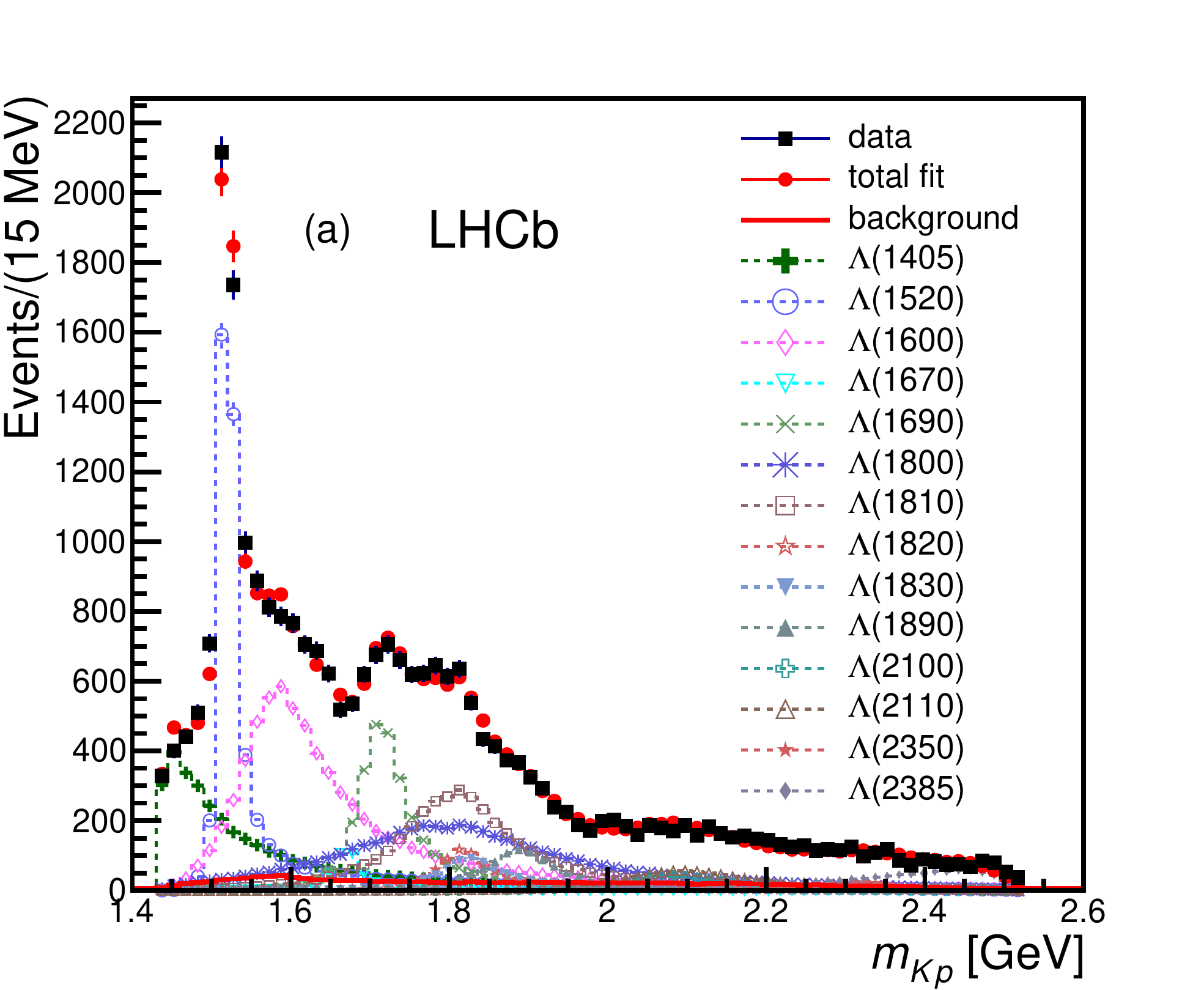}\includegraphics[width=0.50\textwidth]{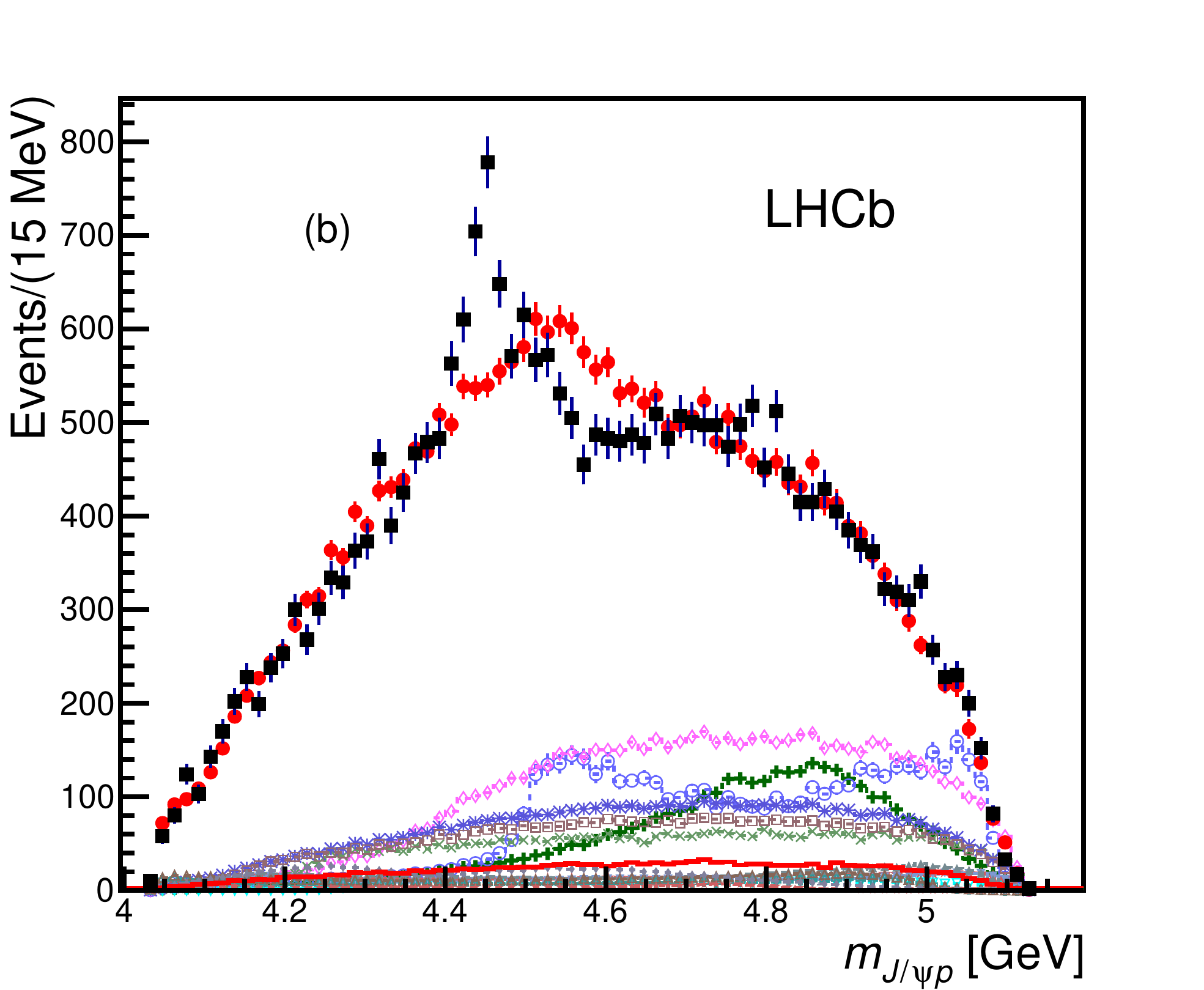}
\end{center}
\vskip -0.5cm
\caption{Results for (a) $m_{Kp}$  and (b) $m_{\jpsi p}$ for the extended $\Lz^*$  model fit without $P_c^+$ states. The data are shown as (black) squares with error bars, while the (red) circles show the results of the fit. The error bars on the points showing the fit results are due to simulation statistics.}
\label{Mall}
\end{figure}

The cFit results without any $P_c^+$ component are shown in Fig.~\ref{Mall}. While the $m_{Kp}$ distribution is reasonably well fitted, the peaking structure in $m_{\jpsi p}$ is not reproduced. The same result is found using sFit.
The speculative
addition of $\PSigma^*$ resonances to the states decaying to $K^-p$ does not change this conclusion.

We will demonstrate that introducing two $P_c^+\to\jpsi p$ resonances leads to a satisfactory description of the data. When determining parameters of the $P_c^+$ states, we use a more 
restrictive model of the $K^-p$ states (hereafter referred to as the ``reduced''  model) 
 that includes only the $\Lz^*$ resonances that are well motivated, and has fewer than half the number of free parameters. 
As the minimal $L_{\Lb}^{\LambdaStar}$ for the spin $9/2$ $\Lz(2350)$ equals $J_{\LambdaStar}-J_{\Lb}-J_{\jpsi}=3$, it is extremely unlikely that this state can be produced so close to the phase space limit.  In fact $L=3$ is the highest orbital angular momentum observed, with a very small rate, in decays of $B$ mesons \cite{Aaij:2014xza,*Aaij:2014baa} with much larger phase space available 
($Q=2366$~MeV, while here $Q=173$~MeV), and without additional suppression from the spin counting factors present in  $\Lz(2350)$ production (all three $\vec{J}_{\LambdaStar}$, $\vec{J}_{\Lb}$ and $\vec{J}_{\jpsi}$ vectors have to line up in the same direction to produce the minimal $L_{\Lb}^{\LambdaStar}$ value). Therefore, we eliminate it from the reduced $\Lz^*$ model. We also eliminate the $\Lz(2585)$ state, which peaks beyond the kinematic limit and has unknown spin. The other resonances are kept but high $L_{\Lb}^{\LambdaStar}$ amplitudes are removed; only the lowest values are kept for the high mass resonances, with a smaller reduction for the lighter ones. The number of $LS$ amplitudes used for each resonance is listed in Table~\ref{tab:Lstar}. With this model we reduce the number of parameters needed to describe the $\Lz^*$ decays from 146 to 64. 
For the different combinations of $P_c^+$ resonances that we try, there are up to 20 additional free parameters. Using the extended model
including one resonant $P_c^+$ improves the fit quality,  but it is still unacceptable (see supplementary material).  We find acceptable fits  with two $P_c^+$ states. 
We use the reduced $\Lz^*$ model for the central values of our results. The differences in fitted quantities with the extended model are included in the systematic uncertainties. 

\begin{figure}[htb]
\begin{center}
\includegraphics[width=1.0\textwidth]{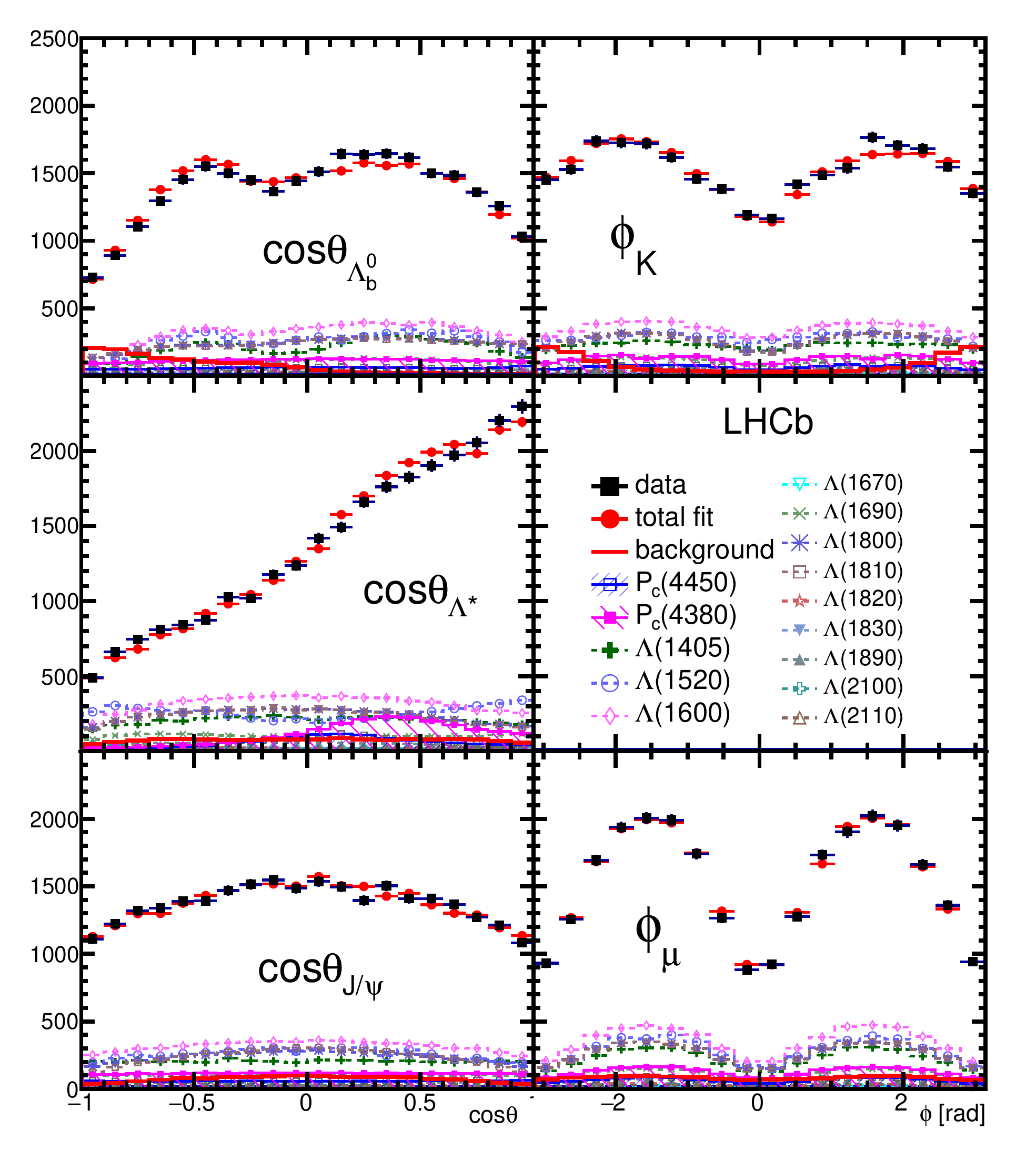}
\end{center}
\vskip -0.1cm
\caption{Various decay angular distributions for the fit with two $P_c^+$ states. The data are shown as (black) squares, while the (red) circles show the results of the fit.  Each fit  component is also shown. The angles are defined in the text.}
\label{TwoZ-angular-cFit}
\end{figure}

\begin{figure}[H!]
\begin{center}
\includegraphics[width=1.0\textwidth]{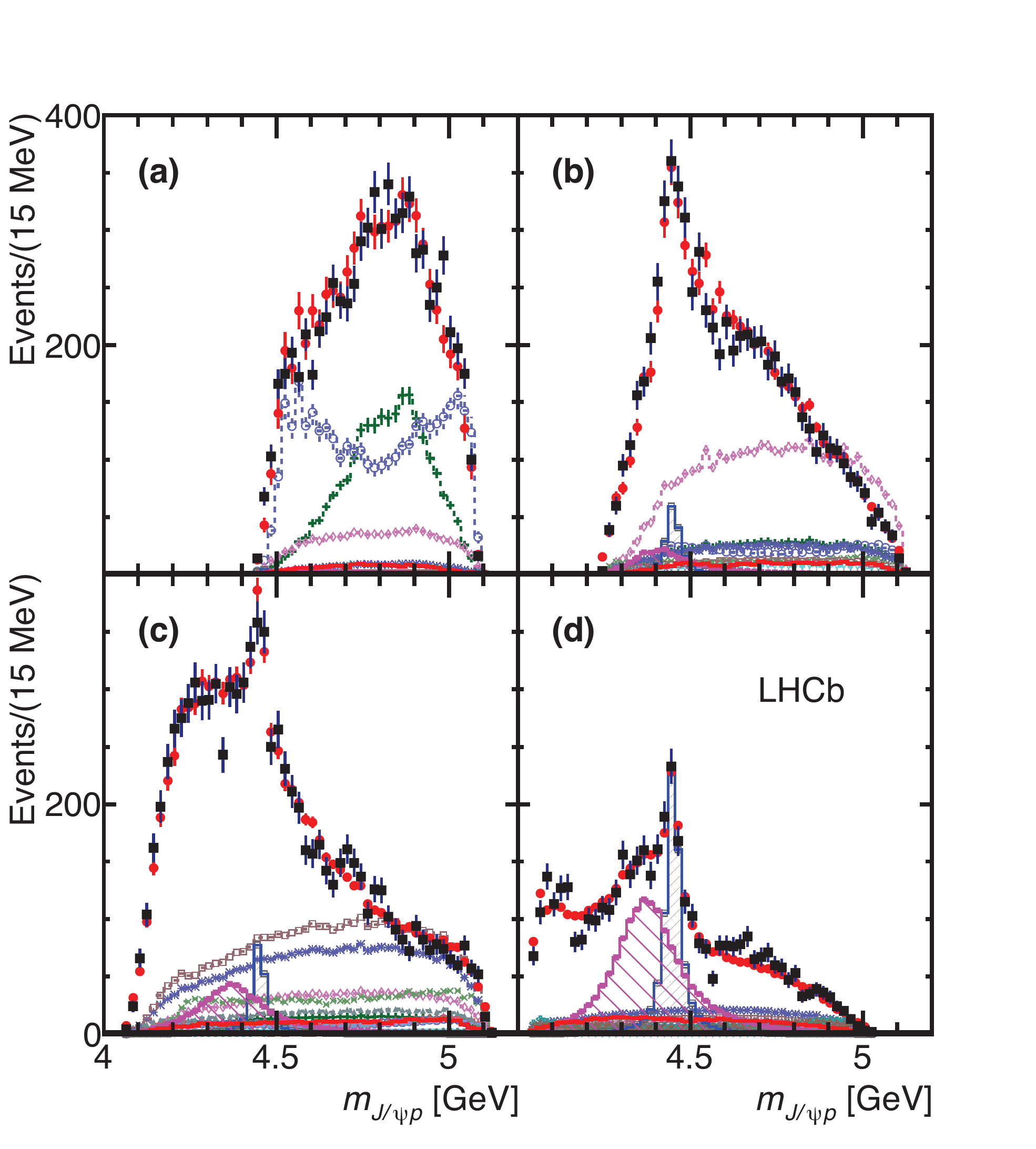}\end{center}
\vskip -0.5cm
\caption{$m_{\jpsi p}$ in various intervals of $m_{Kp}$ for the fit with two $P_c^+$ states: (a) $m_{Kp}<1.55$~GeV, (b) $1.55<m_{Kp}<1.70$~GeV, (c) $1.70<m_{Kp}<2.00$~GeV, and (d) $m_{Kp}>2.00$~GeV.  The data are shown as  (black) squares with error bars, while the (red) circles show the results of the fit. The blue and purple histograms show the two $P_c^+$ states. See Fig.~\ref{TwoZ-angular-cFit} for the legend.}
\label{TwoZ-mjpsip-bins-cFit}
\end{figure}

The best fit combination finds two $P_c^+$ states with
$J^P$ values of $3/2^-$ and $5/2^+$, for the lower and higher mass states, respectively. The  $-2\ln{\cal{L}}$  values differ by only 1 unit between the best fit and the parity reversed combination ($3/2^+$,  $5/2^-$). Other combinations are less likely, 
although the ($5/2^+$,  $3/2^-$) pair changes $-2\ln{\cal{L}}$ by only $2.3^2$ units and therefore cannot be ruled out. All combinations $1/2^{\pm}$ through $7/2^{\pm}$ were tested, and all others  are disfavored by changes of more than $5^2$ in the $-2\ln{\cal{L}}$ values. 
The cFit results for the ($3/2^-$, $5/2^+$)  fit are shown in Fig.~\ref{Pc2d}. Both  distributions of $m_{Kp}$ and $m_{\jpsi p}$ are reproduced. 
The  lower mass $3/2^-$ state has mass 4380$\pm$8~MeV and width 205$\pm$18~MeV, while the $5/2^+$ state has a mass of 4449.8$\pm$1.7~MeV and width 39$\pm$5~MeV; these errors are statistical only, systematic uncertainties are discussed later.  The mass resolution is approximately 2.5~MeV and does not affect the width determinations. The sFit approach gives comparable results.
The angular distributions  are reasonably well reproduced,  as shown in Fig.~\ref{TwoZ-angular-cFit}, and the comparison with the data in $m_{Kp}$ intervals is also satisfactory as can be seen in Fig.~\ref{TwoZ-mjpsip-bins-cFit}. Interference effects between the two $P_c^+$ states are particularly evident in Fig.\ref{TwoZ-mjpsip-bins-cFit}(d), where there is a large destructive contribution (not explicitly shown in the figure) to the total rate.
(A fit fraction comparison between cFit and sFit is given in the supplementary material.)
The addition of further $P_c^+$ states does not significantly improve the fit.

\afterpage{\clearpage}
Adding a single $5/2^+~P_c^+$ state to the  fit with only $\Lz^*$ states reduces $-2\ln\Like$ by  $14.7^2$ using the extended model  and  adding a second lower mass $3/2^-$ $P_c^+$ state results in a further reduction of $11.6^2$. 
The combined reduction of $-2\ln\Like$ by the two states taken together is $18.7^2$. 
Since taking $\sqrt{\Delta2\ln\Like}$ overestimates significances, we perform simulations to obtain more accurate evaluations. 
We generate pseudoexperiments using the null hypotheses having amplitude parameters 
determined by the fits to the data with no or one $\ZP^+$ state. 
We fit each pseudoexperiment with the null hypothesis and with $\ZP^+$ states added to the model.
The $-2\ln{\cal{L}}$ distributions obtained from many pseudoexperiments are consistent with $\chi^2$ distributions with the number of degrees of freedom approximately equal to twice the number of extra parameters in the fit. Comparing these distributions with the ${\Delta2\ln\Like}$ values from the fits to the data,  p-values can be calculated. These studies show reduction of the significances relative to $\sqrt{\Delta2\ln\Like}$ by about 20\%, giving overall significances of $9\,\sigma$ and $12\,\sigma$, for the lower and higher mass $P_c^+$ states, respectively. The combined significance of two $P_c^+$ states is  $15\,\sigma$.
Use of the extended model to evaluate the significance includes the effect of systematic uncertainties due to the possible presence of additional $\Lz^*$ states or higher $L$ amplitudes.

Systematic uncertainties are evaluated for the masses, widths and fit fractions of the $P_c^+$ states, and for the fit fractions of the two lightest and most significant $\Lz^*$ states. Additional sources of modeling uncertainty that we have not considered may affect the fit fractions of the heavier $\Lz^*$ states. The sources of systematic uncertainties  are listed in Table~\ref{tab:syssum}.
They include differences between the results of the extended versus reduced model, varying the $\Lz^*$ masses and widths, uncertainties in the identification requirements for the proton, and restricting its momentum, inclusion of a nonresonant amplitude in the fit, use of separate higher and lower \Lb mass sidebands, alternate $J^P$ fits, varying the Blatt-Weisskopf barrier factor, $d$,  between 1.5 and 4.5~GeV$^{-1}$, changing the  angular momentum $L$ used in Eq.~({\ref{eq:resshape}) by one or two units, and accounting for potential mismodeling of the efficiencies. For the $\Lz(1405)$ fit fraction we also added an uncertainty for the Flatt\'e couplings, determined by both halving and doubling their ratio, and taking the maximum deviation as the uncertainty. 

\begin{table}[b]
\centering
\caption{Summary of systematic uncertainties on $P_c^+$ masses, widths and fit fractions, and $\Lz^*$ fit fractions. 
A fit fraction is the ratio of the phase space integrals of the matrix element squared for a single resonance and for the total amplitude. 
The terms ``low" and ``high" correspond to the lower and higher mass $P_c^+$ states. The sFit/cFit difference is listed as a cross-check and not included as an uncertainty.}
\vspace{0.2cm}
\resizebox{\textwidth}{!}{ 
\begin{tabular}{lrrrrrrrcc}
\hline
~~~Source & \multicolumn{2}{c}{$M_0$ (MeV)$\!\!\!\!$} &\multicolumn{2}{c}{$\Gamma_0$ (MeV)$\!\!\!\!$} &\multicolumn{4}{c}{Fit fractions (\%)}$\!\!\!\!$\\
 &low& high$\!\!$ & low& high$\!\!$& low& high&$\Lz(1405)$&$\Lz(1520)$\\
\hline
Extended vs. reduced &$21$ &0.2&$54$&10 &$3.14$&0.32&1.37~~~&0.15\\
$\Lz^*$ masses \& widths &7 &0.7 &20&4&0.58&0.37&2.49~~~&2.45\\
Proton ID &2& 0.3 & $1$ & $2$&0.27&0.14&0.20~~~&0.05\\
$10<p_p<100$~GeV &  0&1.2&$1$&$1$&0.09&0.03&0.31~~~&0.01\\
Nonresonant & 3&0.3&$34$&~2 &$2.35$&$0.13$&3.28~~~&0.39\\
Separate sidebands & 0 &~~0&~5&~0&$0.24$&$0.14$&0.02~~~&0.03\\
$J^P$ ($3/2^+$, $5/2^-$) or ($5/2^+$, $3/2^-$) &$10$&$1.2$&34&10&$0.76$&0.44&& \\
$d=1.5-4.5~$GeV$^{-1}$&$9$&0.6&19&$3$&0.29&0.42&0.36~~~&1.91\\
$L_{\Lb}^{\ZP}$ $\Lb\to P_c^+~{\rm (low/high)} K^-$ &6&0.7&4&8&$0.37$&0.16&&\\
$L_{\ZP}$ $P_c^+~{\rm (low/high)}\to\jpsi p$&4&$0.4$&31&7&$0.63$&0.37&&\\
$L_{\Lb}^{\LambdaStarn}$ $\Lb\to \jpsi \Lz^*$&11&0.3&20&2&0.81&0.53&3.34~~~&2.31\\
Efficiencies &1&0.4&4&0&0.13&0.02&0.26~~~&0.23\\
Change $\Lz(1405)$ coupling&0 &0&0&0&0&0&1.90~~~&~~~0\\
\hline
Overall & 29&2.5&86&19&4.21&1.05&5.82~~~&3.89\\\hline
{sFit/cFit} cross check &5&1.0&11&3&0.46&0.01&0.45~~~&0.13\\
\hline
\end{tabular}
}
\label{tab:syssum}
\end{table}

The stability of the results is cross-checked by comparing the data recorded in 2011/2012, with the LHCb dipole magnet polarity in up/down configurations, \Lb/\Lbbar decays, and \Lb produced with low/high values of \pt.  Extended model fits without including $P_c^+$ states were tried with the addition of two high mass $\Lz^*$ resonances of freely varied mass and width, or four nonresonant components up to spin 3/2; these do not explain the data. The fitters were tested on simulated pseudoexperiments and no biases were found. In addition, selection requirements are varied, and the vetoes of \Bsb and \Bdb  are removed and explicit models of those backgrounds added to the fit; all give consistent results. 

Further evidence for the resonant character of the higher mass, narrower, $P_c^+$ state is obtained by viewing the evolution of the complex amplitude in the Argand diagram  \cite{PDG}.
In the amplitude fits discussed above, the $P_c(4450)^+$ is represented by a Breit-Wigner amplitude, where
the magnitude and phase vary with $m_{\jpsi p}$ according to
an approximately circular trajectory in the (Re$\,A^{P_c}$, Im$\,A^{P_c}$) plane, where
$A^{P_c}$ is the $m_{\jpsi p}$ dependent part of the $P_c(4450)^+$ amplitude.
We perform an additional fit to the data using the reduced $\Lz^*$ model, in which
we represent the $P_c(4450)^+$ amplitude as the combination of independent
complex amplitudes at six equidistant points in the range $\pm \Gamma_0=39\,$MeV around $M_0=4449.8\,$MeV as determined in the default fit.
Real and imaginary parts of the amplitude are interpolated in mass between the fitted points.
The resulting Argand diagram, shown in Fig.~\ref{DoubleArgand}(a),
is consistent with a rapid counter-clockwise change of the $P_c(4450)^+$ phase when its magnitude
reaches the maximum, a behavior characteristic of a resonance. A similar study for the wider state is shown in Fig.~\ref{DoubleArgand}(b); although the fit does show a large phase change, the amplitude values are sensitive to the details of the $\Lz^*$ model  and so this latter study is not conclusive. 
\begin{figure}[t]
\begin{center}
\includegraphics[width=0.9\textwidth]{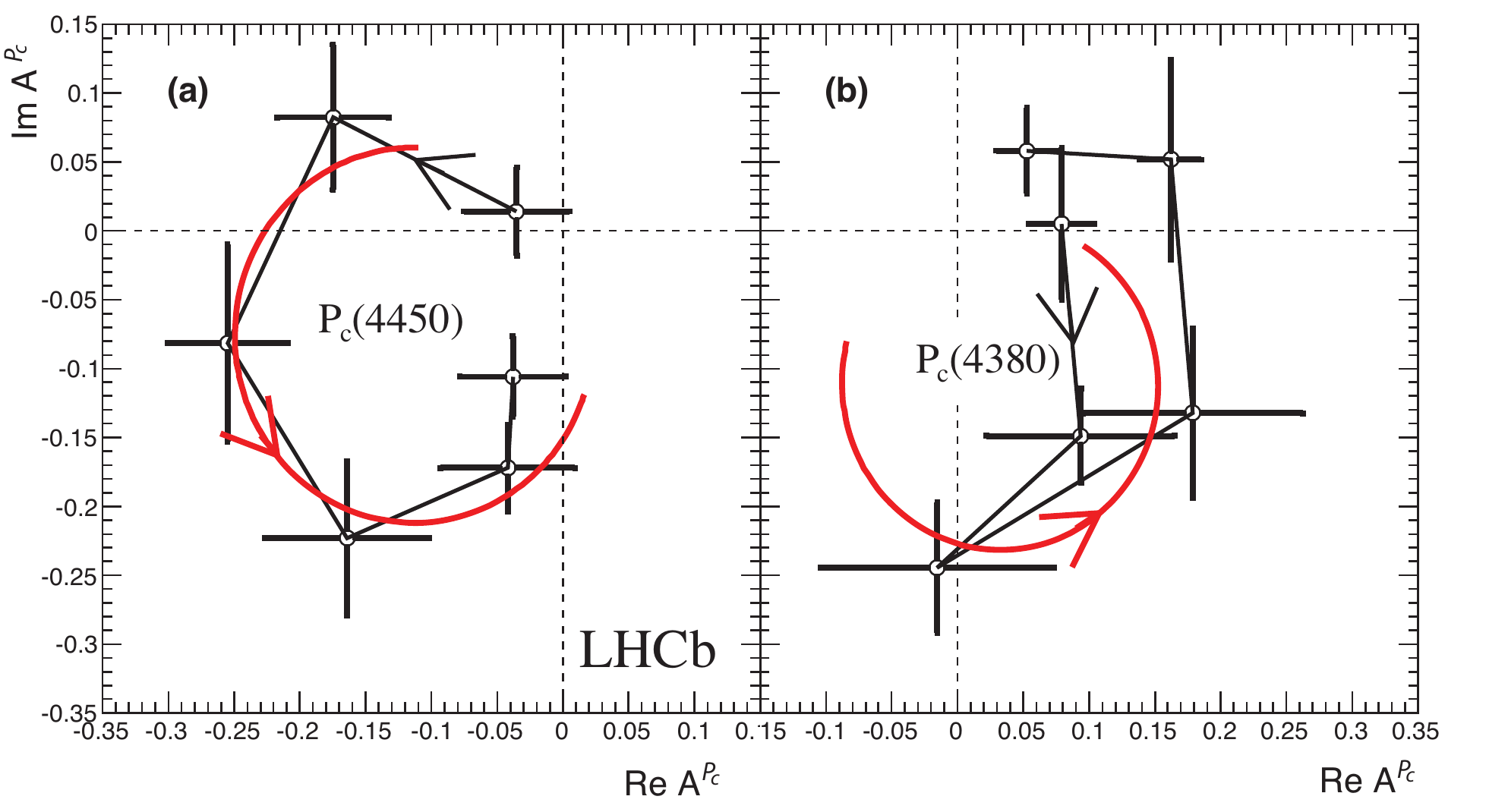}
\vskip -0.2cm
\caption{
Fitted values of the real and imaginary parts of the amplitudes for the baseline ($3/2^-$, $5/2^+$) fit for a) the $P_c(4450)^+$ state and b) the $P_c(4380)^+$ state, each divided into six $m_{\jpsi p}$ bins of equal width between $-\Gamma_0$ and $+\Gamma_0$ shown in the Argand diagrams as connected points with error bars ($m_{\jpsi p}$ increases counterclockwise).
The solid (red) curves are the predictions
from the Breit-Wigner formula for the same mass ranges
with  $M_0$ ($\Gamma_0$) of
 4450 (39) \mev and 4380 (205) \mev, respectively,
with the phases and magnitudes at the resonance masses set to the
average values between the two points around $M_0$.  
The phase convention sets $B_{0,\frac{1}{2}}=(1,0)$ for $\Lz(1520)$. Systematic uncertainties are not included.}
\label{DoubleArgand}
\end{center}
\end{figure}

 Different binding mechanisms of pentaquark states are possible.  Tight-binding was envisioned originally \cite{Jaffe:1976ig,Strottman:1979qu,*Hogaasen:1978jw,Rossi:1977cy}. A possible explanation is heavy-light diquarks \cite{Maiani:2004vq}. Examples of other mechanisms include a  diquark-diquark-antiquark model  \cite{Jaffe:2003sg,Chandra:2012zz},  a diquark-triquark model  \cite{Karliner:2003dt}, and a coupled channel model \cite{Wu:2010jy}. Weakly bound ``molecules" of a baryon plus a meson have been also discussed \cite{Voloshin:1976ap, *DeRujula:1976qd, *Tornqvist:1991ks, *Tornqvist:1993ng,
 *Yang:2011wz,*Wang:2011rga,*Karliner:2015ina}.
 
Models involving  thresholds or ``cusps" have been invoked to explain some exotic meson candidates via nonresonant scattering mechanisms \cite{Swanson:2015bsa,Swanson:2014tra,Bugg:2011jr}. There are certain obvious difficulties with the use of this approach to explain our results. The closest threshold to the high mass state is at 4457.1$\pm$0.3~MeV resulting from a $\Lz_c(2595)^+\Dzb$ combination, which is somewhat higher than the peak mass value and would produce a structure with quantum numbers $J^P=1/2^+$ which are disfavored by our data.  There is no threshold close to the lower mass state.

In conclusion, we have presented a full amplitude fit to the $\Lb\to\jpsi K^- p$ decay. We observe significant $\Lz^*$ production recoiling against the $\jpsi$ 
with the lowest mass contributions, the $\Lz(1405)$ and $\Lz(1520)$ states having fit fractions of $(15\pm 1\pm 6)$\% and $(19\pm 1\pm 4$)\%, respectively. 
The data cannot be satisfactorily described without including two  Breit-Wigner shaped resonances in the $\jpsi p$ invariant mass distribution.  The significances of the lower mass and higher mass states are 9 and 12 standard deviations, respectively. These structures cannot be accounted for by reflections from $\jpsi \Lz^*$ resonances or other known sources.  Interpreted as resonant states they must have minimal quark content of $c\overline{c} uud$, and would therefore be called charmonium-pentaquark states. The lighter state $\ZP(4380)^+$ has a mass of $4380\pm 8\pm 29$~MeV and a width of $205\pm 18\pm 86$ MeV, while the heavier state $\ZP(4450)^+$ has a mass of $4449.8\pm 1.7\pm 2.5$~MeV and a width of $39\pm 5\pm 19$ MeV.
A model-independent representation of the $\ZP(4450)^+$ contribution in the fit shows a phase change in amplitude consistent with that of a resonance. The parities of the two states are opposite with the preferred spins being 3/2 for one state and 5/2 for the other.
The higher mass state has a fit fraction of  ($4.1\pm0.5\pm 1.1)$\%, and the lower mass state of  ($8.4\pm0.7\pm4.2$)\%, of the total $\Lb\to\jpsi K^-p$ sample.

We express our gratitude to our colleagues in the CERN
accelerator departments for the excellent performance of the LHC. We
thank the technical and administrative staff at the LHCb
institutes. We acknowledge support from CERN and from the national
agencies: CAPES, CNPq, FAPERJ and FINEP (Brazil); NSFC (China);
CNRS/IN2P3 (France); BMBF, DFG, HGF and MPG (Germany); INFN (Italy); 
FOM and NWO (The Netherlands); MNiSW and NCN (Poland); MEN/IFA (Romania); 
MinES and FANO (Russia); MinECo (Spain); SNSF and SER (Switzerland); 
NASU (Ukraine); STFC (United Kingdom); NSF (USA).
The Tier1 computing centres are supported by IN2P3 (France), KIT and BMBF 
(Germany), INFN (Italy), NWO and SURF (The Netherlands), PIC (Spain), GridPP 
(United Kingdom).
We are indebted to the communities behind the multiple open 
source software packages on which we depend. We are also thankful for the 
computing resources and the access to software R\&D tools provided by Yandex LLC (Russia).
Individual groups or members have received support from 
EPLANET, Marie Sk\l{}odowska-Curie Actions and ERC (European Union), 
Conseil g\'{e}n\'{e}ral de Haute-Savoie, Labex ENIGMASS and OCEVU, 
R\'{e}gion Auvergne (France), RFBR (Russia), XuntaGal and GENCAT (Spain), Royal Society and Royal
Commission for the Exhibition of 1851 (United Kingdom).
\newpage

\section*{Appendix: Supplementary material}
\tableofcontents
\newpage
\section{Variables used in the BDTG}

Muon identification uses information from several parts of the detector, including the RICH detectors, the calorimeters and the muon system. Likelihoods are formed for the muon and pion hypotheses. The difference in the logarithms of the likelihoods, DLL$(\mu-\pi)$, is used to distinguish between the two 
\cite{LHCb-DP-2012-003}. The smaller value of the two discriminants DLL$(\mu^+ -\pi^+)$ and DLL$(\mu^- -\pi^-)$ is used as one of the BDTG variables.  

The next set of variables uses the kaon and proton tracks.
The $\chi^2_{\rm IP}$ is defined as the difference in $\chi^2$ of the primary vertex reconstructed with and without the considered track. The smaller $\chi^2_{\rm IP}$ of the $K^-$ and $p$ is used in the BDTG. The scalar \pt sum of the $K^-$ and $p$  is another variable. 

The last set of variables uses the \Lb candidate.  The cosine of the angle between a vector from the primary vertex to the  \Lb  vertex and the \Lb momentum vector is one input variable. In addition the $\chi^2_{\rm IP}$, the flight distance, the \pt and the  vertex $\chi^2$ of the \Lb candidate are used.

\section{Additional fit results}
\subsection [Reduced model fit projections for $m_{\jpsi K^-}$ ]{\boldmath Reduced model fit projections for $m_{\jpsi K}$ }

The Dalitz plots for the other two possible projections are shown in Fig.~\ref{dlztwo}. There is no obvious resonance structure in the $\jpsi K^-$ mass-squared distribution. 
\begin{figure}[b]
\begin{center}
\includegraphics[width=0.49\textwidth]{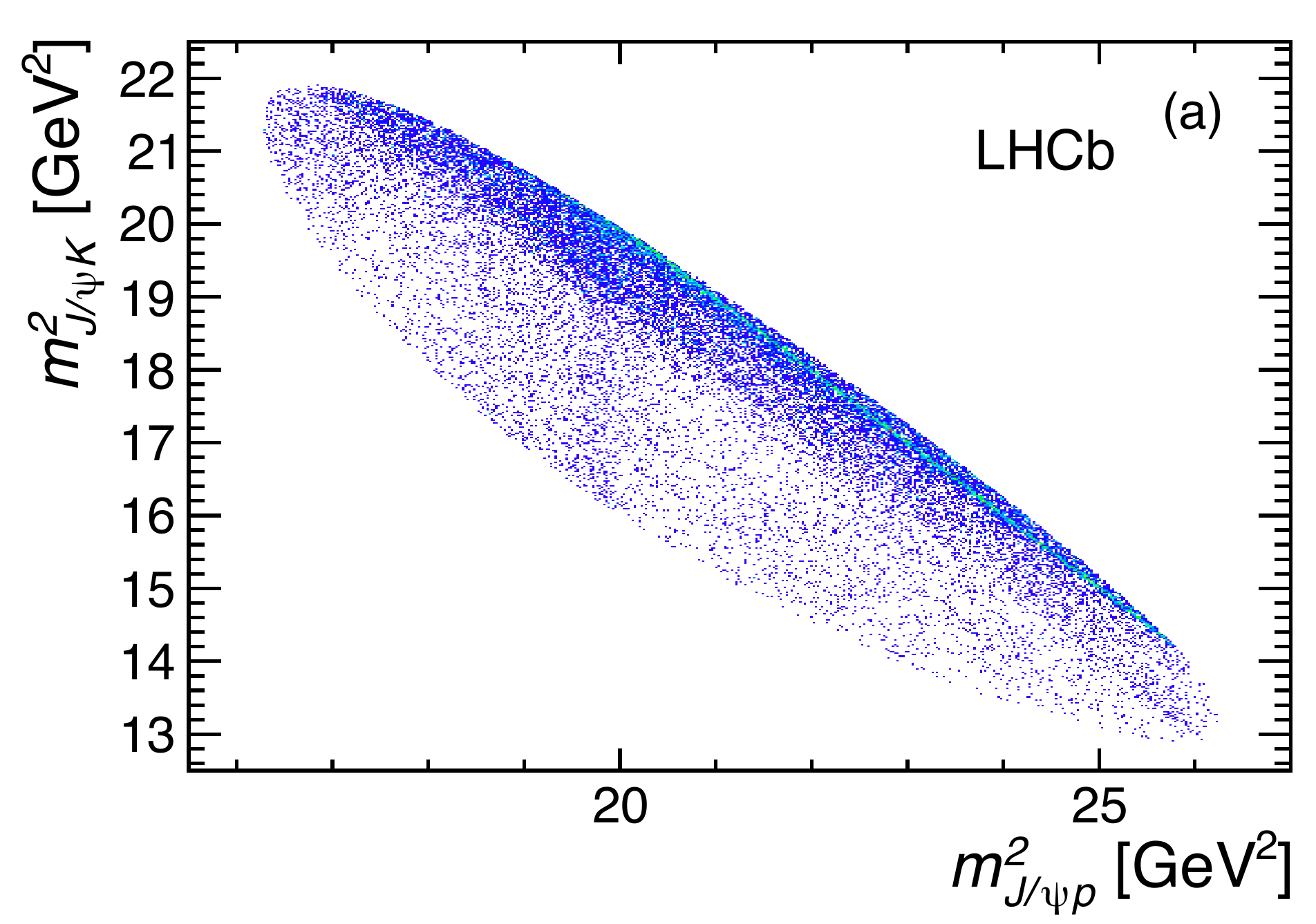}\includegraphics[width=0.49\textwidth]{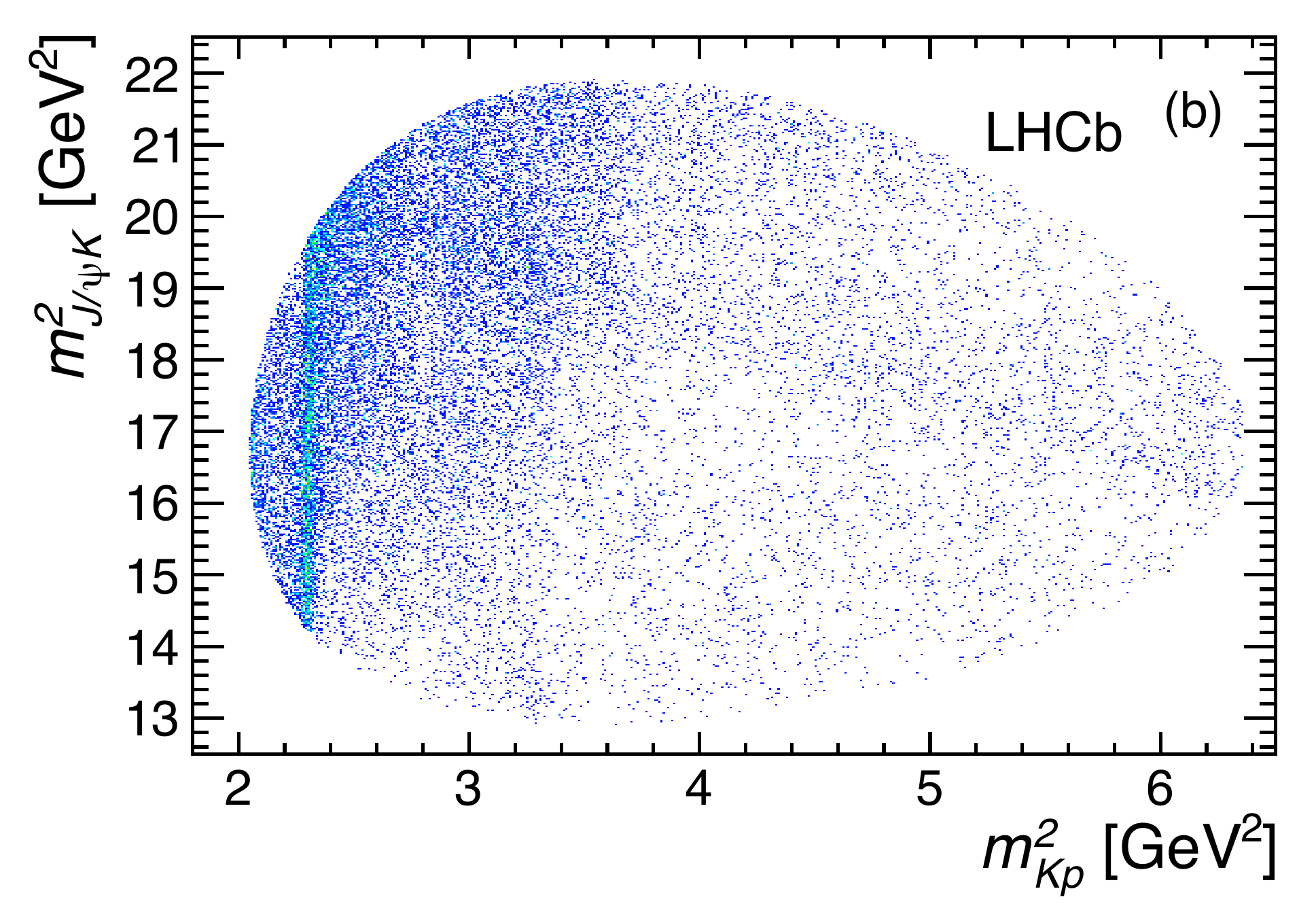}
\end{center}
\vskip -0.5cm
\caption{(a) Invariant mass squared of $\jpsi K^-$ versus $\jpsi p$ and (b) of $\jpsi K^-$ versus $K^-p$ for candidates within $\pm15$~MeV of the \Lb mass.}
\label{dlztwo}
\end{figure}

Our fit describes well the $m_{\jpsi K}$  distribution as shown by viewing the projections of the reduced model fit. They 
are shown for different slices of $m_{Kp}$  in Fig.~\ref{h_mjpsik}.

\begin{figure}[t!]
\begin{center}
\includegraphics[width=0.9\textwidth]{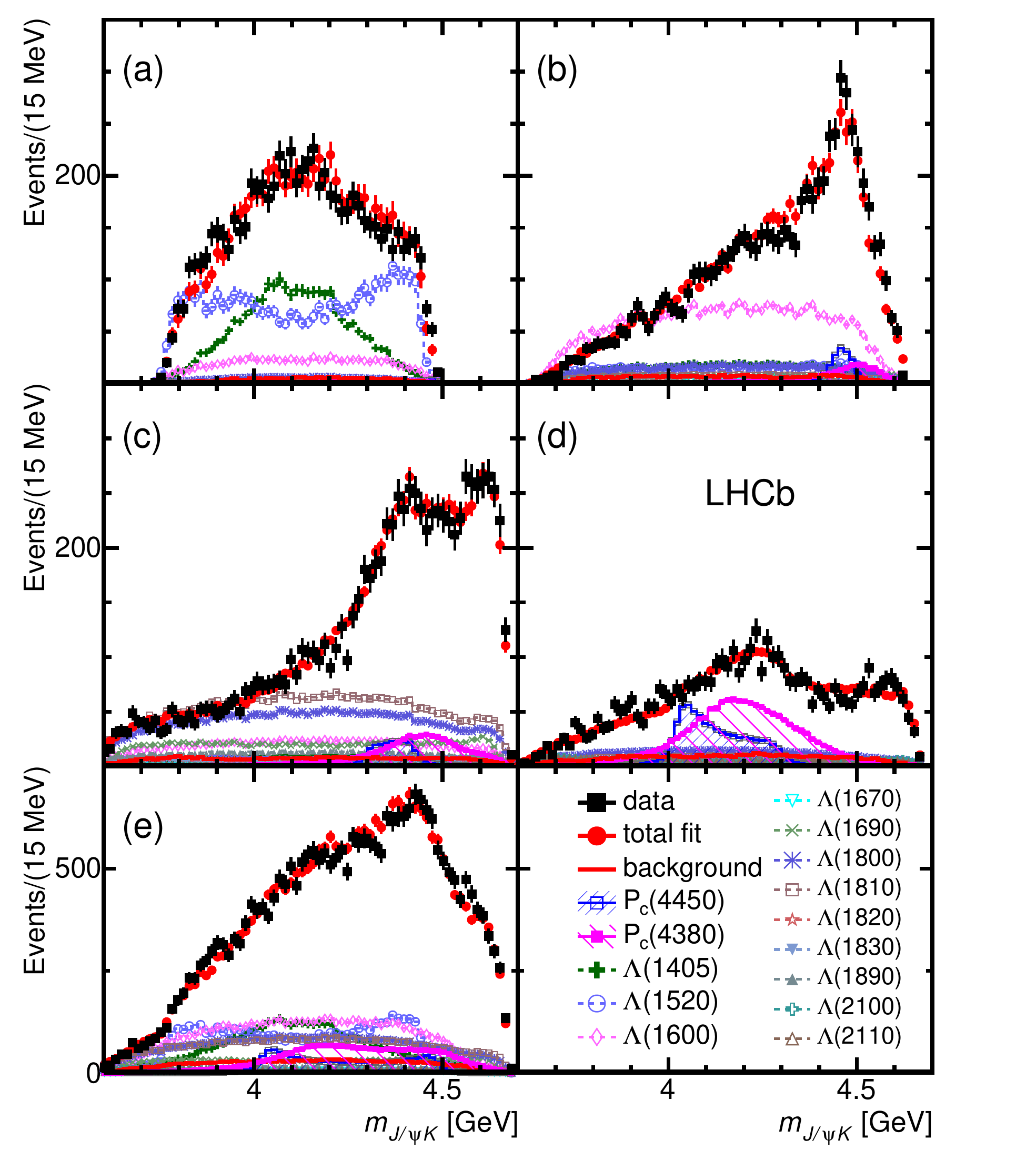}
\end{center}
\vskip -0.5cm
\caption{Projections onto $m_{\jpsi K}$ in various intervals of $m_{Kp}$ for the reduced model fit (cFit) with two $P_c^+$ states of $J^P$ equal to $3/2^-$ and $5/2^+$: (a) $m_{Kp}<1.55$~GeV, (b) $1.55<m_{Kp}<1.70$~GeV, (c) $1.70<m_{Kp}<2.00$~GeV,  (d) $m_{Kp}>2.00$~GeV, and (e) all $m_{Kp}$.  The data are shown as (black) squares with error bars, while the (red) circles show the results of the fit. The individual resonances are given in the legend.}
\label{h_mjpsik}
\end{figure}

\subsection[Reduced model angular fits with two $P_c^+$ states for $m(K^-p)>2$~GeV]{\boldmath Reduced model angular fits with two $P_c^+$ states for $m(K^-p)>2$~GeV}
In Fig.~\ref{TwoZ-angular-cFit-largeMKp} we show the result of the reduced model fit to the angular distributions for $m(K^-p)>2$~GeV. The data is well described by the fits.
\begin{figure}[htb]
\begin{center}
\includegraphics[width=0.98\textwidth]{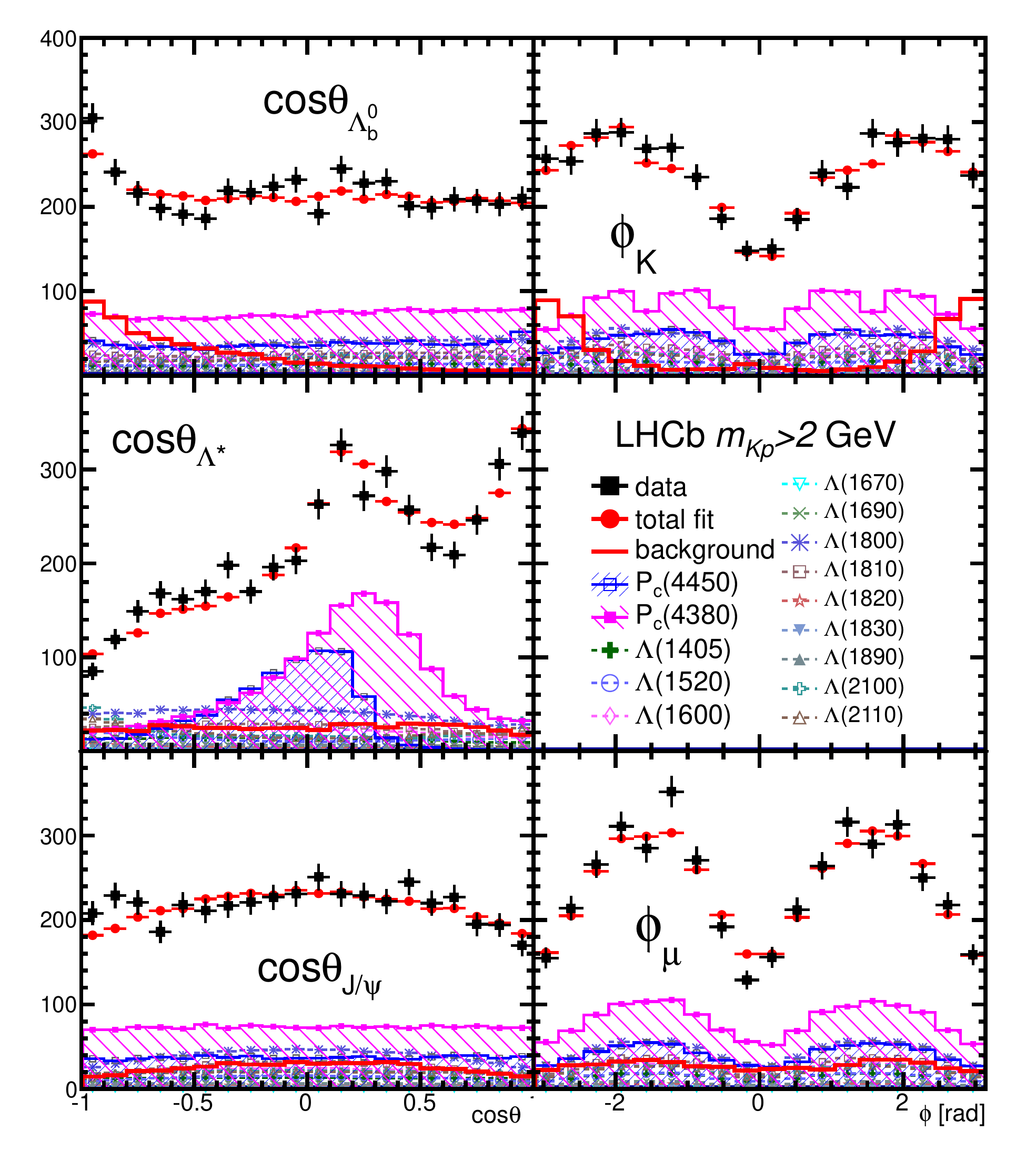}
\end{center}
\vskip -0.1cm
\caption{Various decay angular distributions for the fit with two $P_c^+$ states for $m(K^-p)>2$~GeV. The data are shown as (black) squares, while the (red) circles show the results of the fit.  Each fit  component is also shown. The angles are defined in the text.}
\label{TwoZ-angular-cFit-largeMKp}
\end{figure}

\subsection[Extended model fit with one $P_c^+$ state ]{\boldmath Extended model fit with one $P_c^+$ }
In the fits with one $P_c^+$ amplitude, we test
$J^P$ values of $1/2^{\pm}$, $3/2^{\pm}$ and $5/2^{\pm}$. The mass and width of the putative $P_c^+$ state are allowed to vary. There are a total of 146 free parameters for the $\Lz^*$ states to which we add either three complex couplings for $1/2^{\pm}$ or four for higher spins. The best fit is with a $5/2^+$ state, which improves $-2\ln {\cal{L}}$ by 215. 
Figure~\ref{Extended-1Pc} shows the projections for this fit.  While the $m_{Kp}$ projection is well described, clear discrepancies in $m_{\jpsi p}$ remain visible.

\begin{figure}[t]
\begin{center}
\includegraphics[width=0.48\textwidth]{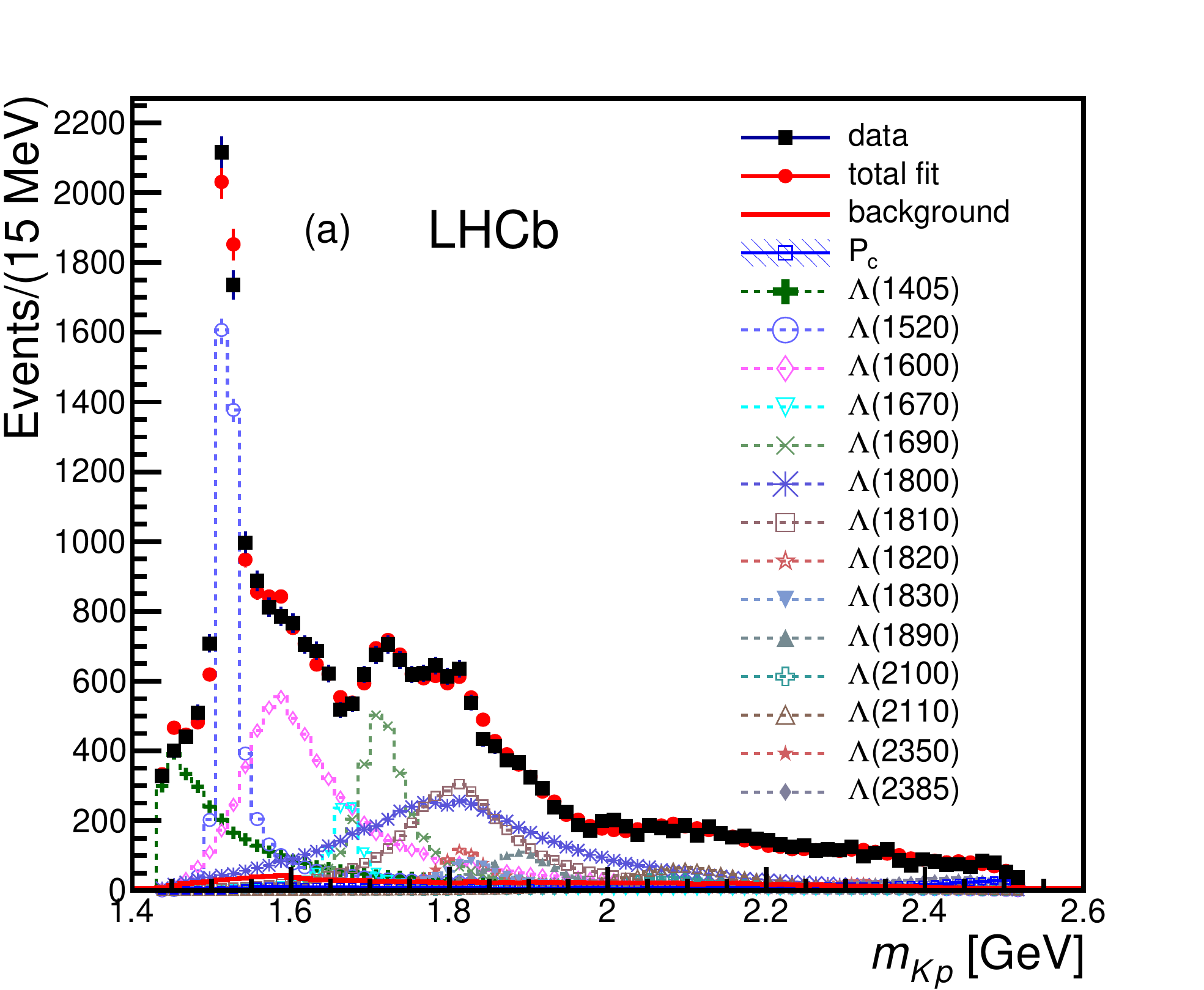}
\includegraphics[width=0.48\textwidth]{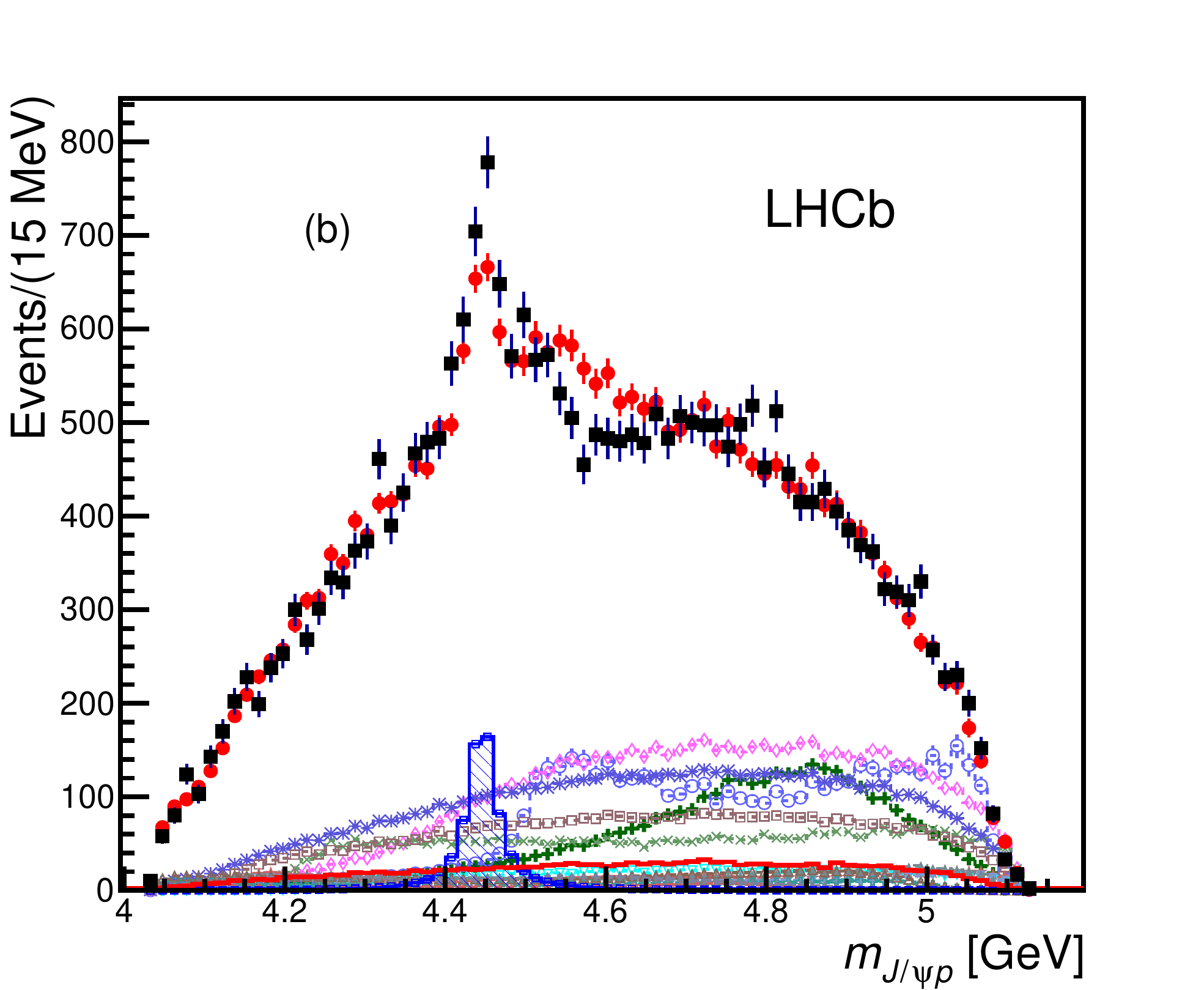}
\end{center}
\vskip -0.5cm
\caption{Results of the fit with one $J^P=5/2^+$ $P_c^+$ candidate. (a) Projection of the invariant mass of $K^-p$ combinations from $\Lb\to\jpsi K^-p$ candidates. The data are shown as (black) squares with error bars, while the  (red) circles show the results of the fit;  (b) the corresponding $\jpsi p$ mass projection. The (blue) shaded plot shows the $P_c^+$ projection, the other curves represent individual  $\Lz^*$ states. }
\label{Extended-1Pc}
\end{figure}

\subsection[Results of extended model fit with two ${P_c}^+$ states]{\boldmath Results of extended model fit with two ${P_c}^+$ states}

For completeness we include here the results of the extended model fit with two ${P_c}^+$ states using cFit.  We find acceptable fits for several combinations. For a lower mass $J^P=3/2^-$ state and a higher mass $5/2^+$ state, the masses (widths) are 4358.9$\pm$6.6~MeV (151.1$\pm$13.7~MeV), and 4450.1$\pm$1.7~MeV (48.6$\pm$4.0~MeV), respectively.
The uncertainties are statistical only.
The results for this two $P_c^+$ fit are shown in Fig.~\ref{M2all}. Both the $m_{Kp}$ distribution and the peaking structure in $m_{\jpsi p}$ are reproduced. 
\begin{figure}[htb]
\begin{center}
\includegraphics[width=0.49\textwidth]{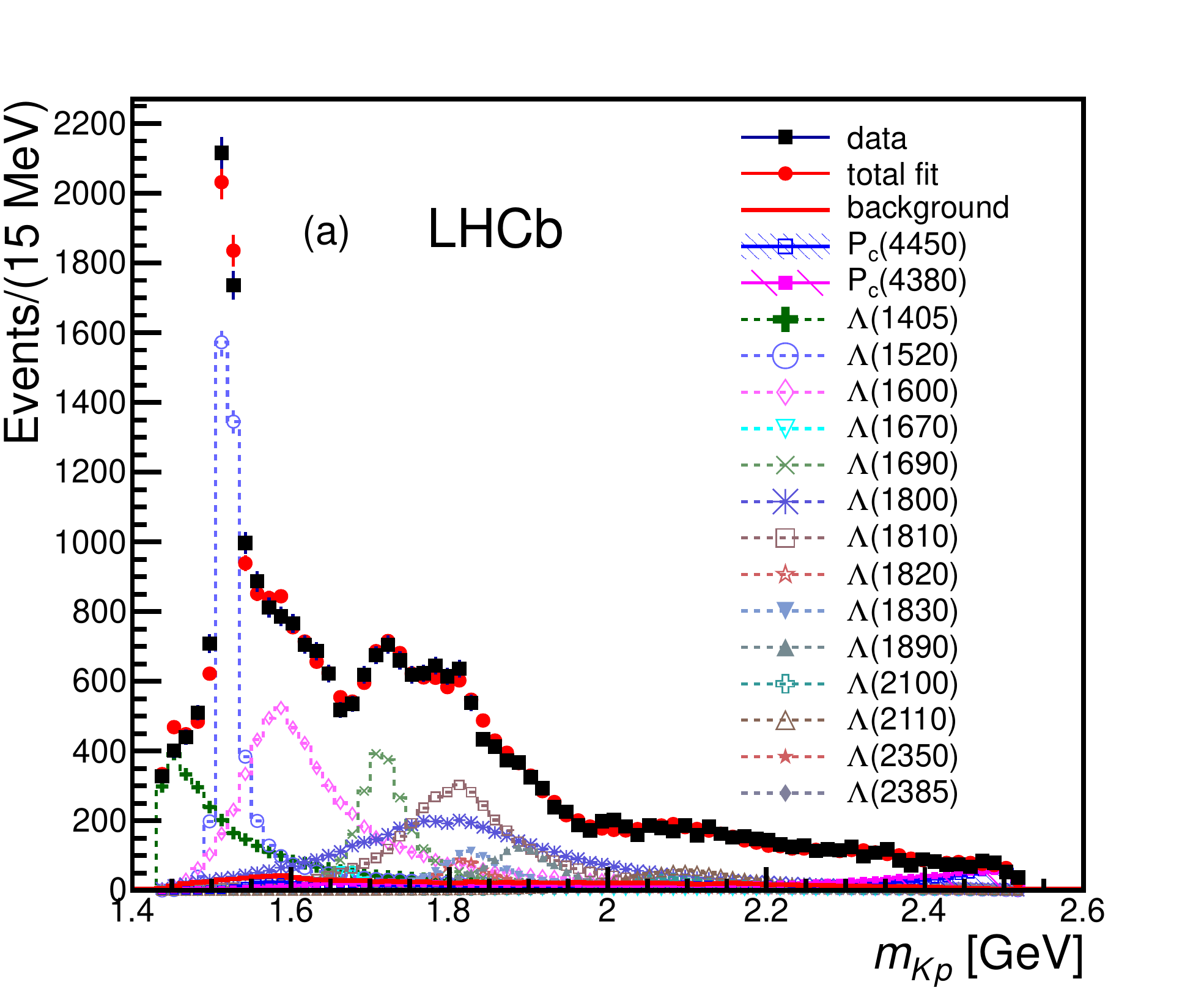}\includegraphics[width=0.49\textwidth]{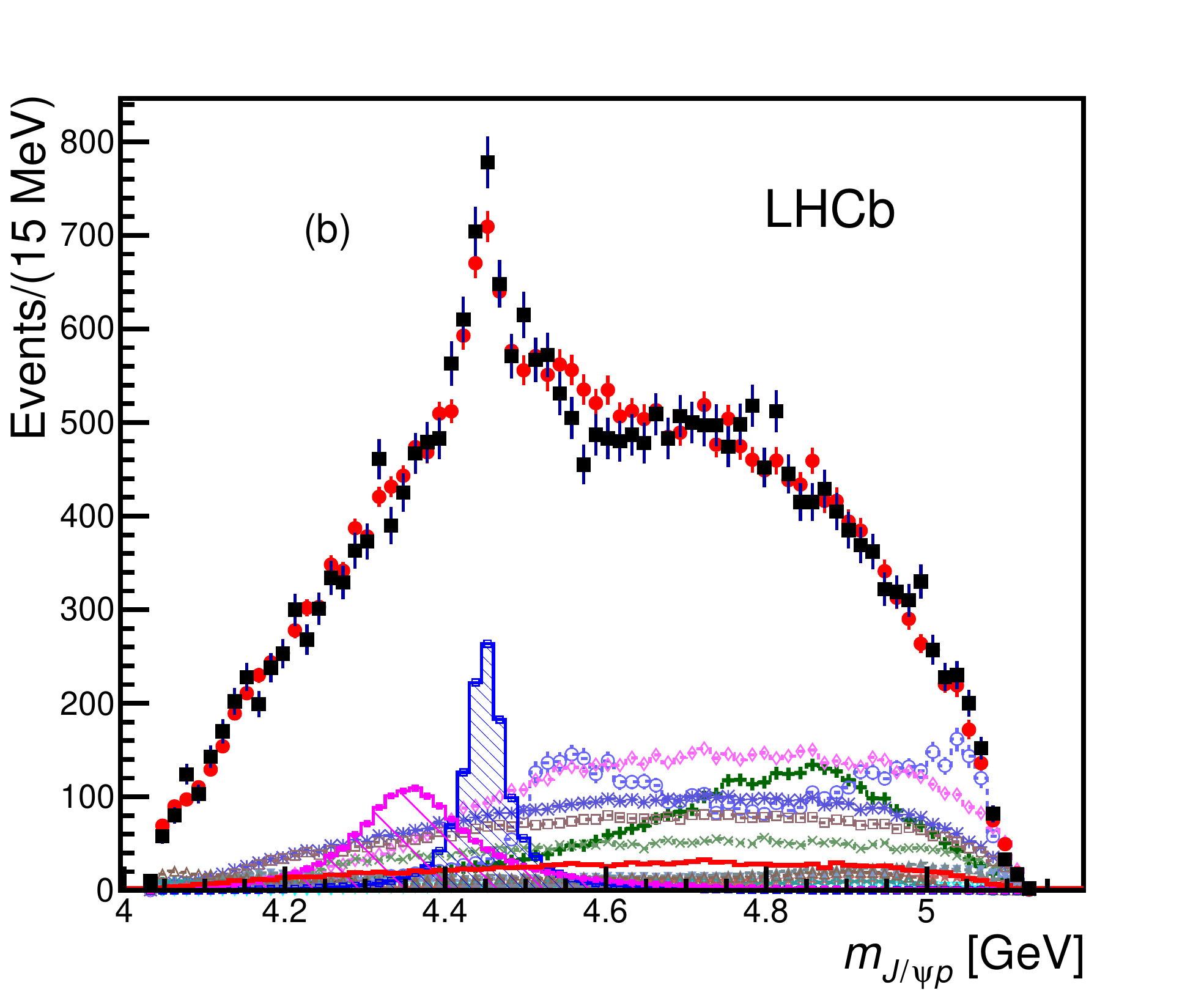}
\end{center}
\vskip -0.5cm
\caption{Results from cFit for (a) $m_{Kp}$  and (b) $m_{\jpsi p}$ for the extended model with two $P_c^+$ states. The data are shown (black) squares with error bars, while the (red) circles show the results of the fit.  Each $\Lz^*$ component is also shown. The (blue) open squares and (purple) solid squares show the two $P_c^+$ states. }
\label{M2all}
\end{figure}

\section{Fit fraction comparison between cFit and sFit}
The fit fraction for a given resonance is a ratio of the phase space integrals of the matrix element squared calculated for the resonance amplitude taken alone and for the total matrix element summing over all contributions.
The fit fractions are listed in Table~\ref{tab:fitfrac}.  The $P_c^+$ states have well determined fit fractions. There is good agreement between cFit and sFit. Note that the results for the $\Lz(1405)$ resonance are based on our use of a particular Flatt\'e amplitude model.
\begin{table}[htp]
\centering
\caption{Fit fractions of the different components from cFit and sFit for the default ($3/2^-$, $5/2^+$) model. Uncertainties are statistical only. }
\vspace{0.2cm}
\begin{tabular}{lrrrr}
\hline
~Particle & Fit fraction (\%) cFit& Fit fraction (\%) sFit\\
\hline
$P_c(4380)^+$& $8.42\pm 0.68 \;\;\;\;$ &$7.96\pm 0.67 \;\;\;\;$ \\
$P_c(4450)^+$& $4.09\pm 0.48 \;\;\;\;$ &$4.10\pm 0.45 \;\;\;\;$ \\ 
$\Lz(1405)$&$14.64\pm 0.72 \;\;\;\;$ & $14.19\pm 0.67 \;\;\;\;$\\
$\Lz(1520)$& $18.93\pm 0.52 \;\;\;\;$ & $19.06\pm 0.47 \;\;\;\;$\\
$\Lz(1600)$& $23.50\pm 1.48 \;\;\;\;$ & $24.42\pm 1.36 \;\;\;\;$\\
$\Lz(1670)$ &  $1.47\pm 0.49 \;\;\;\;$ &$1.53\pm 0.50 \;\;\;\;$ \\
$\Lz(1690)$& $8.66\pm 0.90 \;\;\;\;$ &$8.60\pm 0.85 \;\;\;\;$ \\
$\Lz(1800)$& $18.21\pm 2.27 \;\;\;\;$ &$16.97\pm 2.20 \;\;\;\;$ \\
$\Lz(1810)$&  $17.88\pm 2.11 \;\;\;\;$ &$17.29\pm 1.85 \;\;\;\;$ \\
$\Lz(1820)$& $2.32\pm 0.69 \;\;\;\;$ & $2.32\pm 0.65 \;\;\;\;$\\
$\Lz(1830)$& $1.76\pm 0.58 \;\;\;\;$ &$ 2.00\pm 0.53 \;\;\;\;$\\
$\Lz(1890)$& $3.96\pm 0.43 \;\;\;\;$ &$ 3.97\pm 0.38 \;\;\;\;$\\
$\Lz(2100)$& $1.65\pm 0.29 \;\;\;\;$ &$ 1.94\pm 0.28 \;\;\;\;$\\
$\Lz(2110)$& $1.62\pm 0.32 \;\;\;\;$ &$ 1.44\pm 0.28 \;\;\;\;$\\
\hline
\end{tabular}
\label{tab:fitfrac}
\end{table}

\section{Details of the matrix element for the decay amplitude}
\label{SUPPsec:matrixelement}

\def\ZP{P_c}
\def\LambdaStar{{\Lz^*}}
\def\LambdaStarn{{\Lz^*_{\!n}}}
\def\H{{\cal H}}
\def\F#1{\{#1\}}
\def\BA#1#2#3{{#1}_{{#2}}^{\,\,\F{\!#3\!}}}
\def\Mat{\mathcal{M}}

The matrix element for $\Lb\to \psi K^- p$, $\psi\to\mu^+\mu^-$ 
decays\footnote{We denote $\jpsi$ as $\psi$ for efficiency of the notation.} 
must allow for various conventional $\LambdaStar\to K^- p$ resonances  
and exotic pentaquark states $\ZP^+\to\psi p$ that could interfere with each other.

We use the helicity formalism to write down the matrix element.
To make the derivation of the matrix element easier to comprehend we start with a
brief outline of this formalism and our notation. 
Then we discuss the application to the 
$\Lb\to \LambdaStar\psi$, $\LambdaStar\to K^- p$, $\psi\to\mu^+\mu^-$ decay sequence, called 
hereafter the $\LambdaStar$ decay chain matrix element. 
Next we discuss construction of the
$\Lb\to \ZP^+ K^-$, $\ZP^+\to \psi p$, $\psi\to\mu^+\mu^-$ decay sequence, called
hereafter the $\ZP$ decay chain matrix element, which can be 
coherently added to that for the $\LambdaStar$ decay chain.  We also discuss a possible reduction of the
number of helicity couplings to be determined from the data using their relationships to 
the $LS$ couplings.

\subsection{Helicity formalism and notation}
\label{SUPPsec:hformalism}
 
For each two-body decay $A\to B\,C$, 
a coordinate system is set up in the rest frame of $A$, 
with $\hat{z}$ being\footnote{The ``hat'' symbol denotes a unit vector in a given direction.}
 the direction of quantization for its spin. 
We denote this coordinate system as
$(\BA{{x}}{0}{A},\BA{{y}}{0}{A},\BA{{z}}{0}{A})$,
where the superscript ``$\{A\}$'' means ``in the rest frame of $A$'', while
the subscript ``0'' means the initial coordinates.
For the first particle in the decay chain ($\Lb$), the choice of these
coordinates is arbitrary.\footnote{When designing an analysis to be sensitive (or insensitive) to a particular case of polarization, the choice is not arbitrary, but this does not change the fact that one can quantize the $\Lb$ spin along any well-defined direction.  The $\Lb$ polarization may be different for different choices.}
However, once defined, these coordinates must be used
consistently between all decay sequences described by the matrix element.
For subsequent decays, \eg\ $B\to D\,E$, the choice of these coordinates is
already fixed by the transformation from the $A$ to the $B$ rest frames, as discussed below.
Helicity is defined as the projection of the spin of the particle onto 
the direction of its momentum. When the $z$ axis coincides with the particle momentum,
we denote its spin projection onto it (\ie\ the $m_z$ quantum number) as $\lambda$.  
To use the helicity formalism, the initial coordinate system must be rotated to
align the $z$ axis with the direction of the momentum of one of the daughter particles, \eg\ the $B$. 
A generalized rotation operator can be formulated
in three-dimensional space, ${\cal R}(\alpha,\beta,\gamma)$, that uses Euler angles. 
Applying this operator results in a sequence of rotations: first
 by the angle $\alpha$ about the $\hat{z}_0$ axis,
followed by the angle $\beta$ about the rotated $\hat{y}_1$ axis
and then finally by the angle $\gamma$ about the rotated $\hat{z}_2$ axis.
We use a subscript denoting the axes, to specify the rotations which have been already 
performed on the coordinates.  
The spin eigenstates of particle $A$,  $|J_A,m_A\rangle$, 
in the $(\BA{x}{0}{A},\BA{y}{0}{A},\BA{z}{0}{A})$ coordinate system can be expressed in the basis of its spin eigenstates, $|J_A,m_A'\rangle$, 
in the rotated $(\BA{x}{3}{A},\BA{y}{3}{A},\BA{z}{3}{A})$ coordinate system 
with the help of Wigner's $D-$matrices
\begin{equation}
|J_A,m_A\rangle = \sum\limits_{m_A'} D^{\,J_A}_{m_A,\,m_A'}(\alpha,\beta,\gamma)^*\, |J_A,m_A'\rangle,   
\end{equation}
where
\begin{equation}
D^{\,J}_{m,\,m'}(\alpha,\beta,\gamma)^* =
\langle J,m|{\cal R}(\alpha,\beta,\gamma)|J,m'\rangle^*=
e^{i\,m\alpha}\,\,d^{\,J}_{m,m'}(\beta)\,\,e^{i\,m'\gamma},
\label{SUPeq:dmatrix}
\end{equation}
and where the small-$d$ Wigner matrix contains known functions of $\beta$ that depend on $J,m,m'$.  
To achieve the rotation of the original $\BA{\hat{z}}{0}{A}$ axis onto the $B$ momentum 
($\BA{\vec{p}}{B}{A}$), it
is sufficient to rotate by $\alpha=\BA{\phi}{B}{A}$, $\beta=\BA{\theta}{B}{A}$,
where $\BA{\phi}{B}{A}$, $\BA{\theta}{B}{A}$ are the azimuthal and polar angles of the $B$ momentum 
vector in the original coordinates \ie\ $(\BA{\hat{x}}{0}{A},\BA{\hat{y}}{0}{A},\BA{\hat{z}}{0}{A})$.
This is depicted in Fig.~\ref{fig:helicitygeneric}, for the case when the quantization axis for the
spin of $A$ is  its momentum in some other reference frame.
Since the third rotation is not necessary, we set $\gamma=0$.\footnote{An alternate convention is
to set $\gamma=-\alpha$. The two conventions lead to equivalent formulae.}
The angle $\BA{\theta}{B}{A}$ is usually called ``the $A$ helicity angle'', thus to simplify the notation
we will denote it as $\theta_A$. 
For compact notation, we will also denote $\BA{\phi}{B}{A}$ as $\phi_B$.  
These angles can be determined from\footnote{The function atan2$(x,y)$ is the $\tan^{-1}(y/x)$ function with two arguments. The purpose of using two arguments
instead of one is to gather information on the signs of the inputs in order to return the appropriate
quadrant of the computed angle.}
\begin{align}
\phi_B & =  {\rm atan2}\left( {\BA{p}{B}{A}}_{\!y},\,{\BA{p}{B}{A}}_{\!x} \right) \notag\\
       & =  {\rm atan2}\left( \BA{\hat{y}}{0}{A} \cdot\BA{\vec{p}}{B}{A} ,\, \BA{\hat{x}}{0}{A}\cdot\BA{\vec{p}}{B}{A} \right) \notag\\
       & =  {\rm atan2}\left( (\BA{\hat{z}}{0}{A} \times\BA{\hat{x}}{0}{A}) \cdot\BA{\vec{p}}{B}{A} ,\, \BA{\hat{x}}{0}{A} \cdot \BA{\vec{p}}{B}{A} \right),
   \label{SUPeq:ph}\\
\cos\theta_A & = \BA{\hat{z}}{0}{A} \cdot \BA{\hat{p}}{B}{A}. \label{SUPeq:theta} 
\end{align}
\begin{figure}[b]
\begin{center}
\includegraphics[width=0.7\textwidth]{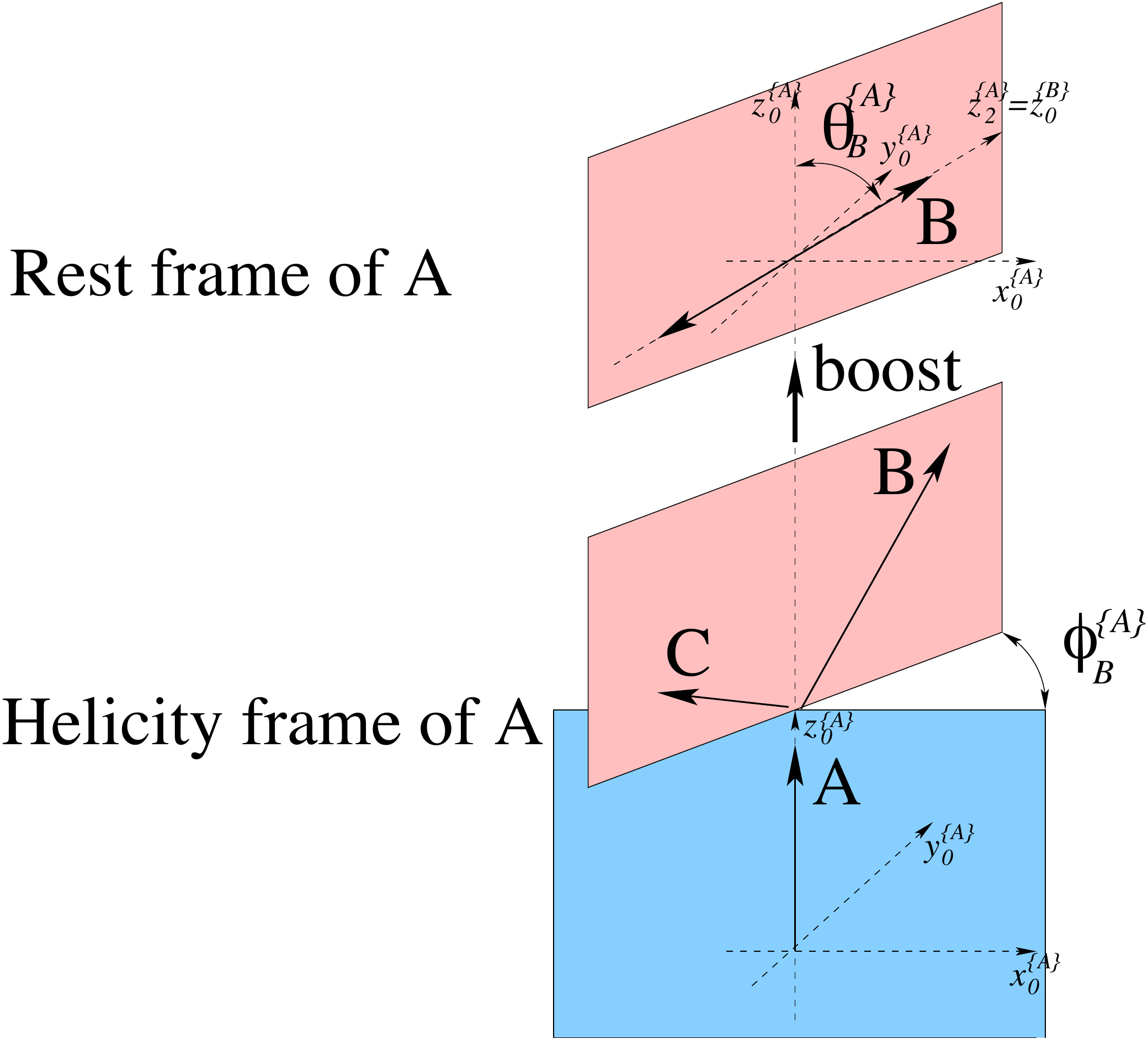}
\end{center}
\vskip -0.5cm
\caption{
Coordinate axes for the spin quantization of particle $A$ (bottom part), 
chosen to be the helicity frame of $A$ 
($\hat{z}_{0}||\vec{p}_{A}$ in the rest frame of its mother particle or in the laboratory frame),
together with the polar ($\BA{\theta}{B}{A}$) and azimuthal ($\BA{\phi}{B}{A}$) angles
of the momentum of its daughter $B$ in the $A$ rest frame (top part).
Notice that the directions of these coordinate axes, denoted as 
$\BA{\hat{x}}{0}{A}$, 
$\BA{\hat{y}}{0}{A}$, and 
$\BA{\hat{z}}{0}{A}$,  
do not change when boosting from the helicity frame
of $A$ to its rest frame. 
After the Euler 
rotation ${\cal R}(\alpha=\BA{\phi}{B}{A},\beta=\BA{\theta}{B}{A},\gamma=0)$
(see the text),
the rotated $z$ axis, $\BA{\hat{z}}{2}{A}$, 
is aligned with the $B$ momentum; thus the rotated coordinates
become the helicity frame of $B$. 
If $B$ has a sequential decay, then the same boost-rotation process
is repeated to define the helicity frame for its daughters.
}
\label{fig:helicitygeneric}
\end{figure}

Angular momentum conservation requires $m_A'=m_B'+m_C'=\lambda_B - \lambda_C$
(since $\BA{\vec{p}}{C}{A}$ points in the opposite direction to $\BA{\hat{z}}{3}{A}$, $m_C'=-\lambda_C$).
Each two-body decay adds a multiplicative term to the matrix element
\begin{equation}
 \H_{\lambda_B,\,\lambda_C}^{A\to B\,C}\,D^{\,J_A}_{m_A,\,\lambda_B-\lambda_C}(\phi_B,\theta_A,0)^*.
\label{SUPeq:ABCterm}
\end{equation}
The helicity couplings $\H_{\lambda_B,\,\lambda_C}^{A\to B\,C}$ are complex constants.
Their products from subsequent decays are to be determined by the fit to the data
(they represent the decay dynamics).
If the decay is strong or electromagnetic, it conserves parity which reduces the number of independent helicity couplings via the relation\begin{equation}
\H_{-\lambda_B,-\lambda_C}^{A\to B\,C} = P_A\,P_B\,P_C\,(-1)^{J_B+J_C-J_A}\, \H_{\lambda_B,\,\lambda_C}^{A\to B\,C},
\label{SUPeq:parity}
\end{equation}
where $P$ stands for the intrinsic parity of a particle. 

After multiplying terms given by Eq.~(\ref{SUPeq:ABCterm}) for all decays in the decay sequence, they must be summed up coherently over the helicity states of intermediate particles, and incoherently over  the helicity states of the initial and final-state particles. Possible helicity values of $B$ and $C$ particles are constrained by 
$|\lambda_B|\le J_B$, $|\lambda_C|\le J_C$ and $|\lambda_B-\lambda_C|\le J_A$.

When dealing with the subsequent decay of the daughter, $B\to D\, E$, 
four-vectors of all particles must be first Lorentz boosted to the rest frame of $B$, along 
the $\BA{\vec{p}}{B}{A}$ \ie\ $\BA{\hat{z}}{3}{A}$ direction (this is the $z$ axis in the rest frame of 
$A$ after the Euler rotations; we use the subscript ``3'' for the number of rotations performed on the
coordinates, because of the three Euler angles, however, since we use the $\gamma=0$ convention 
these coordinates are the same as after the first two rotations). 
This is visualized in Fig.~\ref{fig:helicitygeneric}, 
with $B\to D\, E$ particle labels replaced by $A\to B\, C$ labels. 
This transformation does not change vectors that are perpendicular to the boost direction.
The transformed coordinates become the initial coordinate system quantizing the spin of $B$ in its
rest frame,
\begin{align}
\BA{\hat{x}}{0}{B} & =\BA{\hat{x}}{3}{A}, \notag\\  
\BA{\hat{y}}{0}{B} & =\BA{\hat{y}}{3}{A}, \notag\\  
\BA{\hat{z}}{0}{B} & =\BA{\hat{z}}{3}{A}.       
\label{SUPeq:xbfromxa}
\end{align}
The processes of rotation and subsequent boosting can be repeated until the final-state particles are reached.
In practice, there are two equivalent ways to determine the $\BA{\hat{z}}{0}{B}$ direction.
Using Eq.~(\ref{SUPeq:xbfromxa}) we can set it to the direction of  the $B$ momentum in the $A$ rest frame
\begin{equation}
\BA{\hat{z}}{0}{B} =\BA{\hat{z}}{3}{A} = \BA{\hat{p}}{B}{A}.
\label{SUPeq:zboost}
\end{equation}
Alternatively, we can make use of the fact that $B$ and $C$ are 
back-to-back in the rest frame of $A$, $\BA{\vec{p}}{C}{A}= - \BA{\vec{p}}{B}{A}$.
Since the momentum of $C$ is antiparallel to the boost direction from the $A$ to $B$ rest frames,
the $C$ momentum in the $B$ rest frame will be different, but it will still be antiparallel to this
boost direction
\begin{equation}
\BA{\hat{z}}{0}{B} = - \BA{\hat{p}}{C}{B}.
\label{SUPeq:zrecoil}
\end{equation}
To determine $\BA{\hat{x}}{0}{B}$
from Eq.~(\ref{SUPeq:xbfromxa}), we need to find  $\BA{\hat{x}}{3}{A}$.
After the first rotation by $\phi_B$ about $\BA{\hat{z}}{0}{A}$, 
the $\BA{\hat{x}}{1}{A}$ axis is along the component of 
$\BA{\vec{p}}{B}{A}$ which is perpendicular to the $\BA{\hat{z}}{0}{A}$ axis
\begin{align}
\BA{\vec{a}}{B\perp z_0}{A} & \equiv (\BA{\vec{p}}{B}{A})_{\perp \BA{\hat{z}}{0}{A}}                   
                    = {\BA{\vec{p}}{B}{A}} - (\BA{\vec{p}}{B}{A})_{|| \BA{\hat{z}}{0}{A}}, \notag\\
                   & = {\BA{\vec{p}}{B}{A}} - ({\BA{\vec{p}}{B}{A}}\cdot\BA{\hat{z}}{0}{A})\,\BA{\hat{z}}{0}{A}, \notag\\
\BA{\hat{x}}{1}{A} & = \BA{\hat{a}}{B\perp z_0}{A} = \frac{\BA{\vec{a}}{B\perp z_0}{A}}{|\,\BA{\vec{a}}{B\perp z_0}{A}\,|}.
\label{SUPeq:xonea}
\end{align}
After the second rotation by $\theta_A$ about $\BA{\hat{y}}{1}{A}$,
$\BA{\hat{z}}{2}{A}\equiv \BA{\hat{z}}{3}{A} = \BA{\hat{p}}{B}{A}$, and 
$\BA{\hat{x}}{2}{A}= \BA{\hat{x}}{3}{A}$ is antiparallel to 
the component of the $\BA{\hat{z}}{0}{A}$ vector that is perpendicular to 
the new $z$ axis \ie\ $\BA{\hat{p}}{B}{A}$. Thus
\begin{align}
\BA{\vec{a}}{z_0\perp B}{A} & \equiv
(\BA{\hat{z}}{0}{A})_{\perp \BA{\vec{p}}{B}{A}}                   
                    = {\BA{\hat{z}}{0}{A}} - ({\BA{\hat{z}}{0}{A}}\cdot\BA{\hat{p}}{B}{A})\,\BA{\hat{p}}{B}{A}, \notag\\
\BA{\hat{x}}{0}{B} &=
\BA{\hat{x}}{3}{A}  = \,-\,\, \BA{\hat{a}}{z_0\perp B}{A} = \,-\,\,\frac{\BA{\vec{a}}{z_0\perp B}{A}}{|\,\BA{\vec{a}}{z_0\perp B}{A}\,|}.
\label{SUPeq:xaxis}
\end{align}
Then we obtain $\BA{\hat{y}}{0}{B}=\BA{\hat{z}}{0}{B}\times\BA{\hat{x}}{0}{B}$. 

If $C$ also decays, $C\to F\, G$, then the coordinates for the quantization of $C$ spin
in the $C$ rest frame are defined by
\begin{align}
\BA{\hat{z}}{0}{C} & = - \BA{\hat{z}}{3}{A} = \BA{\hat{p}}{C}{A} = - \BA{\hat{p}}{B}{C}, \\        
\BA{\hat{x}}{0}{C} & =  \BA{\hat{x}}{3}{A} = \,-\, \BA{\hat{a}}{z_0\perp B}{A} = + \BA{\hat{a}}{z_0\perp C}{A}, \label{SUPeq:xcaxis} \\  
\BA{\hat{y}}{0}{C} & = \BA{\hat{z}}{0}{C}\times\BA{\hat{x}}{0}{C}, 
\label{SUPeq:xcfromxa}
\end{align}
\ie\ the $z$ axis is reflected compared to the system used for the decay of particle $B$
(it must point in the direction of $C$ momentum in the $A$ rest frame),
but the $x$ axis is kept the same, since we chose particle $B$ for the rotation used 
in Eq.~(\ref{SUPeq:ABCterm}).

\subsection{Matrix element for the $\LambdaStar$ decay chain}
\label{SUPPsec:mlambdastar}
    
We first discuss the part of the matrix element 
describing conventional $\Lb\to\LambdaStarn\psi$, 
$\LambdaStarn\to K p$ decays (\ie\ $\LambdaStar$ decay chain),
where $\LambdaStarn$ denotes various possible excitations of the $\Lz$, \eg\ $\Lz(1520)$.
For simplicity we often refer to $\LambdaStarn$ as $\LambdaStar$, 
unless we label an $n$-dependent quantity. 

The weak decay $\Lb\to\LambdaStarn\psi$ is described by 
\begin{equation}
\label{SUPeq:lbtolpsi}
\H^{\Lb\to \LambdaStarn \psi}_{\lambda_{\LambdaStar},\,\lambda_{\psi}} \,\,
D^{\,\,\frac{1}{2}}_{\lambda_{\Lb},\,\lambda_\LambdaStar-\lambda_\psi}(
\phi_{\LambdaStar},\theta_{\Lb},0)^*,
\end{equation}
where $\H^{\Lb\to \LambdaStarn \psi}_{\lambda_{\LambdaStar},\,\lambda_{\psi}}$ 
are resonance (\ie\ $n$) dependent helicity couplings to be determined by a fit to the data.
There are 4 different complex values of these 
couplings to be determined for each $\LambdaStarn$ resonance
with spin $J_{\LambdaStarn}=\frac{1}{2}$, and 6 values for higher spins.
The couplings are complex parameters; thus each independent coupling contributes 
2 free parameters (taken to be real and imaginary parts) to the fit.
Since the $\psi$ and $\LambdaStar$ are intermediate particles in the decay chain, the
matrix element terms for different values of $\lambda_\psi$ and $\lambda_\LambdaStar$ must be added
coherently. 

The choice of the $\BA{\hat{z}}{0}{\Lb}$ direction for the $\Lb$ spin quantization is arbitrary.
We choose the $\Lb$ momentum in the {\rm lab} frame 
to define the $\BA{\hat{z}}{0}{\Lb}$ direction, 
giving its spin projection onto this axis
the meaning of the $\Lb$ helicity ($\lambda_{\Lb}$).
In the $\Lb$ rest frame, this direction is defined by the direction
of the boost from the {\rm lab} frame (Eq.~(\ref{SUPeq:zboost})),
\begin{equation}
\BA{\hat{z}}{0}{\Lb}=\BA{\hat{p}}{\Lb}{{\rm lab}},
\label{SUPeq:zlambdab}
\end{equation}
as depicted in Fig.~\ref{fig:helicitylambdastar}.
With this choice, $\theta_{\Lb}$ is the $\Lb$
helicity angle and can be calculated as
\begin{equation}
\label{SUPeq:lbhel}
\cos\theta_{\Lb}= \BA{\hat{p}}{\Lb}{{\rm lab}} \cdot \BA{\hat{p}}{\LambdaStar}{\Lb}.
\end{equation}
Longitudinal polarization of  the $\Lb$ via strong production mechanisms is forbidden due to parity conservation in strong interactions,
causing $\lambda_{\Lb}=+\frac{1}{2}$ and $-\frac{1}{2}$ to be equally likely.
Terms with different $\lambda_{\Lb}$ values must be added incoherently.   
The choice of $\BA{\hat{x}}{0}{\Lb}$ direction in the $\Lb$ rest frame is also arbitrary.
We use the $\Lb\to\LambdaStar\psi$ decay plane in the {\rm lab} frame to define it, which makes
the $\phi_\LambdaStar$ angle zero by definition.

\begin{figure}[htb]
\begin{center}
\includegraphics[width=.9\textwidth]{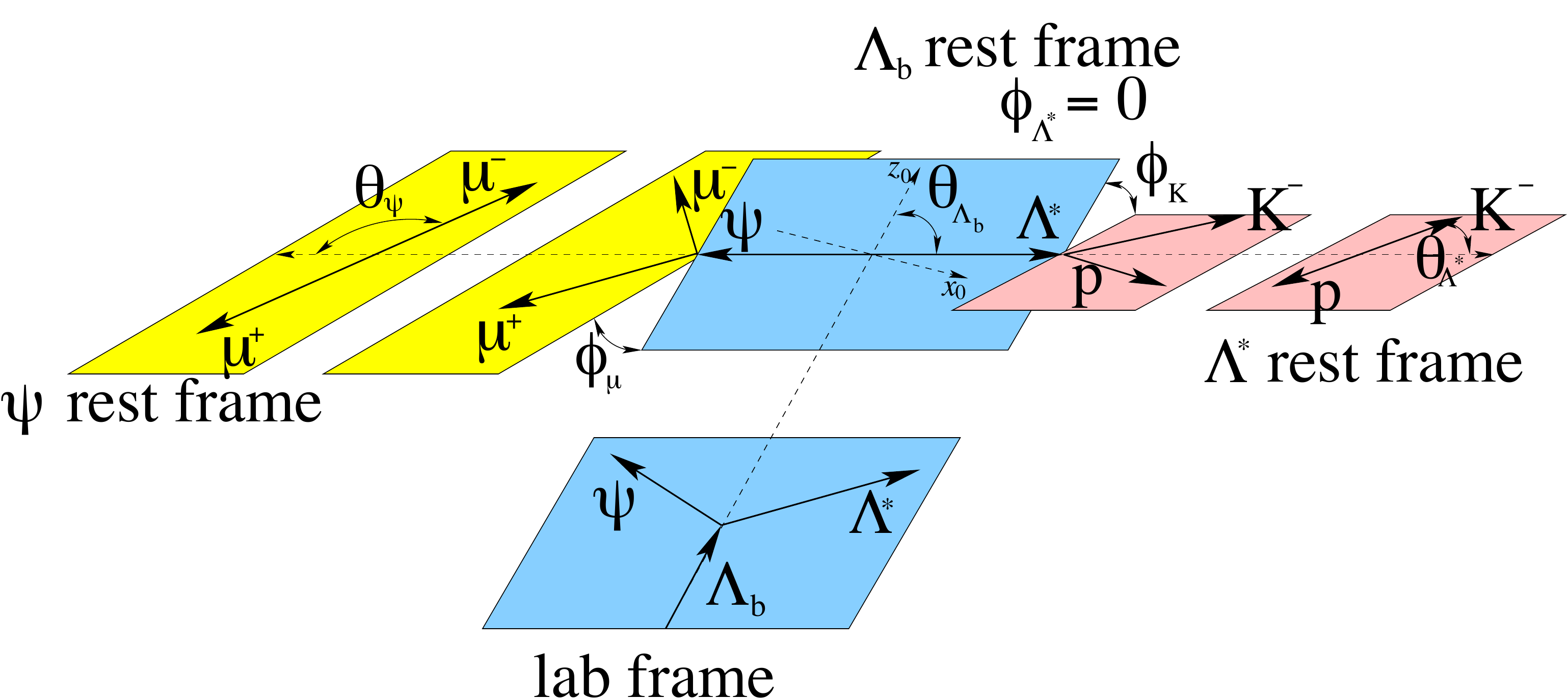}
\end{center}
\vskip -0.5cm
\caption{
Definition of the decay angles in the $\Lz^*$ decay chain. 
}
\label{fig:helicitylambdastar}
\end{figure} 

The strong decay $\LambdaStarn\to K p$ is described by a term
\begin{equation}
\label{SUPeq:ltopk}
\H^{\LambdaStarn\to K p}_{\lambda_p} \,\,
D^{\,\,J_{\LambdaStarn}}_{\lambda_{\LambdaStar},\,\lambda_p}(
\phi_{K},\theta_{\LambdaStar},0)^*\,\,
R_{\LambdaStarn}(m_{Kp}).
\end{equation}
Since the $K^-$ meson is spinless, the resonance-dependent 
helicity coupling $\H^{\LambdaStarn\to K p}_{\lambda_p}$ depends only 
on proton helicity, $\lambda_p=\pm\frac{1}{2}$.
As strong decays conserve parity, the two helicity couplings are related
\begin{equation}
\H^{\LambdaStarn\to K p}_{-\lambda_p} = - P_{\LambdaStarn}\,(-1)^{J_{\LambdaStarn}-\frac{1}{2}}\,\, \H^{\LambdaStarn\to K p}_{\lambda_p},
\end{equation}
where $P_{\LambdaStarn}$ is the parity of $\LambdaStarn$.
Since the overall magnitude and phase of $\H^{\LambdaStarn\to K p}_{+\frac{1}{2}}$ 
can be absorbed into a redefinition of the
$\H^{\Lb\to \LambdaStarn \psi}_{\lambda_{\LambdaStar},\,\lambda_{\psi}}$
couplings, we set $\H^{\LambdaStarn\to K p}_{+\frac{1}{2}} = (1,0)$ and
$\H^{\LambdaStarn\to K p}_{-\frac{1}{2}} = ( P_{\LambdaStarn}\,(-1)^{J_{\LambdaStarn}-\frac{3}{2}} , 0 )$,
where the values in parentheses give the real and imaginary parts of the couplings.
 
The angles $\phi_{K}$ and $\theta_{\LambdaStar}$ are the azimuthal and polar angles of the kaon in the $\LambdaStar$ rest frame
(see Fig.~\ref{fig:helicitylambdastar}).
The $\BA{\hat{z}}{0}{\LambdaStar}$ direction is defined by the boost direction from the $\Lb$ rest frame, which coincides
with the $-\BA{\vec{p}}{\psi}{\LambdaStar}$ direction in this 
frame (Eq.~(\ref{SUPeq:zrecoil})). This leads to
\begin{equation}
\label{SUPeq:thl}
\cos\theta_{\LambdaStar}= - \BA{\hat{p}}{\psi}{\LambdaStar} \cdot \BA{\hat{p}}{K}{\LambdaStar},
\end{equation}
with both vectors in the $\LambdaStar$ rest frame.
As explained in Sec.~\ref{SUPPsec:hformalism},
the $\BA{\hat{x}}{0}{\LambdaStar}$ direction 
is defined by the choice of coordinates in the $\Lb$ rest frame discussed above.
Following Eq.~(\ref{SUPeq:xaxis}) and (\ref{SUPeq:zlambdab}), we have
\begin{align}
\BA{\vec{a}}{z_0\perp\LambdaStar}{\Lb}
                   & = \BA{\hat{p}}{\Lb}{{\rm lab}} - (\BA{\hat{p}}{\Lb}{{\rm lab}}\cdot\BA{\hat{p}}{\LambdaStar}{\Lb})\,\BA{\hat{p}}{\LambdaStar}{\Lb}, \notag\\
\BA{\hat{x}}{0}{\LambdaStar}& =
\BA{\hat{x}}{3}{\Lb}  =\, -\,\, \frac{\BA{\vec{a}}{z_0\perp\LambdaStar}{\Lb}}{
                               |\,\BA{\vec{a}}{z_0\perp\LambdaStar}{\Lb}\,|}~.
\label{SUPeq:xlambda}
\end{align}
The azimuthal angle of the $K^-$ can now be determined in the $\LambdaStar$ 
rest frame from (Eq.~(\ref{SUPeq:ph}))
\begin{equation}
\label{SUPeq:phk}
\phi_K = {\rm atan2} \left( -(\BA{\hat{p}}{\psi}{\LambdaStar}\times\BA{\hat{x}}{0}{\LambdaStar})\cdot\BA{\hat{p}}{K}{\LambdaStar},\, \BA{\hat{x}}{0}{\LambdaStar}\cdot\BA{\hat{p}}{K}{\LambdaStar} \right).
\end{equation}

The term $R_{\LambdaStarn}(m_{Kp})$ describes the $\LambdaStarn$ 
resonance that appears in the invariant mass distribution of the kaon-proton system,
\begin{equation}
R_{\LambdaStarn}(m_{Kp}) =
B'_{L_{\Lb}^{\LambdaStarn}}(p,p_0,d) \left(\frac{p}{M_{\Lb}}\right)^{L_{\Lb}^{\LambdaStarn}} \, 
{\rm BW}(m_{Kp} | M_{0}^{\LambdaStarn}, \Gamma_{0}^{\LambdaStarn} ) \, 
B'_{L_{\LambdaStarn}}(q,q_0,d) 
\left(\frac{q}{M_{0}^{\LambdaStarn}}\right)^{L_{\LambdaStarn}}.
\label{SUPeq:resshape}
\end{equation}
Here, $p$ is the $\LambdaStar$ momentum in the $\Lb$ rest frame ($p=|\BA{\vec{p}}{\LambdaStar}{\Lb}|$).
Similarly, $q$ is the $K^-$ momentum in the $\LambdaStar$ rest frame ($q=|\BA{\vec{p}}{K}{\LambdaStar}|$).
The symbols $p_0$ and $q_0$ denote values of these quantities at the resonance peak ($m_{Kp}=M_0^{\LambdaStarn}$). 
The orbital angular momentum between the $\LambdaStar$ and $\psi$ particles 
in the $\Lb$ decay is denoted as $L_{\Lb}^{\LambdaStarn}$.
Similarly, $L_{\LambdaStarn}$ is the orbital angular momentum between the $p$ and $K^-$ in the $\LambdaStarn$ decay.
For the orbital angular momentum barrier factors, $p^L\,B'_{L}(p,p_0,d)$, we use the
Blatt-Weisskopf functions \cite{Blatt-Weisskopf-1979},
{
\def\1{p\phantom{_0}}
\def\2{p_0}
\small
\begin{align}
\label{SUPeq:blattw}
B'_{0}(p,p_0,d)=&1 \,,\nonumber \\
B'_{1}(p,p_0,d)=&\sqrt{ \frac{1+(\2\,d)^2}{1+(\1\,d)^2} } \,,\nonumber \\
B'_{2}(p,p_0,d)=&\sqrt{ \frac{9+3(\2\,d)^2+(\2\,d)^4}{9+3(\1\,d)^2+(\1\,d)^4} } \,,\\
B'_{3}(p,p_0,d)=&\sqrt{ \frac{225+45(\2\,d)^2+6(\2\,d)^4+(\2\,d)^6}{225+45(\1\,d)^2+6(\1\,d)^4+(\1\,d)^6} } \,,\nonumber\\
B'_{4}(p,p_0,d)=&\sqrt{ \frac{11025+1575(\2\,d)^2+135(\2\,d)^4+10(\2\,d)^6+(\2\,d)^8}{11025+1575(\1\,d)^2+135(\1\,d)^4+10(\1\,d)^6+(\1\,d)^8} } \,,\nonumber \\
B'_{5}(p,p_0,d)=&\sqrt{ \frac{893025+99225(\2\,d)^2+6300(\2\,d)^4+315(\2\,d)^6+15(\2\,d)^8+(\2\,d)^{10}}{
                            893025+99225(\1  \,d)^2+6300(\1  \,d)^4+315(\1  \,d)^6+15(\1  \,d)^8+(\1  \,d)^{10}} } \,\nonumber,
\end{align}
}
$\!\!\!$ to account for the difficulty in creating the orbital angular momentum $L$, which depends on the momentum of 
the decay products $p$ (in the rest frame of the decaying particle) and on the size of the decaying particle 
given by the constant $d$. 
We set $d=3.0~{\rm GeV}^{-1}$ $\sim$0.6~fm.  
The relativistic Breit-Wigner amplitude is given by
\begin{equation}   
{\rm BW}(m | M_0, \Gamma_0 ) = \frac{1}{M_0^2-m^2 - i M_0 \Gamma(m)} \,,
\label{SUPeq:breitwigner}
\end{equation}
where
\begin{equation}   
\Gamma(m)=\Gamma_0 \left(\frac{q}{q_0}\right)^{2\,L_{\LambdaStar}+1} \frac{M_0}{m} B'_{L_{\LambdaStar}}(q,q_0,d)^2 \,.
\label{SUPeq:mwidth}
\end{equation}

In the case of the $\Lz(1405)$ resonance, which peaks below the $K^-p$ threshold, we use a two-component width 
equivalent to the Flatt\'e parameterization. We add to the above width in the $K^-p$ channel, a width for its decay to the dominant 
$\PSigma^+\pi^-$ channel, $\Gamma(m)=\Gamma(m)_{K^- p}+\Gamma(m)_{\PSigma\pi}$, where $q$ in the second term and $q_0$ in both terms 
are calculated assuming the decay to $\PSigma^+\pi^-$. Assuming that both channels are dynamically equally likely and differ only
by the phase space factors we set $\Gamma_0$ to the total width of $\Lz(1405)$ in both terms.

Angular momentum conservation limits $L_{\LambdaStarn}$ to $J_{\LambdaStarn}\pm\frac{1}{2}$, 
which is then uniquely
defined by  parity conservation in the $\LambdaStarn$ decay, $P_{\LambdaStarn}=(-1)^{L_{\LambdaStarn}+1}$.
Angular momentum conservation also requires
${\rm max}(J_{\LambdaStarn}-\frac{3}{2}\,, \,0)\le L_{\Lb}^{\LambdaStar}\le J_{\LambdaStarn}+\frac{3}{2}$.
We assume the minimal value of $L_{\Lb}^{\LambdaStarn}$ in $R_{\LambdaStarn}(m_{Kp})$.

The electromagnetic decay $\psi\to\mu^+\mu^-$ is described by a term
\begin{equation}
\label{SUPeq:psitomm}
D^{\,\,1}_{\lambda_{\psi},\,\Delta\lambda_\mu}(
\phi_{\mu},\theta_{\psi},0)^*,
\end{equation}
where $\Delta\lambda_{\mu}\equiv\lambda_{\mu^+}-\lambda_{\mu^-}=\,\pm1$,
and $\phi_{\mu},\theta_{\psi}$ are the azimuthal and polar angles of $\mu^+$ for $\Lbbar$ ($\mu^-$ for $\Lb$ decays) 
in the $\psi$ rest frame (see Fig.~\ref{fig:helicitylambdastar}).
There are no helicity couplings in Eq.~(\ref{SUPeq:psitomm}), since they are all equal due to conservation of $C$ and $P$ parities.
Therefore, this coupling can be set to unity as its magnitude and phase can be absorbed into the other helicity couplings which are left free in the fit.
The calculation of the $\psi$ decay angles is analogous to that of the $\LambdaStar$ decay angles described above
(Eqs.~(\ref{SUPeq:thl})--(\ref{SUPeq:phk}))
\begin{align}
\cos\theta_{\psi} & =  -\, \BA{\hat{p}}{\LambdaStar}{\psi} \cdot \BA{\hat{p}}{\mu}{\psi}, \label{SUPeq:psihelth} \\
\phi_\mu & =  {\rm atan2} \left( -(\BA{\hat{p}}{\LambdaStar}{\psi}\times\BA{\hat{x}}{0}{\psi})\cdot\BA{\hat{p}}{\mu}{\psi},\, \BA{\hat{x}}{0}{\psi}\cdot\BA{\hat{p}}{\mu}{\psi} \right),
\label{SUPeq:psihelph} 
\end{align}
with
\begin{equation}
\BA{\hat{x}}{0}{\psi}=\BA{\hat{x}}{0}{\LambdaStar}=\BA{\hat{x}}{3}{\Lb}
\end{equation}
and $\BA{\hat{x}}{3}{\Lb}$ given by Eq.~(\ref{SUPeq:xlambda}).

Collecting terms from the subsequent decays together, the matrix element 
connecting different helicity states of the initial and the final-state particles
for the entire $\LambdaStar$ decay chain 
can be written as
\begin{align}  
\Mat_{\lambda_{\Lb},\,\lambda_p,\,\Delta\lambda_\mu}^{\,\,\LambdaStar} =  
&  \sum\limits_{n} 
     R_{\LambdaStarn}(m_{Kp}) \, 
     \H^{\LambdaStarn\to K p}_{\lambda_p}
     \sum\limits_{\lambda_\psi}
        e^{i\,\lambda_\psi\phi_{\mu}} \,\, 
        d^{\,\,1}_{\lambda_\psi,\Delta\lambda_\mu}(\theta_\psi) 
\notag\\
& ~~~\times       \sum\limits_{\lambda_\LambdaStar}
           \H^{\Lb\to \LambdaStarn \psi}_{\lambda_{\LambdaStar},\,\lambda_{\psi}} \,\,
           e^{i\,\lambda_\LambdaStar \phi_K} \,\,
           d^{\,\,\frac{1}{2}}_{\lambda_{\Lb},\,\lambda_{\LambdaStar}-\lambda_{\psi}}(\theta_{\Lb}) \,\,
           d^{\,\,J_{\LambdaStarn}}_{\lambda_{\LambdaStar},\,\lambda_{p}}(\theta_\LambdaStar).
\label{SUPeq:LambdaStar_matrixelement_partial}
\end{align}

\def\Pol{{\rm P}^{\Lb}}
Terms with different helicities of the initial and final-state particles ($\lambda_p$, $\Delta\lambda_\mu$) must be added incoherently
\begin{equation}
\left| \Mat^{\LambdaStar} \right|^2 = 
  \frac{1+\Pol}{2} \sum\limits_{\lambda_p} \sum\limits_{\Delta\lambda_\mu} 
   \left| \Mat_{(\lambda_{\Lb}=+1/2),\,\lambda_p,\,\Delta\lambda_\mu} \right|^2 +  
  \frac{1-\Pol}{2} \sum\limits_{\lambda_p} \sum\limits_{\Delta\lambda_\mu} 
   \left| \Mat_{(\lambda_{\Lb}=-1/2),\,\lambda_p,\,\Delta\lambda_\mu} \right|^2,
\label{SUPeq:eq:LambdaStar_matrixelement}
\end{equation}
where $\Pol$ is the $\Lb$ polarization, 
defined as the difference of probabilities for $\lambda_{\Lb}=+1/2$ and $-1/2$ \cite{Aaij:2013oxa}.
For our choice of the quantization axis for $\Lb$ spin, no polarization is expected
($\Pol=0$) from parity conservation in strong interactions which dominate $\Lb$ production at LHCb. 

\subsection{Matrix element for the $\ZP^+$ decay chain}
\label{SUPPsec:mzp}

\begin{figure}[b]
\begin{center}
\hbox{\includegraphics[width=.95\textwidth]{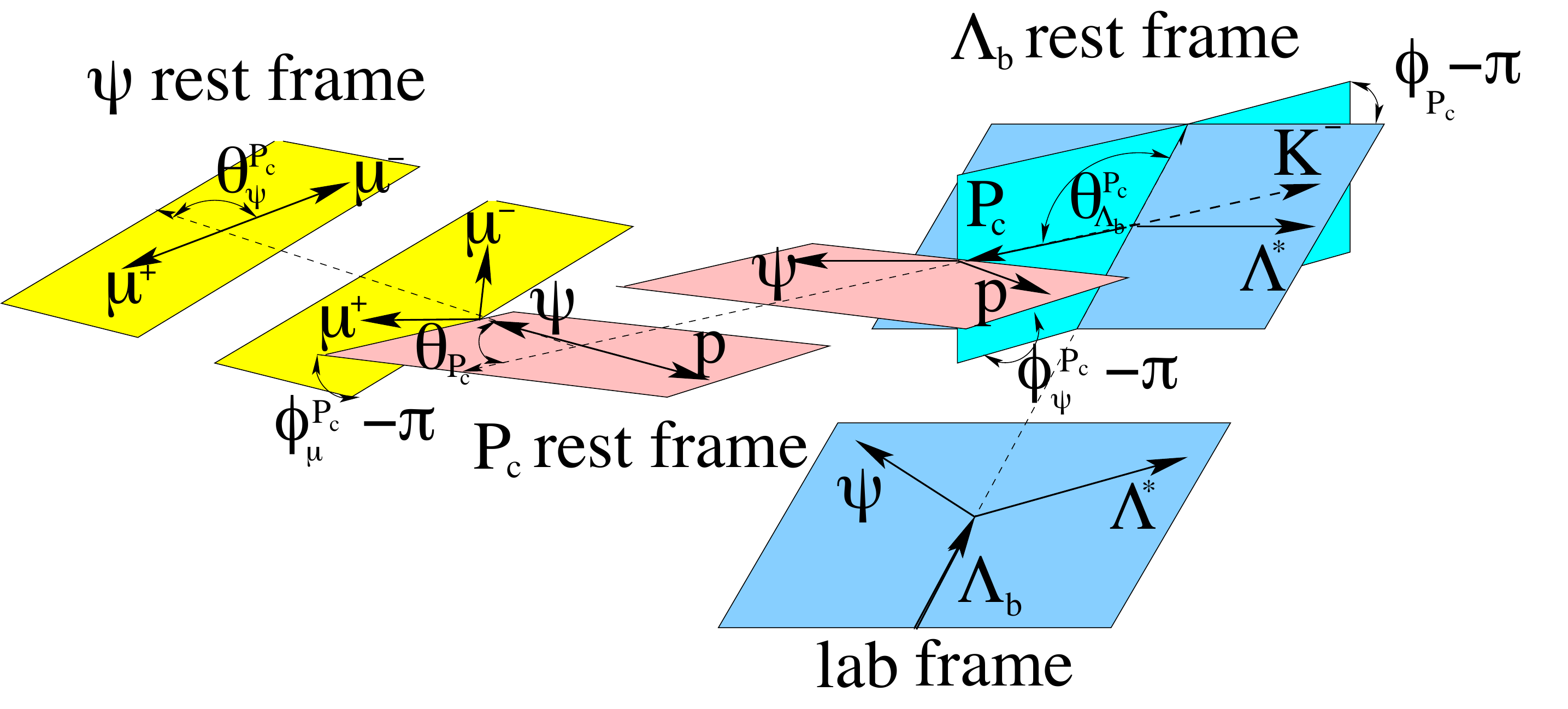}\hskip-4cm\quad}
\end{center}
\vskip -0.8cm
\caption{
Definition of the decay angles in the $\ZP^+$ decay chain. 
}
\label{fig:helicitypc}
\end{figure} 
We now turn to the discussion of $\Lb\to {\ZP}_j K^-$, ${\ZP}_j\to\psi p$ decays, in which we 
allow more than one pentaquark state, $j=1, 2, \ldots\,\,$. 
Superscripts containing the $\ZP$ decay chain name without curly brackets, \eg\ $\phi^{\,\ZP}$, 
 will denote quantities belonging to this decay chain and should not
be confused with the superscript ``$\{\ZP\}$'' denoting the $P_c^+$ rest frame, \eg\ $\phi^{\,\,\F{\ZP}}$.
With only a few exceptions, we omit the $\LambdaStar$ decay chain label.

The weak decay $\Lb\to{\ZP}_j K^-$ 
is described by the term,
\begin{equation}
\label{SUPeq:lbtozk}
\H^{\Lb\to {\ZP}_j K}_{\lambda_{\ZP}} \,\,
D^{\,\,\frac{1}{2}}_{\lambda_{\Lb},\,\lambda_{\ZP}}(
\phi_{\ZP},\theta_{\Lb}^{\ZP},0)^*,
\end{equation}
where $\H^{\Lb\to {\ZP}_j K}_{\lambda_{\ZP}}$ 
are resonance (\ie\ $j$) dependent helicity couplings.
The helicity of the pentaquark state, $\lambda_{\ZP}$, can take
values of $\pm\frac{1}{2}$ independently of its spin, 
$J_{{\ZP}_j}=\frac{1}{2}, \frac{3}{2}, \ldots\,\,$.
Therefore, there are two independent helicity couplings to be determined for
each ${\ZP}_j$ state.
 The above mentioned $\phi_{\ZP}$, $\theta_{\Lb}^{\ZP}$ symbols refer to 
the azimuthal and polar angles of $\ZP$ in the $\Lb$ rest frame
(see Fig.~\ref{fig:helicitypc}).

Similar to Eq.~(\ref{SUPeq:lbhel}), the $\Lb$ helicity angle in the $\ZP$ decay chain can be calculated as,
\begin{equation}
\label{SUPeq:lbhelz}
\cos\theta_{\Lb}^{\ZP}= \BA{\hat{p}}{\Lb}{{\rm lab}} \cdot \BA{\hat{p}}{\ZP}{\Lb}.
\end{equation}

The $\phi_{\ZP}$ angle cannot be set to zero, 
since we have already defined the $\BA{\hat{x}}{0}{\Lb}$ axis in the $\Lb$ rest frame 
by the $\phi_{\LambdaStar}=0$ convention. 
According to Eq.~(\ref{SUPeq:xonea}) ($\BA{x}{0}{\Lb}=\BA{x}{1}{\Lb}$) we have:
\begin{align}
\BA{\vec{a}}{\LambdaStar\perp z_0}{\Lb}
                   & = \BA{\vec{p}}{\LambdaStar}{\Lb} - (\BA{\vec{p}}{\LambdaStar}{\Lb}\cdot\BA{\hat{p}}{\Lb}{{\rm lab}})\,\BA{\hat{p}}{\Lb}{{\rm lab}}, \notag\\
\BA{\hat{x}}{0}{\Lb} 
                   & =  \frac{\BA{\vec{a}}{\LambdaStar\perp z_0}{\Lb}}{
                         |\,\BA{\vec{a}}{\LambdaStar\perp z_0}{\Lb}\,|}.
\label{SUPeq:xlambdab}
\end{align}
The $\phi_{\ZP}$ angle can be determined in the $\Lb$ rest frame from
\begin{equation}
\label{SUPeq:phz}
\phi_{\ZP} = {\rm atan2} \left( (\BA{\hat{p}}{\Lb}{{\rm lab}}\times\BA{\hat{x}}{0}{\Lb})\cdot\BA{\hat{p}}{\ZP}{\Lb},\, \BA{\hat{x}}{0}{\Lb}\cdot\BA{\hat{p}}{\ZP}{\Lb} \right).
\end{equation}

The strong decay ${\ZP}_j\to \psi p$ is described by a term
\begin{equation}
\label{SUPeq:ztopsip}
\H^{{\ZP}_j\to \psi p}_{\lambda_\psi^{\ZP},\lambda_p^{\ZP}} \,\,
D^{\,\,J_{{\ZP}_j}}_{\lambda_{\ZP},\,\lambda_\psi^{\ZP}-\lambda_p^{\ZP}}(
\phi_{\psi},\theta_{\ZP},0)^*\,\,
R_{{\ZP}_j}(M_{\psi p}),
\end{equation}
where $\phi_{\psi}^{\ZP},\theta_{\ZP}$ are the azimuthal and polar angles of the $\psi$ in the $\ZP$ rest frame 
(see Fig.~\ref{fig:helicitypc}).
They are defined analogously to Eqs.~(\ref{SUPeq:thl})$-$(\ref{SUPeq:phk}).  
The $\BA{\hat{z}}{0}{\ZP}$ direction is defined by the boost direction from the $\Lb$ rest frame, which coincides
with the $-\BA{\vec{p}}{K}{\ZP}$ direction. This leads to
\begin{equation}
\cos\theta_{\ZP}= - \BA{\hat{p}}{K}{\ZP} \cdot \BA{\hat{p}}{\psi}{\ZP}.
\end{equation}
The azimuthal angle of the $\psi$ can now be determined in the $\ZP$ rest frame (see Fig.~\ref{fig:helicitypc}) from
\begin{equation}
\label{SUPeq:phpsi}
\phi_\psi^{\ZP} = {\rm atan2} \left( -(\BA{\hat{p}}{K}{\ZP}\times\BA{\hat{x}}{0}{\ZP})\cdot\BA{\hat{p}}{\psi}{\ZP},\, \BA{\hat{x}}{0}{\ZP}\cdot\BA{\hat{p}}{\psi}{\ZP} \right).
\end{equation}

In Eq.~(\ref{SUPeq:ztopsip}),
the $\BA{\hat{x}}{0}{\ZP}$ direction is defined by the convention that we used in the
$\Lb$ rest frame. Thus,
similar to Eq.~(\ref{SUPeq:xlambda}) we have
\begin{align}
\BA{\vec{a}}{z_0\perp\ZP}{\Lb}
                   & = \BA{\hat{p}}{\Lb}{{\rm lab}} - (\BA{\hat{p}}{\Lb}{{\rm lab}}\cdot\BA{\hat{p}}{\ZP}{\Lb})\,\BA{\hat{p}}{\ZP}{\Lb} ,\notag\\
\BA{\hat{x}}{0}{\ZP} 
                   & =\, -\,\, \frac{\BA{\vec{a}}{z_0\perp\ZP}{\Lb}}{
                               |\,\BA{\vec{a}}{z_0\perp\ZP}{\Lb}\,|}.
\label{SUPeq:xz}
\end{align}  
We have labeled the $\psi$ and $p$ helicities, $\lambda_\psi^{\ZP}$ and $\lambda_p^{\ZP}$,
with the $\ZP$ superscript to make it clear that the spin quantization axes are different than in the $\LambdaStar$ decay chain.
Since the $\psi$ is an intermediate particle, this has no consequences after we sum (coherently) over $\lambda_\psi^{\ZP}=-1,0,+1$.
The proton, however, is a final-state particle. Before the $\ZP$ terms in the matrix element can be added coherently to the $\LambdaStar$ terms,
the $\lambda_p^{\ZP}$ states must be rotated to $\lambda_p$ states (defined in the $\LambdaStar$ decay chain).
The proton helicity axes are different, since the proton comes from a decay of different particles in the two decay sequences, 
the $\LambdaStar$ and $\ZP$. 
The quantization axes are along the proton direction in the $\LambdaStar$ and the $\ZP$ rest frames, thus antiparallel 
to the particles recoiling against the proton: the $K^-$ and $\psi$, respectively. These directions are preserved when 
boosting to the proton rest frame (see Fig.~\ref{fig:helicityp}).
Thus, the polar angle between the two proton quantization  axes ($\theta_p$) can be determined from
the opening angle between the $K^-$ and $\psi$ mesons in the $p$ rest frame, 
\begin{equation}
\cos\theta_p=\BA{\hat{p}}{K}{p}\cdot\BA{\hat{p}}{\psi}{p}.
\end{equation}
(A similar problem is discussed in Ref.~\cite{Mizuk:2008me},
where the two different $\chi_{c1}$ helicity frames in $B^0\to K^+\pi^-\chi_{c1}$ decays, 
in the  interference of $B^0\to K^*\chi_{c1}$, $K^*\to K^+\pi^-$ and of $B^0\to Z^- K^+$, $Z^-\to \chi_{c1} \pi^-$ contributions,  
are realigned.)
The dot product above must be calculated by operating on the $\BA{\vec{p}}{K}{p}$ and $\BA{\vec{p}}{\psi}{p}$ vectors 
in the proton rest frame obtained by  the same sequence of  boost transformations, either according to the $\LambdaStar$ or 
$\ZP$ decay chains, or even by a direct boost transformation 
from the {\rm lab} frame.\footnote{Numerical values of momentum vector components, $(p_x,p_y,p_z)$, depend on the boost sequence taken 
and are related between different boosts via the rotation matrix. However, the dot product between the two vectors remains 
independent of the boost sequences.}
%
\begin{figure}[t!]
\begin{center}
\includegraphics[width=\textwidth]{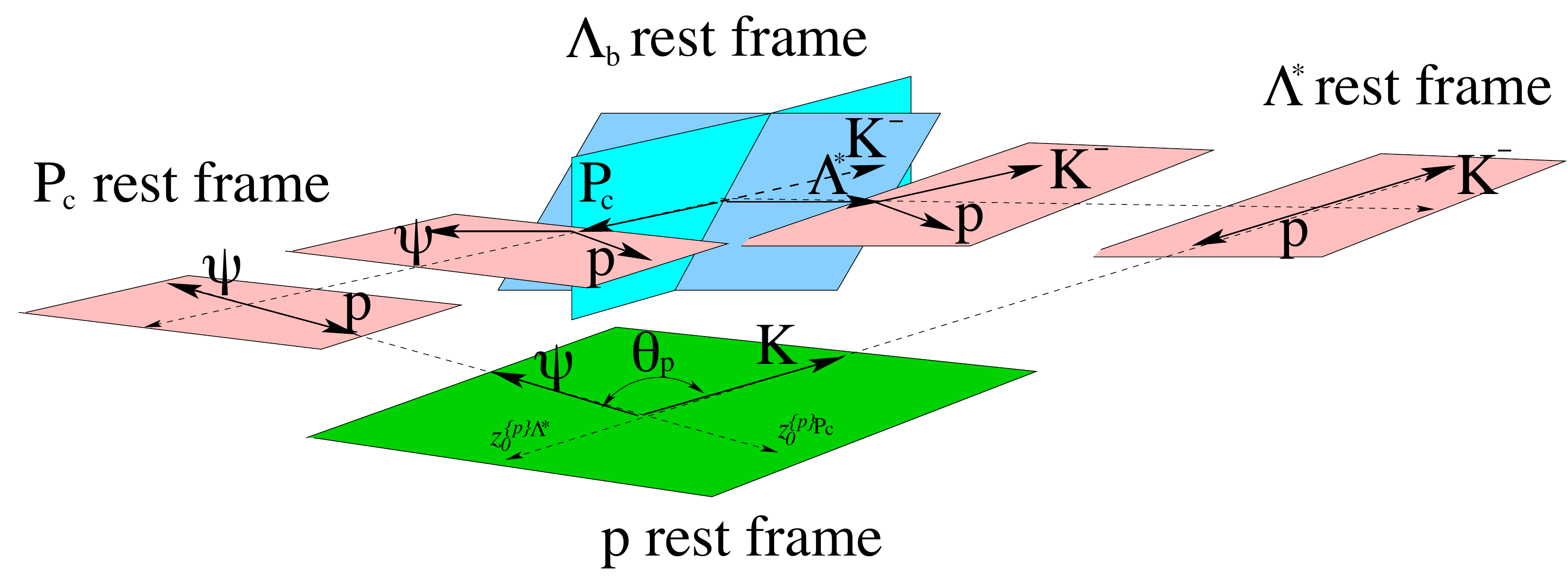}
\end{center}
\vskip-5mm
\caption{
Definition of the $\theta_p$ angle.
}
\label{fig:helicityp}
\end{figure} 

No azimuthal rotation is needed to align the two proton helicity frames, since  the decay planes 
of the $\LambdaStar$ and the $\ZP$ are the same (see Fig.~\ref{fig:helicityp}).
Therefore, the relation between $\lambda_p$ and $\lambda_p^{\ZP}$ states is
\begin{equation}
|\lambda_p\rangle = \sum\limits_{\lambda_p^{\ZP}} D^{\,\,J_p}_{\lambda_p^{\ZP},\,\lambda_p}(0,\theta_p,0)^* |\lambda_p^{\ZP}\rangle
            = \sum\limits_{\lambda_p^{\ZP}} d^{\,\,J_p}_{\lambda_p^{\ZP},\,\lambda_p}(\theta_p) |\lambda_p^{\ZP}\rangle.
\end{equation}
Thus, the term given by Eq.~(\ref{SUPeq:ztopsip}) must be preceded by
\begin{equation}
\label{SUPeq:prot}
\sum\limits_{\lambda_p^{\ZP}=\pm\frac{1}{2}} d^{\,\,J_p}_{\lambda_p^{\ZP},\,\lambda_p}(\theta_p).
\end{equation}

Parity conservation in ${\ZP}_j\to \psi p$ decays leads to the following relation
\begin{align}
\H^{{\ZP}_j\to \psi p}_{-\lambda_\psi^{\ZP},-\lambda_p^{\ZP}} 
& = P_\psi\,P_p\,P_{{\ZP}_j}\,(-1)^{J_\psi+J_p-J_{{\ZP}_j}}\,
   \H^{{\ZP}_j\to \psi p}_{\lambda_\psi^{\ZP},\,\lambda_p^{\ZP}} \notag\\
& = P_{{\ZP}_j}\, (-1)^{\frac{1}{2}-J_{{\ZP}_j}}\,
   \H^{{\ZP}_j\to \psi p}_{\lambda_\psi^{\ZP},\,\lambda_p^{\ZP}}, 
\label{SUPeq:zparity}
\end{align}
where $P_{{\ZP}_j}$ is the parity of the ${\ZP}_j$ state.
This relation reduces the number of independent helicity couplings to be determined from the data
to 2 for $J_{{\ZP}_j}=\frac{1}{2}$ and 3 for $J_{{\ZP}_j}\ge\frac{3}{2}$.
Since the helicity couplings enter the matrix element formula as a product,  
$\H^{\Lb\to {\ZP}_j K}_{\lambda_{\ZP}}\,\H^{{\ZP}_j\to \psi p}_{\lambda_\psi^{\ZP},\,\lambda_p^{\ZP}}$, the
relative magnitude and phase of these two sets must be fixed by a convention.
For example, $\H^{\Lb\to {\ZP}_j K}_{\lambda_{\ZP}=-\frac{1}{2}}$ can be set to $(1,0)$ for
every ${\ZP}_j$ resonance, in which case
$\H^{\Lb\to {\ZP}_j K}_{\lambda_{\ZP}=+\frac{1}{2}}$ develops a meaning of the complex ratio of
$\H^{\Lb\to {\ZP}_j K}_{\lambda_{\ZP}=+\frac{1}{2}}/\H^{\Lb\to {\ZP}_j K}_{\lambda_{\ZP}=-\frac{1}{2}}$,
while all $\H^{{\ZP}_j\to \psi p}_{\lambda_\psi^{\ZP},\,\lambda_p^{\ZP}}$ couplings should have both
real and imaginary parts free in the fit. 

The term $R_{{\ZP}_j}(M_{\psi p})$ describes the $\psi p$  invariant mass distribution of the 
${\ZP}_j$ resonance 
and is given by Eq.~(\ref{SUPeq:resshape}) after appropriate substitutions.
Angular momentum conservation limits $L_{\Lb}^{{\ZP}_j}$ in $\Lb\to{\ZP}_j K^-$ decays
to $J_{{\ZP}_j}\pm\frac{1}{2}$.
The angular momentum conservation also imposes 
${\rm max}(J_{{\ZP}_j}-\frac{3}{2}\,, \,0)\le L_{{\ZP}_j} \le J_{{\ZP}_j}+\frac{3}{2}$,
which is further restricted by the parity conservation in the ${\ZP}_j$ decays,
$P_{{\ZP}_j}=(-1)^{L_{{\ZP}_j}+1}$.
We assume the minimal values of $L_{\Lb}^{{\ZP}_{j}}$ and of $L_{{\ZP}_{j}}$ in $R_{{\ZP}_j}(m_{\psi p})$.

The electromagnetic decay $\psi\to\mu^+\mu^-$ in the $\ZP$ decay chain contributes a term
\begin{equation}
\label{SUPeq:psitommz}
D^{\,\,1}_{\lambda_{\psi}^{\ZP},\,\Delta\lambda_\mu^{\ZP}}(
\phi_{\mu}^{\ZP},\theta_{\psi}^{\ZP},0)^*,
\end{equation}
which is the same as Eq.~(\ref{SUPeq:psitomm}), except that since the $\psi$ meson comes from the decay of different particles 
in the two decay chains, the azimuthal and polar angle of the muon in the $\psi$ rest frame, 
$\phi_{\mu}^{\ZP}$, $\theta_{\psi}^{\ZP}$, are different from 
$\phi_{\mu}$, $\theta_{\psi}$ introduced in the $\LambdaStar$ decay chain.
The $\psi$ helicity axis is along the boost direction from the $\ZP$ to the $\psi$ rest frames, which 
is given by
\begin{equation}
\BA{\hat{z}}{0}{\psi}~^{\ZP}  =  \, - \, \BA{\hat{p}}{p}{\psi}, 
\end{equation}
and so
\begin{equation}
\cos\theta_{\psi}^{\ZP}  =  - \BA{\hat{p}}{p}{\psi} \cdot \BA{\hat{p}}{\mu}{\psi}. \label{SUPeq:psihelthz}
\end{equation}
The $x$ axis is inherited from the $\ZP$ rest frame (Eq.~(\ref{SUPeq:xaxis})),
\begin{align}
\BA{\vec{a}}{z_0\perp \psi}{\ZP} & 
                    = - {\BA{\vec{p}}{K}{\ZP}} + ({\BA{\vec{p}}{K}{\ZP}}\cdot\BA{\hat{p}}{\psi}{\ZP})\,\BA{\hat{p}}{\psi}{\ZP} \notag\\
\BA{\hat{x}}{0}{\psi}~^{\ZP} =
\BA{\hat{x}}{3}{\ZP} & = \,-\,\,\frac{\BA{\vec{a}}{z_0\perp \psi}{\ZP}}{|\,\BA{\vec{a}}{z_0\perp \psi}{\ZP}\,|},
\label{SUPeq:xpsiz}
\end{align}
which leads to
\begin{equation}
\label{SUPeq:psihelphz} 
\phi_\mu^{\ZP}  =  {\rm atan2} \left( -(\BA{\hat{p}}{p}{\psi}\times\BA{\hat{x}}{0}{\psi}~^{\ZP})\cdot\BA{\hat{p}}{\mu}{\psi},\, 
                                        \BA{\hat{x}}{0}{\psi}~^{\ZP}\cdot\BA{\hat{p}}{\mu}{\psi} \right).
\end{equation}
  
Since the muons are final-state particles,
their helicity states in the $\ZP$ decay chain, $|\lambda_\mu^{\ZP}\rangle$, need to be rotated to the muon helicity states
in the $\LambdaStar$ decay chain, $|\lambda_\mu\rangle$, before the $\ZP$ matrix element terms can be coherently added to the 
$\LambdaStar$ matrix element terms. 
The situation is simpler than for the rotation of the proton helicities discussed above, as the muons come from
the $\psi$ decay in both decay chains. 
This makes the polar angle $\theta_\mu$  (analogous to $\theta_p$ in Eq.~(\ref{SUPeq:prot})) equal to zero, which leads to
$d^{\,\frac{1}{2}}_{\lambda_\mu^{\ZP},\,\lambda_\mu}(0)=\delta_{\lambda_\mu^{\ZP},\,\lambda_\mu}$, where $\delta_{i,j}$ is the Kronecker symbol.
However, the muon helicity states are not identical since the $x$ axes are offset by the
azimuthal angle $\alpha_\mu$. 
Since the boost to the $\mu$ rest frame is the same for both decay chains (\ie\ always from the $\psi$ rest frame), 
we can determine $\alpha_\mu$ in the $\psi$ rest frame
\begin{equation}
\label{SUPeq:alphamu}
\alpha_{\mu} = {\rm atan2} \left( (\BA{\hat{z}}{3}{\psi}\times\BA{\hat{x}}{3}{\psi}~^{\ZP})\cdot\BA{\hat{x}}{3}{\psi}~^{\LambdaStar},\, 
                                   \BA{\hat{x}}{3}{\psi}~^{\ZP}\cdot\BA{\hat{x}}{3}{\psi}~^{\LambdaStar} \right),
\end{equation}
where
$\BA{\hat{z}}{3}{\psi}=\BA{\hat{p}}{\mu}{\psi}$,
and from Eq.~(\ref{SUPeq:xaxis})
\begin{align}
\BA{\hat{x}}{3}{\psi}~^{\ZP} & = \, - \, \BA{\hat{a}}{z_0\perp\mu}{\psi}~^{\ZP}, \\
\BA{\vec{a}}{z_0\perp\mu}{\psi}~^{\ZP} & 
   = \, - \, \BA{\hat{p}}{p}{\psi} + (\BA{\hat{p}}{p}{\psi}\cdot\BA{\hat{p}}{\mu}{\psi})\,\BA{\hat{p}}{\mu}{\psi},
\end{align}
as well as
\begin{align}
\BA{\hat{x}}{3}{\psi}~^{\LambdaStar} & = \, - \, \BA{\hat{a}}{z_0\perp\mu}{\psi}~^{\LambdaStar}, \\
\BA{\vec{a}}{z_0\perp\mu}{\psi}~^{\LambdaStar} & 
   = \, - \, \BA{\hat{p}}{\LambdaStar}{\psi} + (\BA{\hat{p}}{\LambdaStar}{\psi}\cdot\BA{\hat{p}}{\mu}{\psi})\,\BA{\hat{p}}{\mu}{\psi}.
\end{align}
The term aligning the muon helicity states between the two reference frames is given by
\begin{equation}
   \sum\limits_{\lambda_\mu^{\ZP}} D^{\,\,J_\mu}_{\lambda_\mu^{\ZP}\,\lambda_\mu}(\alpha_\mu,0,0)^* =
 \sum\limits_{\lambda_\mu^{\ZP}} e^{i\,\lambda_\mu^{\ZP}\alpha_\mu} \delta_{\lambda_\mu^{\ZP},\,\lambda_\mu}
= e^{i\,\lambda_\mu\alpha_\mu}.
\end{equation}
The transformation of $\mu^-$ states will be similar to that of the $\mu^+$ states, except that since $\hat{z}_\psi$ will have the opposite direction, 
$\alpha_{{\mu^+}} = - \alpha_{\mu^-}$. 
The transformation of $|\lambda_{\mu^+}^{\ZP}\rangle|\lambda_{\mu^-}^{\ZP}\rangle$ to $|\lambda_{\mu^+}\rangle|\lambda_{\mu^-}\rangle$ states
will require multiplying the terms for the $\ZP$ decay chain by 
\begin{equation}
  e^{i\,\lambda_\mu\alpha_\mu} e^{i\,\lambda_{\bar{\mu}}\alpha_{\bar{\mu}}} 
=  e^{i\,(\lambda_\mu-\lambda_{\bar{\mu}})\alpha_{\mu}} =  e^{i\,\Delta\lambda_\mu\alpha_{\mu}}.
\label{SUPeq:mutran}
\end{equation}
An alternative derivation of Eq.~(\ref{SUPeq:mutran}) is discussed in 
Ref.~\cite{Chilikin:2013tch} (Eqs.~(20)$-$(22) therein) for the
interference of $B^0\to K^*\psi$, $K^*\to K\pi$ and of 
$B^0\to Z K^-$, $Z\to \psi\pi$ ($\psi\to\ell^+\ell^-$) terms,
which are analogous to the two decay chains discussed here with the 
substitution $\B^0\to\Lb$, $K^*\to\LambdaStar$, $Z\to \ZP$ and $\pi\to p$.
The rotation by $\alpha_{\mu}$ about the $\ell^+$ direction in 
the $\psi$ rest frame in the $Z$ decay chain is 
incorporated by setting $\gamma=\alpha_\mu$, instead of $\gamma=0$  in 
Eq.~(\ref{SUPeq:psitommz}). This leads to the same formulae since
\begin{equation}
   D^{\,\,1}_{\lambda_{\psi}^{\ZP},\,\Delta\lambda_\mu}(\phi_{\mu}^{\ZP},\theta_{\psi}^{\ZP},\alpha_\mu)^* 
=  D^{\,\,1}_{\lambda_{\psi}^{\ZP},\,\Delta\lambda_\mu}(\phi_{\mu}^{\ZP},\theta_{\psi}^{\ZP},0)^*\,e^{i\,\Delta\lambda_\mu\alpha_{\mu}}.
\end{equation}
We use the more generic derivation here to demonstrate that the methods of transforming the muon and proton helicity states
between the two decay chains are the same. 

Collecting terms from the three subsequent decays in the $\ZP$ chain together, 
\begin{align}  
\Mat_{\lambda_{\Lb},\,\lambda_p^{\ZP},\,\Delta\lambda_\mu^{\ZP}}^{\,\,\ZP} =  
&
  e^{i\,\lambda_{\Lb}\phi_{\ZP}}
  \sum\limits_{j} 
     R_{{\ZP}_j}(M_{\psi p}) \,
     \sum\limits_{\lambda_\psi^{\ZP}}
        e^{i\,\lambda_\psi^{\ZP}\phi_{\mu}^{\ZP}} \,\,
        d^{\,\,1}_{\lambda_\psi^{\ZP},\,\Delta\lambda_\mu}(\theta_\psi^{\ZP}) \,\,
	\notag \\
& ~~\times \sum\limits_{\lambda_{\ZP}} 
        \H^{\Lb\to {\ZP}_j K}_{\lambda_{\ZP}} \,\,
        e^{i\,\lambda_{\ZP} \phi_\psi^{\ZP}} \,\,
        d^{\,\,\frac{1}{2}}_{\lambda_{\Lb},\,\lambda_{\ZP}}(\theta_{\Lb}^{\ZP}) 
        \H^{{\ZP}_j\to\psi p}_{\lambda_{\psi}^{\ZP},\,\lambda_{p}^{\ZP}}\,\,
        d^{\,\,J_{{\ZP}_j}}_{\lambda_{\ZP},\,\lambda_{\psi}^{\ZP}-\lambda_p^{\ZP}}(\theta_{\ZP}),
\label{SUPeq:Pc_matrixelement_partial}
\end{align}
and adding them coherently to the $\LambdaStar$ matrix element, 
via appropriate relation of $|\lambda_p\rangle|\lambda_{\mu^+}\rangle|\lambda_{\mu^-}\rangle$ to 
$|\lambda_p^{\ZP}\rangle|\lambda_{\mu^+}^{\ZP}\rangle|\lambda_{\mu^-}^{\ZP}\rangle$ states as discussed above,
leads to the final matrix element squared
\begin{equation} 
\left| \Mat \right|^2 = 
\sum\limits_{\lambda_{\Lb}=\pm\frac{1}{2}}
\sum\limits_{\lambda_{p}=\pm\frac{1}{2}}
\sum\limits_{\Delta\lambda_{\mu}=\pm1}
\left|
\Mat_{\lambda_{\Lb},\,\lambda_p,\,\Delta\lambda_\mu}^{\LambdaStar} 
+ 
e^{i\,{\Delta\lambda_\mu}\alpha_{\mu}}\,
\sum\limits_{\lambda_p^{\ZP}} 
d^{\,\,\frac{1}{2}}_{\lambda_p^{\ZP},\,\lambda_p}(\theta_p)\,
\Mat_{\lambda_{\Lb},\,\lambda_p^{\ZP},\,\Delta\lambda_\mu}^{\ZP} 
\right|^2,
\label{SUPPeq:total_matrixelement}
\end{equation}  
where we set $\Pol=0$. 
As a cross-check, fitting the $\Lb$ polarization to the data 
with the default $\Lz^*$ and $P_c^+$ model yields a value consistent 
with zero, $\Pol=(-2.0\pm2.3)\%$ (statistical error only). 

Assuming approximate $\CP$ symmetry, the helicity couplings for $\Lb$ and $\Lbbar$ can be made equal, but the calculation 
of the angles requires some care, since  parity ($P$) conservation does not change polar (i.e. helicity) angles, but does change azimuthal angles. 
Thus, not only must $\vec{p}_{\mu^+}$ be used instead of $\vec{p}_{\mu^-}$ for $\Lbbar$ candidates (with $K^+$ and $\bar p$ 
in the final-state) in Eqs.~(\ref{SUPeq:psihelth}), (\ref{SUPeq:psihelph}), (\ref{SUPeq:psihelthz}), (\ref{SUPeq:psihelphz}) and (\ref{SUPeq:alphamu}), 
but also all azimuthal angles must be reflected before entering the matrix element formula:
$\phi_K\to -\phi_K$,  
$\phi_\mu\to -\phi_\mu$,
$\phi_{\ZP}\to -\phi_{\ZP}$,  
$\phi_\psi^{\ZP}\to -\phi_\psi^{\ZP}$,  
$\phi_\mu^{\ZP}\to -\phi_\mu^{\ZP}$
and
$\alpha_\mu\to -\alpha_\mu$ \cite{Chilikin:2013tch}.

It is clear from Eq.~(\ref{SUPPeq:total_matrixelement}) that various $\LambdaStarn$ 
and $\ZP$ resonances
interfere in the differential distributions.
By integrating the matrix element squared over the entire phase space 
the interferences cancel in the integrated rates unless
the resonances belong to the same decay chain 
and have the same quantum numbers.\footnote{For $\LambdaStarn-\ZP$, 
the $\lambda_{\Lb}=+1/2$ interference terms have the opposite effect to  
the $\lambda_{\Lb}=-1/2$ interference terms.}

\subsection{Reduction of the number of helicity couplings}

A possible reduction of the helicity couplings can be achieved by relating them 
to the $LS$ couplings ($B_{L,S}$) using Clebsch-Gordan coefficients
\begin{equation}
\H_{\lambda_B,\lambda_C}^{A\to B\,C}=\sum_{L} \sum_{S} 
\sqrt{ \tfrac{2L+1}{2J_A+1} } B_{L,S} 
\left( 
\begin{array}{cc|c}
 J_{B} & J_{C} & S \\
 \lambda_{B} & -\lambda_{C} & \lambda_{B}-\lambda_{C} 
\end{array}
\right)
\times
\left( 
\begin{array}{cc|c}
 L  & S & J_A \\
 0 & \lambda_{B}-\lambda_{C} & \lambda_{B}-\lambda_{C}   
\end{array}
\right), 
\label{SUPeq:LS}
\end{equation}
and then restricting the $L$ values. 
Here $L$ is the orbital angular momentum in the decay, and 
$S$ is the total spin of the daughters, $\vec{S}=\vec{J}_B+\vec{J}_C$
($|J_B-J_C|\le S \le J_B+J_C$).
If the energy release in the decay, $Q=M_A-M_B-M_C$, is small, $Q/M_A \ll  1$,
then higher values of $L$ should be suppressed; this effect is usually called ``the angular momentum barrier."
Applying this approach to $\Lb\to\psi\LambdaStarn$ decays, 
the lowest $L_{\Lb}^{\LambdaStarn}$ value ($L_{\rm min}$) corresponds to a single possible value of $S$, thus 
reducing the number of couplings to fit, from 4 ($J_{\LambdaStarn}=\frac{1}{2}$) 
or 6 ($J_{\LambdaStarn}\ge\frac{3}{2}$), to just one $B_{L,S}$ coupling per resonance.
Accepting also $L_{\rm min}+1$ values, gives three $B_{L,S}$ couplings to fit per resonance.
    
In $\Lb\to{\ZP}_j K^-$ decays, $S=J_{{\ZP}_j}$ and $L_{\Lb}^{{\ZP}_j}=J_{{\ZP}_j}\pm\frac{1}{2}$.
Taking only the lower $L_{\Lb}^{{\ZP}_j}$ value reduces the number of couplings from 2 to 1.
Since its magnitude and phase convention can be absorbed into     
$\H^{{\ZP}_j\to \psi p}_{\lambda_\psi^{\ZP},\,\lambda_p^{\ZP}}$ (see the discussion in Sec.~\ref{SUPPsec:mzp}), 
one can simply set $B_{J_{{\ZP}_j}-\frac{1}{2},J_{{\ZP}_j}}^{\Lb\to{\ZP}_j K}=(1,0)$ in this approach.

The reduction of couplings to fit for ${\ZP}_j\to \psi p$ decays depends on the spin and parity of
the ${\ZP}_j$ state. $S$ can take values of $\frac{1}{2}$ and $\frac{3}{2}$.
Values of $L_{{\ZP}_j}$ must be odd (even) for even (odd) $P_{{\ZP}_j}$.
For  a $J_{{\ZP}_j}^P=\frac{1}{2}^+$ state, only $L_{{\ZP}_j}=1$ is allowed with the two possible values of $S$.
Therefore, no reduction of couplings is possible.
For a $J_{{\ZP}_j}^P=\frac{1}{2}^-$ state, $L_{{\ZP}_j}=0, 2$ are allowed, each corresponding to one $S$ value.
Therefore, the number of couplings to fit can be reduced from $2$ to $1$ when taking $L_{{\ZP}_j}=0$. 
Gains can be larger for $J_{{\ZP}_j}\ge\frac{3}{2}$ states.

Even if no reduction in parameters is achieved, expressing the helicity couplings
via corresponding $B_{L,S}$ couplings using Eq.~(\ref{SUPeq:LS})
is useful, since it automatically implements 
the parity constraints (Eq.~(\ref{SUPeq:zparity})) by restricting possible $L$ values. 
Since the overall magnitude of the matrix element does not affect the normalized 
signal PDF, and because its overall phase also drops out when taking its modulus, we fix 
the magnitude and phase convention by setting $B_{0,\frac{1}{2}}^{\Lb\to\Lz(1520)\jpsi}=(1,0)$.

\section{Details of fitting techniques}

The matrix element given by Eq.~(\ref{SUPPeq:total_matrixelement}), 
is a 6-dimensional (6D) function of $m_{Kp}$ and $\POmega$ and
depends on the fit parameters, $\Pars$, which represent 
independent helicity or $LS$ couplings, and possibly masses and widths of resonances,
$\Mat(m_{Kp},\POmega|\Pars)$.
We perform an unbinned maximum likelihood fit of these parameters 
to the 6D data by minimizing $-2\ln\Like(\Pars)=-2\ln\sum_i\PDF( m_{Kp~i},\POmega_i|\Pars)$
with respect to $\Pars$.
The signal PDF is obtained
by multiplying the matrix element squared 
with the selection efficiency, $\epsilon(m_{Kp},\POmega)$,
\begin{equation}
\label{SUPPeq:sigpdf}
\PDF_{\rm sig}(m_{Kp},\POmega|\Pars)
=\frac{1}{I(\Pars)}\left|\Mat(m_{Kp},\POmega|\Pars)\right|^2
\Phi(m_{Kp})
\epsilon(m_{Kp},\POmega),
\end{equation}
where 
$\Phi(m_{Kp})$ is the phase space function equal to $p\,q$, where $p$ is the momentum of
the $Kp$ system (\ie\ $\LambdaStar$) in the $\Lb$ rest frame, and $q$ is the momentum of
$K^-$ in the $\LambdaStar$ rest frame,
and $I(\Pars)$ is the normalization integral.

We use two fit algorithms that were independently coded and that differ in the approach used for background subtraction.
The sFit procedure uses the sPlot technique described in Ref.~\cite{Pivk:2004ty} 
to subtract background from the log-likelihood sum. It has been used in previous LHCb analyses, \eg\ measurement of $\phi_s$ in $\Bs\to\jpsi\phi$ decays \cite{Aaij:2014zsa}.
The data in the entire $5480-5760$~MeV range are passed to the fitter.
Candidates are assigned ``sWeights'' ($W_i$) according to their $m_{\jpsi p K}$ value with the signal and background probabilities 
determined by the fit to the $m_{\jpsi p K}$ distribution,
\begin{equation}
-2\ln\Like(\Pars) = -2s_W \sum_i W_i \ln \PDF(m_{Kp\, i},\POmega_i|\Pars),
\end{equation}
where $s_W\equiv \sum_i W_i/\sum_i {W_i}^2$ is a constant factor 
accounting for the effect of the background subtraction on the statistical uncertainty.
Candidates in the sideband region have negative sWeights
which on average compensate for the background events present in the peak region, where events 
have positive sWeights.
From signal simulations, 
we see significant variations of $\Lb\to \jpsi p K^-$ invariant mass resolution as functions of $\cos\theta_{\Lb}$ and $\cos\theta_{\jpsi}$.
The background distributions also vary in intensity and shape with these two variables.
No strong variation is seen for the other fitting observables. 
To determine the sWeights, the events are divided 
into 4  $|\cos\theta_{\jpsi}|$ $\times$ 8 $\cos\theta_{\Lb}$ bins.
A separate fit to the $\Lb\to \jpsi p K^-$ invariant mass distribution of each bin is performed. 

In the sFit approach the total PDF is equal to the signal PDF, as the background
is subtracted from the log-likelihood sum using sWeights,
\begin{align}
-2\ln\Like(\Pars)=
& -2 s_W\,\sum_i W_i\ln \PDF_{\rm sig}(m_{Kp~i},\POmega_i|\Pars)  \notag\\
= & -2 s_W\, \sum_i W_i \ln |\Mat(m_{Kp~i},\POmega_i|\Pars)|^2 
+ 2 s_W\, \ln I(\Pars) \sum_i W_i \notag\\
&~~~~~~~~~ -2 s_W\, \sum_i W_i \ln[ \Phi(m_{Kp~i})\epsilon(m_{Kp~i},\POmega_i) ].
\end{align}
The last term does not depend on the fitted parameters $\Pars$ and is therefore dropped. 
The efficiency still appears in the normalization integral.  The integration is done without the need to parameterize the efficiency, 
by summing the matrix element squared 
over the simulated events that are
generated uniformly in phase space and passed
through the detector modeling and the data selection procedure,
\begin{eqnarray}
I(\Pars)& \equiv & \int \left|\Mat(m_{Kp},\POmega|\Pars)\right|^2
\Phi(m_{Kp})
\epsilon(m_{Kp},\POmega)
\,dm_{Kp}\,d\POmega \notag\\
&\propto & \sum\limits_{j}^{N_{\rm MC}} w_j^{\rm MC} \left|\Mat(m_{Kp~j},\POmega_{j}|\Pars)\right|^2,
\label{SUPPeq:signor}
\end{eqnarray}
where $w_j^{\rm MC}$ are the weights given to the simulated events 
to  improve the agreement between data and simulations.
They include particle identification weights obtained from 
calibration samples of $\Lz \to p \pi^-$ and $\Dz\to K^-\pi^+$
as functions of momentum and pseudorapidity of the protons and kaons.
They also include a weight correcting the overall efficiency for
the dependence on the charged track multiplicity of events,
determined from the ratio of the background-subtracted data and 
the signal simulations. 
Imperfect description of the $\Lb$ production kinematics in our simulation
is corrected in a similar way via a weight that depends on the $p$ and $\pt$ of  the $\Lb$ baryon.
The final weights are the ratio of the data and the simulations as 
a function of proton and kaon momenta. 

In the cFit approach, candidates are not weighted ($W_i=1$).
The data that are fitted are restricted to be within a $\pm2\sigma$ 
mass window around the \Lb mass peak, in the interval
$5605.7< m_{\jpsi pK} < 5635.8$~MeV. 
The background is represented explicitly in the fitted PDF,
with the integrated background probability set to $\beta=5.4\%$
as determined from the fit to the $\jpsi K^-p$ mass distribution,
\begin{equation}
 \PDF(m_{Kp},\POmega|\Pars) =
 (1-\beta)\, \PDF_{\rm sig}(m_{Kp},\POmega|\Pars)
+ \beta\, \PDF_{\rm bkg}(m_{Kp},\POmega).
\end{equation}
The 6-dimensional background parameterization $\PDF_{\rm bkg}(m_{Kp},\POmega)$
is developed using the background sample
in which the \Lb  candidate invariant mass is
more than $5\sigma$ away from the peak, specifically 
within the intervals $5480.0-5583.2$~MeV and $5658.3-5760.0$~MeV.
We minimize the negative log-likelihood defined as,
\begin{align}
& \!\!\!\!-2\ln\Like(\Pars)= & \nonumber\\
&~~~~~ -2\sum_i\ln \left[ (1-\beta)\,\frac{\left|\Mat(m_{Kp~i},\POmega_i|\Pars)\right|^2 \Phi(m_{Kp~i})\epsilon(m_{Kp~i},\POmega_i)}{I(\Pars)}
+\beta\, \frac{\PDF_{\rm bkg}^{u}(m_{Kp~i},\POmega_i)}{I_{\rm bkg}} \right] & \nonumber \\
&~~~=-2\sum_i\ln \left[\left|\Mat(m_{Kp~i},\POmega_i|\Pars)\right|^2  
+ \frac{\beta\,I(\Pars)}{(1-\beta) I_{\rm bkg}}\,\frac{\PDF_{\rm bkg}^{u}(m_{Kp~i},\POmega_i)}{\Phi(m_{Kp~i})\epsilon(m_{Kp~i},\POmega_i)}\right] & \nonumber\\
&  \quad\quad\quad\quad\quad +2N\ln I(\Pars) +{\rm constant}, &
\end{align}
where $N$ is the number of candidates, and
$\PDF_{\rm bkg}^{u}(m_{Kp},\POmega)$ is the unnormalized background density proportional to the density of sideband candidates,
with its normalization determined by\footnote{Note that the distribution of MC events 
includes both the $\Phi(m_{Kp})$ and $\epsilon(m_{Kp},\POmega)$ factors, which cancel their product in the numerator.}
\begin{equation}
I_{\rm bkg}\equiv \int \PDF_{\rm bkg}^{u}(m_{Kp})\,d m_{Kp}\,d\POmega \, \propto \,
\sum_j w_j^{\rm MC} \frac{\PDF_{\rm bkg}^{u}(m_{Kp~j},\POmega_j)}{\Phi(m_{Kp~j})\epsilon(m_{Kp~j},\POmega_j)}.
\label{SUPPeq:bkgnor}
\end{equation}
The background term is then efficiency-corrected so it can be added
to the efficiency-independent signal probability expressed by $\left|\Mat\right|^2$.
This way the  efficiency parametrization, $\epsilon(m_{Kp},\POmega)$, 
influences only the background component which affects only a tiny part 
of the total PDF, while
the efficiency corrections to the signal part enter Eq.~(\ref{SUPPeq:signor}).

The efficiency in the background term is assumed to factorize as
\begin{equation}
\epsilon(m_{Kp},\POmega)=\epsilon_1(m_{Kp},\cos\theta_\Lz^*)\epsilon_2(\cos\theta_{\Lb}|m_{Kp})\epsilon_3(\cos\theta_{\jpsi}|m_{Kp}) \epsilon_4(\phi_K|m_{Kp})\epsilon_5(\phi_{\mu}|m_{Kp}).
\end{equation}
The $\epsilon_1(m_{Kp},\cos\theta_\Lz^*)$ term is obtained from 
a 2D histogram of the simulated events weighted by $1/(p\, q)$. 
A bi-cubic interpolation is used to interpolate between bin centers.
The other terms are again built from 2D histograms, but with each bin divided by the number of simulated events in the corresponding $m_{Kp}$ slice to remove the  leading dependence on this mass, which is already taken into account in the first term.
 The leading variation of the efficiency is in the $\epsilon_1(m_{Kp},\cos\theta_\Lz^*)$ term
which is visualized in the normal Dalitz variables $(m_{\jpsi p}^2,m_{Kp}^2$) in Fig.~\ref{SUPPeffbkgpara}(a) 
instead of the ``rectangular Dalitz plane'' variables $(m_{Kp},\cos\theta_\Lz^*)$ used to parameterize this variation.

The background PDF, $\PDF_{\rm bkg}^{u}(m_{Kp},\POmega)/\Phi(m_{Kp})$, 
is built using the same approach,
\begin{eqnarray}
\frac{{\PDF_{\rm bkg}^{u}}(m_{Kp},\POmega)}{\Phi(m_{Kp})}&=& 
{P_{\rm bkg}}_1(m_{Kp},\cos\theta_\Lz^*){P_{\rm bkg}}_2(\cos\theta_{\Lb}|m_{Kp})
\notag\\
&&~~\times{P_{\rm bkg}}_3(\cos\theta_{\jpsi}|m_{Kp}) {P_{\rm bkg}}_4(\phi_K|m_{Kp}){P_{\rm bkg}}_5(\phi_{\mu}|m_{Kp}).
\end{eqnarray}
A visualization of ${P_{\rm bkg}}_1(m_{Kp},\cos\theta_\Lz^*)$ on the Dalitz plane 
is shown in Fig.~\ref{SUPPeffbkgpara}(b).

\begin{figure}[htb]
\begin{center}
\includegraphics[width=0.49\textwidth]{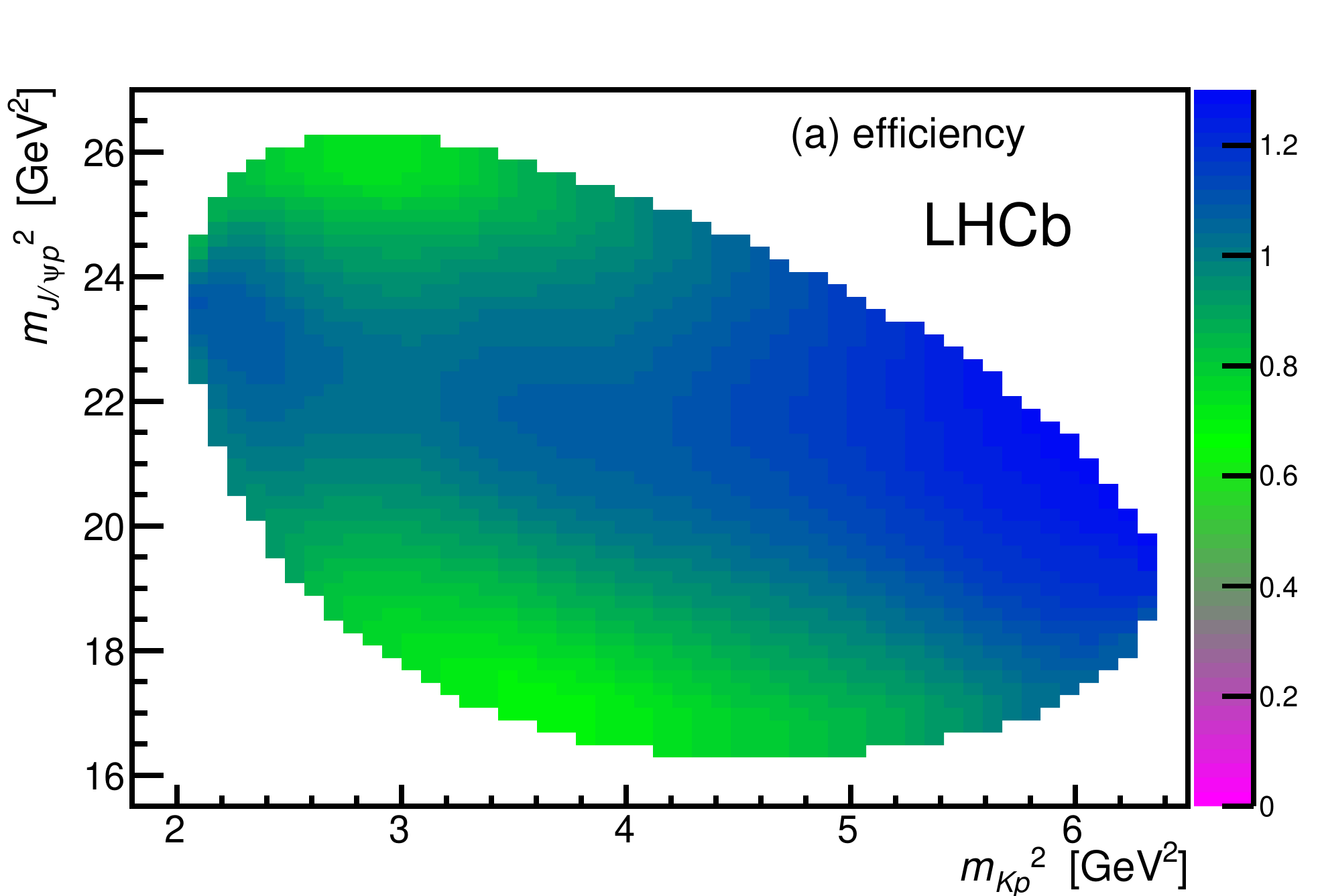}
\includegraphics[width=0.49\textwidth]{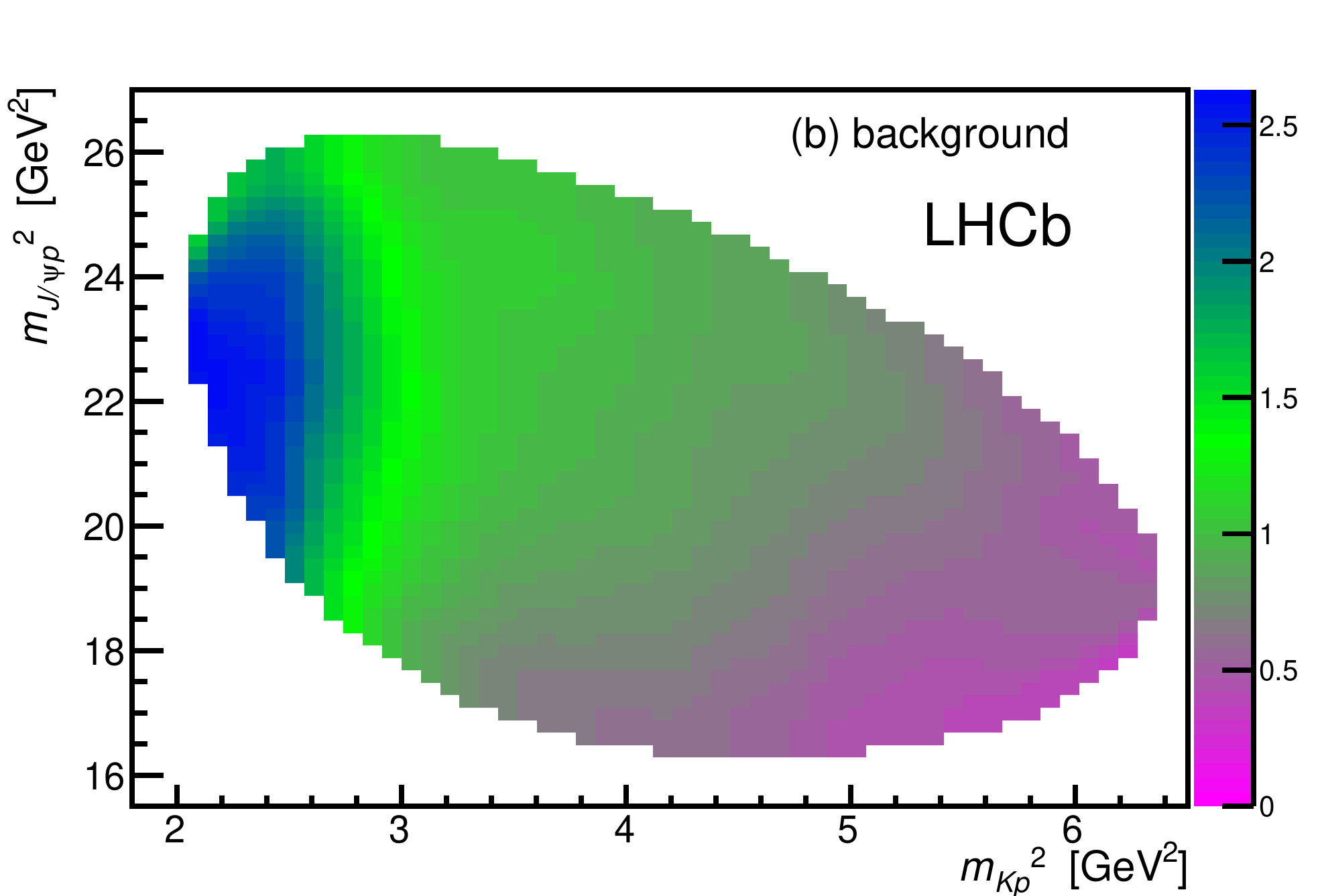}
\end{center}
\vskip -0.5cm
\caption{Parameterized dependence of (a) the relative signal efficiency  and of (b) the background density 
on the Dalitz plane. The units of the relative efficiency and of the relative background density are arbitrary.}
\label{SUPPeffbkgpara}
\end{figure}

\ifx\mcitethebibliography\mciteundefinedmacro
\PackageError{LHCb.bst}{mciteplus.sty has not been loaded}
{This bibstyle requires the use of the mciteplus package.}\fi
\providecommand{\href}[2]{#2}

\newpage
\centerline{\large\bf LHCb collaboration}
\begin{flushleft}
\small
R.~Aaij$^{38}$, 
B.~Adeva$^{37}$, 
M.~Adinolfi$^{46}$, 
A.~Affolder$^{52}$, 
Z.~Ajaltouni$^{5}$, 
S.~Akar$^{6}$, 
J.~Albrecht$^{9}$, 
F.~Alessio$^{38}$, 
M.~Alexander$^{51}$, 
S.~Ali$^{41}$
% ...

{\footnotesize \it
% ...
$ ^{5}$Clermont Universit\'{e}, Universit\'{e} Blaise Pascal, CNRS/IN2P3, LPC, Clermont-Ferrand, France\\
$ ^{6}$CPPM, Aix-Marseille Universit\'{e}, CNRS/IN2P3, Marseille, France\\
% ...
$ ^{9}$Fakult\"{a}t Physik, Technische Universit\"{a}t Dortmund, Dortmund, Germany\\
% ...
$ ^{37}$Universidad de Santiago de Compostela, Santiago de Compostela, Spain\\
$ ^{38}$European Organization for Nuclear Research (CERN), Geneva, Switzerland\\
% ...
$ ^{41}$Nikhef National Institute for Subatomic Physics, Amsterdam, The Netherlands\\
% ...
$ ^{46}$H.H. Wills Physics Laboratory, University of Bristol, Bristol, United Kingdom\\
% ...
$ ^{51}$School of Physics and Astronomy, University of Glasgow, Glasgow, United Kingdom\\
$ ^{52}$Oliver Lodge Laboratory, University of Liverpool, Liverpool, United Kingdom\\
% ...
}
\end{flushleft}

\end{document}